\documentclass[nonacm=true, acmsmall]{acmart}
\usepackage[utf8]{inputenc}
\usepackage[T1]{fontenc}

\newcommand{\packageGraphicx}{\usepackage{graphicx}}
\newcommand{\packageHyperref}{\usepackage{hyperref}}
\newcommand{\renewrmdefault}{\renewcommand{\rmdefault}{ptm}}
\newcommand{\packageRelsize}{\usepackage{relsize}}
\newcommand{\packageMathabx}{\usepackage{mathabx}}
\newcommand{\packageWasysym}{
  \let\leftmoon\relax \let\rightmoon\relax \let\fullmoon\relax \let\newmoon\relax \let\diameter\relax
  \usepackage{wasysym}}
\newcommand{\packageTextcomp}{\usepackage{textcomp}}
\newcommand{\packageFramed}{\usepackage{framed}}
\newcommand{\packageHyphenat}{\usepackage[htt]{hyphenat}}
\newcommand{\packageColor}{\usepackage[usenames,dvipsnames]{color}}
\newcommand{\doHypersetup}{\hypersetup{bookmarks=true,bookmarksopen=true,bookmarksnumbered=true}}
\newcommand{\packageTocstyle}{\IfFileExists{tocstyle.sty}{\usepackage{tocstyle}\usetocstyle{standard}}{}}
\newcommand{\packageCJK}{\IfFileExists{CJK.sty}{\usepackage{CJK}}{}}
\renewcommand\packageColor\relax
\renewcommand\packageTocstyle\relax
\renewcommand\packageMathabx{\ifx\bigtimes\undefined \usepackage{mathabx} \else \relax \fi}

\renewcommand{\renewrmdefault}{}

\packageGraphicx
\packageHyperref
\renewrmdefault
\packageRelsize
\packageMathabx
\packageWasysym
\packageTextcomp
\packageFramed
\packageHyphenat
\packageColor
\doHypersetup
\packageTocstyle
\packageCJK

\newcommand{\sectionNewpage}{}

\newcommand{\preDoc}{}
\newcommand{\postDoc}{}

\newcommand{\BookRef}[2]{\emph{#2}}
\newcommand{\ChapRef}[2]{\SecRef{#1}{#2}}
\newcommand{\SecRef}[2]{section~#1}
\newcommand{\PartRef}[2]{part~#1}
\newcommand{\BookRefUC}[2]{\BookRef{#1}{#2}}
\newcommand{\ChapRefUC}[2]{\SecRefUC{#1}{#2}}
\newcommand{\SecRefUC}[2]{Section~#1}
\newcommand{\PartRefUC}[2]{Part~#1}

\newcommand{\BookRefLocal}[3]{\hyperref[#1]{\BookRef{#2}{#3}}}
\newcommand{\ChapRefLocal}[3]{\hyperref[#1]{\ChapRef{#2}{#3}}}
\newcommand{\SecRefLocal}[3]{\hyperref[#1]{\SecRef{#2}{#3}}}
\newcommand{\PartRefLocal}[3]{\hyperref[#1]{\PartRef{#2}{#3}}}
\newcommand{\BookRefLocalUC}[3]{\hyperref[#1]{\BookRefUC{#2}{#3}}}
\newcommand{\ChapRefLocalUC}[3]{\hyperref[#1]{\ChapRefUC{#2}{#3}}}
\newcommand{\SecRefLocalUC}[3]{\hyperref[#1]{\SecRefUC{#2}{#3}}}
\newcommand{\PartRefLocalUC}[3]{\hyperref[#1]{\PartRefUC{#2}{#3}}}

\newcommand{\BookRefUN}[1]{\BookRef{}{#1}}

\newcommand{\SecRefUN}[1]{``#1''}

\newcommand{\BookRefLocalUN}[2]{\hyperref[#1]{\BookRefUN{#2}}}

\newcommand{\SecRefLocalUN}[2]{\hyperref[#1]{\SecRefUN{#2}}}

\newcommand{\SectionNumberLink}[2]{\hyperref[#1]{#2}}

\newcommand{\Scribtexttt}[1]{{\texttt{#1}}}

\newcommand{\intextrgbcolor}[2]{\textcolor[rgb]{#1}{#2}}

\newcommand{\inrgbcolorbox}[2]{{\fboxrule=0pt\fboxsep=0pt\colorbox[rgb]{#1}{#2}}}

\newcommand{\Smaller}[1]{\textsmaller{#1}}

\newcommand{\planetName}[1]{PLane\hspace{-0.1ex}T}

\newcommand{\Stttextmore}{{\fontencoding{T1}\selectfont>}}
\newcommand{\Stttextless}{{\fontencoding{T1}\selectfont<}}

\makeatletter
\def\empty@finalstrut#1{%
  \unskip\ifhmode\nobreak\fi\vrule\@width\z@\@height\z@\@depth\z@}
\def\no@strut{\global\setbox\@arstrutbox\hbox{%
    \vrule \@height\z@
           \@depth\z@
           \@width\z@}%
    \gdef\@endpbox{\empty@finalstrut\@arstrutbox\par\egroup\hfil}%
}%
\def\yes@strut{\global\setbox\@arstrutbox\hbox{%
    \vrule \@height\arraystretch \ht\strutbox
           \@depth\arraystretch \dp\strutbox
           \@width\z@}%
    \gdef\@endpbox{\@finalstrut\@arstrutbox\par\egroup\hfil}%
}%
\def\@mkpream#1{\@firstamptrue\@lastchclass6
  \let\@preamble\@empty\def\empty@preamble{\add@ins}%
  \let\protect\@unexpandable@protect
  \let\@sharp\relax\let\add@ins\relax
  \let\@startpbox\relax\let\@endpbox\relax
  \@expast{#1}%
  \expandafter\@tfor \expandafter
    \@nextchar \expandafter:\expandafter=\reserved@a\do
       {\@testpach\@nextchar
    \ifcase \@chclass \@classz \or \@classi \or \@classii \or \@classiii
      \or \@classiv \or\@classv \fi\@lastchclass\@chclass}%
  \ifcase \@lastchclass \@acol
      \or \or \@preamerr \@ne\or \@preamerr \tw@\or \or \@acol \fi}
\def\@addamp{%
  \if@firstamp
    \@firstampfalse
    \edef\empty@preamble{\add@ins}%
  \else
    \edef\@preamble{\@preamble &}%
    \edef\empty@preamble{\expandafter\noexpand\empty@preamble &\add@ins}%
  \fi}
\newif\iftw@hlines \tw@hlinesfalse
\def\@xhline{\ifx\reserved@a\hline
               \tw@hlinestrue
             \else\ifx\reserved@a\Hline
               \tw@hlinestrue
             \else
               \tw@hlinesfalse
             \fi\fi
      \iftw@hlines
        \aftergroup\do@after
      \fi
      \ifnum0=`{\fi}%
}
\def\do@after{\emptyrow[\the\doublerulesep]}
\def\emptyrow{\noalign\bgroup\@ifnextchar[\@emptyrow{\@emptyrow[\z@]}}
\def\@emptyrow[#1]{\no@strut\gdef\add@ins{\vrule \@height\z@ \@depth#1 \@width\z@}\egroup%
\empty@preamble\\
\noalign{\yes@strut\gdef\add@ins{\vrule \@height\z@ \@depth\z@ \@width\z@}}%
}
\def\tabrow#1{\noalign\bgroup\@ifnextchar[{\@tabrow{#1}}{\@tabrow{#1}[]}}
\def\@tabrow#1[#2]{\no@strut\egroup#1\ifx.#2.\\\else\\[#2]\fi\noalign{\yes@strut}}
\def\endpltstabular{\crcr\egroup\egroup \egroup}
\expandafter \let \csname endpltstabular*\endcsname = \endpltstabular
\def\pltstabular{\let\@halignto\@empty\@pltstabular}
\@namedef{pltstabular*}#1{\def\@halignto{to#1}\@pltstabular}
\def\@pltstabular{\leavevmode \bgroup \let\@acol\@tabacol
   \let\@classz\@tabclassz
   \let\@classiv\@tabclassiv \let\\\@tabularcr\@stabarray}
\def\@stabarray{\m@th\@ifnextchar[\@sarray{\@sarray[c]}}
\def\@sarray[#1]#2{%
  \bgroup
  \setbox\@arstrutbox\hbox{%
    \vrule \@height\arraystretch\ht\strutbox
           \@depth\arraystretch \dp\strutbox
           \@width\z@}%
  \@mkpream{#2}%
  \edef\@preamble{%
    \ialign \noexpand\@halignto
      \bgroup \@arstrut \@preamble \tabskip\z@skip \cr}%
  \let\@startpbox\@@startpbox \let\@endpbox\@@endpbox
  \let\tabularnewline\\%
    \let\@sharp##%
    \set@typeset@protect
    \lineskip\z@skip\baselineskip\z@skip
    \@preamble}

\makeatother

\newenvironment{bigtabular}{\begin{pltstabular}}{\end{pltstabular}}

\newlength{\stabLeft}
\newcommand{\bigtableleftpad}{\hspace{\stabLeft}}
\newcommand{\atItemizeStart}[0]{\addtolength{\stabLeft}{\labelsep}
                                \addtolength{\stabLeft}{\labelwidth}}

\newenvironment{SingleColumn}{\begin{list}{}{\topsep=0pt\partopsep=0pt%
\listparindent=0pt\itemindent=0pt\labelwidth=0pt\leftmargin=0pt\rightmargin=0pt%
\itemsep=0pt\parsep=0pt}\item}{\end{list}}

\newcommand{\Sendabbrev}[1]{#1\@}

\newenvironment{SInsetFlow}{\begin{quote}}{\end{quote}}

\newcommand{\SCodePreSkip}{\vskip\abovedisplayskip}
\newcommand{\SCodePostSkip}{\vskip\belowdisplayskip}
\newenvironment{SCodeFlow}{\SCodePreSkip\begin{list}{}{\topsep=0pt\partopsep=0pt%
\listparindent=0pt\itemindent=0pt\labelwidth=0pt\leftmargin=2ex\rightmargin=2ex%
\itemsep=0pt\parsep=0pt}\item}{\end{list}\SCodePostSkip}

\newcommand{\SVInsetPreSkip}{\vskip\abovedisplayskip}
\newcommand{\SVInsetPostSkip}{\vskip\belowdisplayskip}

\newenvironment{SCentered}{\begin{trivlist}\item \centering}{\end{trivlist}}

\newcommand{\titleAndVersionAndAuthors}[3]{\title{#1\\{\normalsize \SVersionBefore{}#2}}\author{#3}\maketitle}

\newcommand{\titleAndEmptyVersionAndAuthors}[3]{\title{#1}\author{#3}\maketitle}

\newcommand{\SAuthor}[1]{#1}
\newcommand{\SAuthorSep}[1]{\qquad}
\newcommand{\SVersionBefore}[1]{Version }

\newcommand{\SNumberOfAuthors}[1]{}

\let\SOriginalthesubsection\thesubsection
\let\SOriginalthesubsubsection\thesubsubsection

\newcommand{\Ssection}[2]{\section[#1]{#2}\let\thesubsection\SOriginalthesubsection}
\newcommand{\Ssubsection}[2]{\subsection[#1]{#2}\let\thesubsubsection\SOriginalthesubsubsection}

\newcommand{\Ssectionstar}[1]{\section*{#1}\renewcommand*\thesubsection{\arabic{subsection}}\setcounter{subsection}{0}}

\newcommand{\Ssubsubsectionstar}[1]{\subsubsection*{#1}}

\newcommand{\Ssectionstarx}[2]{\Ssectionstar{#2}\phantomsection\addcontentsline{toc}{section}{#1}}

\newcommand{\Ssubsubsectionstarx}[2]{\Ssubsubsectionstar{#2}\phantomsection\addcontentsline{toc}{subsubsection}{#1}}

\newcounter{GrouperTemp}

\newcommand{\notitlesection}{\vspace{2ex}\phantomsection\noindent}

\newcommand{\Sincsubsection}{\stepcounter{subsection}}
\newcommand{\Sincsubsubsection}{\stepcounter{subsubsection}}

\newcommand{\Snolinkurl}[1]{\nolinkurl{#1}}

\newcommand{\SAuthorinfo}[4]{#1}
\newcommand{\SAuthorPlace}[1]{#1}
\newcommand{\SAuthorEmail}[1]{#1}

\newcommand{\SConferenceInfo}[2]{}
\newcommand{\SCopyrightYear}[1]{}
\newcommand{\SCopyrightData}[1]{}
\newcommand{\Sdoi}[1]{}

\newcommand{\SCategory}[3]{}
\newcommand{\SCategoryPlus}[4]{}
\newcommand{\STerms}[1]{}
\newcommand{\SKeywords}[1]{}

\newcommand{\AutobibLink}[1]{\color{ACMPurple}{#1}}
\usepackage{ccaption}

\makeatletter
\@ifundefined{belowcaptionskip}{}{}
\makeatother

\newcommand{\Legend}[1]{~

                        \hrule width \hsize height .33pt
                        \vspace{4pt}
                        \legend{#1}}

\newcommand{\FigureTarget}[2]{#1}
\newcommand{\FigureRef}[2]{#1}

\newlength{\FigOrigskip}
\FigOrigskip=\parskip

\newcommand{\FigureSetRef}{\refstepcounter{figure}}

\newenvironment{Figure}{\begin{figure}\FigureSetRef}{\end{figure}}
\newenvironment{FigureMulti}{\begin{figure*}[t!p]\FigureSetRef}{\end{figure*}}

\newenvironment{Centerfigure}{\begin{Xfigure}\centering\item}{\end{Xfigure}}
\newenvironment{Leftfigure}{\begin{Xfigure}\item}{\end{Xfigure}}

\newenvironment{Xfigure}{\begin{list}{}{\leftmargin=0pt\topsep=0pt\parsep=\FigOrigskip\partopsep=0pt}}{\end{list}}

\newenvironment{FigureInside}{}{}

\newcommand{\Centertext}[1]{\begin{center}#1\end{center}}

\newcommand{\SColorize}[2]{\color{#1}{#2}}

\newcommand{\SHyphen}[1]{#1}

\newcommand{\inColor}[2]{{\SHyphen{\Scribtexttt{\SColorize{#1}{#2}}}}}
\definecolor{PaleBlue}{rgb}{0.90,0.90,1.0}
\definecolor{LightGray}{rgb}{0.90,0.90,0.90}
\definecolor{CommentColor}{rgb}{0.76,0.45,0.12}
\definecolor{ParenColor}{rgb}{0.52,0.24,0.14}
\definecolor{IdentifierColor}{rgb}{0.15,0.15,0.50}
\definecolor{ResultColor}{rgb}{0.0,0.0,0.69}
\definecolor{ValueColor}{rgb}{0.13,0.55,0.13}
\definecolor{OutputColor}{rgb}{0.59,0.00,0.59}

\newcommand{\RktCmt}[1]{\inColor{CommentColor}{#1}}
\newcommand{\RktPn}[1]{\inColor{ParenColor}{#1}}

\newcommand{\RktSym}[1]{\inColor{IdentifierColor}{#1}}

\newcommand{\RktVal}[1]{\inColor{ValueColor}{#1}}

\newcommand{\RktModLink}[1]{\inColor{blue}{#1}}

\newcommand{\RktMeta}[1]{\inColor{IdentifierColor}{#1}}
\newcommand{\RktMod}[1]{\inColor{black}{#1}}

\newcommand{\RktVarCol}[1]{\inColor{IdentifierColor}{#1}}
\newcommand{\RktVar}[1]{{\RktVarCol{\textsl{#1}}}}

\newenvironment{RktBlk}{}{}

\newcommand{\RBackgroundLabel}[1]{}

\newcommand{\NoteBox}[1]{\footnote{#1}}
\newcommand{\NoteContent}[1]{#1}

\newcommand{\FootnoteRef}[1]{}
\newcommand{\FootnoteTarget}[1]{}

\newcommand{\FootnoteBlockContent}[1]{}

\newenvironment{AutoBibliography}{\begin{small}}{\end{small}}
\newcommand{\Autobibentry}[1]{\hspace{0.05\linewidth}\parbox[t]{0.95\linewidth}{\parindent=-0.05\linewidth#1\vspace{1.0ex}}}

\usepackage{calc}
\newlength{\ABcollength}

\newcommand{\Autobibref}[1]{#1}

\providecommand{\AutobibLink}[1]{#1}

\renewcommand{\titleAndVersionAndAuthors}[3]{\title{#1}#3\maketitle}
\renewcommand{\titleAndEmptyVersionAndAuthors}[3]{\titleAndVersionAndAuthors{#1}{#2}{#3}}

\def\SAuthor#1{\SAutoAuthor#1\SAutoAuthorDone{#1}}
\def\SAutoAuthorDone#1{}
\def\SAutoAuthor{\futurelet\next\SAutoAuthorX}
\def\SAutoAuthorX{\ifx\next\SAuthorinfo \let\Snext\relax \else \let\Snext\SToAuthorDone \fi \Snext}
\def\SToAuthorDone{\futurelet\next\SToAuthorDoneX}
\def\SToAuthorDoneX#1{\ifx\next\SAutoAuthorDone \let\Snext\SAddAuthorInfo \else \let\Snext\SToAuthorDone \fi \Snext}
\newcommand{\SAddAuthorInfo}[1]{\SAuthorinfo{#1}{}{}}

\renewcommand{\SAuthorinfo}[4]{\author{#1}{#2}{#3}{#4}}
\renewcommand{\SAuthorSep}[1]{}

\renewcommand{\SAuthorPlace}[1]{\affiliation{#1}}
\renewcommand{\SAuthorEmail}[1]{\email{#1}}

\renewcommand{\SConferenceInfo}[2]{\conferenceinfo{#1}{#2}}
\renewcommand{\SCopyrightYear}[1]{\copyrightyear{#1}}
\renewcommand{\SCopyrightData}[1]{\copyrightdata{#1}}

\renewcommand{\SCategory}[3]{\category{#1}{#2}{#3}}
\renewcommand{\SCategoryPlus}[4]{\category{#1}{#2}{#3}[#4]}
\renewcommand{\STerms}[1]{\terms{#1}}
\renewcommand{\SKeywords}[1]{\keywords{#1}}

\renewcommand{\SColorize}[2]{\color{black}{#2}}

\newcommand{\identity}[1]{#1}
\newcommand{\goAway}[1]{}

\usepackage{booktabs}
\usepackage{multirow}
\usepackage{color}
\definecolor{gold}{rgb}{0.83, 0.69, 0.22}
\definecolor{darkred}{rgb}{0.55, 0.0, 0.0}

\renewenvironment{SCodeFlow}{\begin{small}}{\par\end{small}}
\renewenvironment{RktBlk}{\begin{small}}{\par\end{small}}

\begin{document}
\preDoc

\begin{abstract} While high{-}level languages come with significant readability and
maintainability benefits, their performance remains difficult to
predict. For example, programmers may unknowingly use language features
inappropriately, which cause their programs to run slower than expected. To
address this issue, we introduce \textit{feature{-}specific profiling}, a
technique that reports performance costs in terms of linguistic
constructs. Festure{-}specific profilers help programmers find expensive uses
of specific features of their language. We describe the architecture of a
profiler that implements our approach, explain prototypes of the profiler
for two languages with different characteristics and
implementation strategies, and provide empirical evidence for the
approach{'}s general usefulness as a performance debugging tool.\end{abstract}\titleAndEmptyVersionAndAuthors{Feature{-}Specific Profiling}{}{\SNumberOfAuthors{4}\SAuthor{\SAuthorinfo{Leif Andersen}{}{\SAuthorPlace{\department{PLT}\department{CCIS}\institution{PLT @ Northeastern University}\city{Boston}\state{Massachusetts}\country{United States of America}}}{\SAuthorEmail{leif@ccs.neu.edu}}}\SAuthorSep{}\SAuthor{\SAuthorinfo{Vincent St{-}Amour}{}{\SAuthorPlace{\department{PLT}\department{Department of Electrical Engineering and Computer Science}\institution{PLT @ Northwestern University}\city{Evanston}\state{Illinois}\country{United States of America}}}{\SAuthorEmail{stamourv@eecs.northwestern.edu}}}\SAuthorSep{}\SAuthor{\SAuthorinfo{Jan Vitek}{}{\SAuthorPlace{\institution{Northeastern University}}\SAuthorPlace{\institution{Czech Technical University}}}{\SAuthorEmail{j.vitek@neu.edu}}}\SAuthorSep{}\SAuthor{\SAuthorinfo{Matthias Felleisen}{}{\SAuthorPlace{\department{PLT}\department{CCIS}\institution{PLT @ Northeastern University}\city{Boston}\state{Massachusetts}\country{United States of America}}}{\SAuthorEmail{matthias@ccs.neu.edu}}}}
\label{t:x28part_x22Featurex2dSpecificx5fProfilingx22x29}

\noindent 

\noindent 

\noindent

\sectionNewpage

\Ssection{Profiling with Actionable Advice}{Profiling with Actionable Advice}\label{t:x28part_x22Profilingx5fwithx5fActionablex5fAdvicex22x29}

When programs take too long to run, programmers tend to
reach for profilers to diagnose the problem. Most profilers
attribute the run{-}time costs during a program{'}s execution
to \textit{cost centers} such as function calls or statements
in source code. Then they rank all of a program{'}s cost
centers in order to identify and eliminate key
bottlenecks\Autobibref{~(\hyperref[t:x28autobib_x22Gene_Mx2e_AmdahlValidity_of_the_Single_Processor_Approach_to_Achieving_Large_Scale_Computing_CapabilitiesIn_Procx2e_Spring_Joint_Computer_Conference1967x22x29]{\AutobibLink{Amdahl}} \hyperref[t:x28autobib_x22Gene_Mx2e_AmdahlValidity_of_the_Single_Processor_Approach_to_Achieving_Large_Scale_Computing_CapabilitiesIn_Procx2e_Spring_Joint_Computer_Conference1967x22x29]{\AutobibLink{1967}})}. If such a profile helps programmers optimize
their code, we call it \textit{actionable} because
it points to inefficiencies that can be remedied with
changes to the program.

The advice of conventional profilers fails the actionable
standard in some situations, mostly because their conventional
choice of cost centers{---}e.g. lines or functions{---}does not
match programming language concepts. For example, their
advice is misleading in a context where a performance
problem has a unique cause that manifests itself as a cost
at many locations. Similarly, when a language allows the
encapsulation of syntactic features in libraries,
conventional profilers often misjudge the source of related
performance bottlenecks.

\goAway{\Autobibref{~(\hyperref[t:x28autobib_x22Vincent_Stx2dAmourx2c_Leif_Andersenx2c_and_Matthias_FelleisenFeaturex2dspecific_ProfilingIn_Procx2e_International_Conference_on_Compiler_Construction2015httpsx3ax2fx2fdoix2eorgx2f10x2e1007x2f978x2d3x2d662x2d46663x2d6x5f3x22x29]{\AutobibLink{St{-}Amour et al\Sendabbrev{.}}} \hyperref[t:x28autobib_x22Vincent_Stx2dAmourx2c_Leif_Andersenx2c_and_Matthias_FelleisenFeaturex2dspecific_ProfilingIn_Procx2e_International_Conference_on_Compiler_Construction2015httpsx3ax2fx2fdoix2eorgx2f10x2e1007x2f978x2d3x2d662x2d46663x2d6x5f3x22x29]{\AutobibLink{2015}})}}
\textit{Feature{-}specific profiling} (FSP) addresses these issues
with the introduction of linguistic features as cost
centers. By {``}features{''} we specifically mean syntactic
constructs with operational costs: functions and linguistic
elements, such as pattern matching, keyword{-}based function
calls, or behavioral contracts. This paper, an expansion of
St{-}Amour et al.{'}s (2015) original report on this idea,
explains its principles, describes how to turn them into
reasonably practical prototypes, and presents evaluation
results. While the original paper introduced the idea and
used a Racket\Autobibref{~(\hyperref[t:x28autobib_x22Matthew_Flatt_and_PLTReferencex3a_RacketPLT_Incx2ex2c_PLTx2dTRx2d2010x2d12010httpx3ax2fx2fracketx2dlangx2eorgx2ftr1x2fx22x29]{\AutobibLink{Flatt and PLT}} \hyperref[t:x28autobib_x22Matthew_Flatt_and_PLTReferencex3a_RacketPLT_Incx2ex2c_PLTx2dTRx2d2010x2d12010httpx3ax2fx2fracketx2dlangx2eorgx2ftr1x2fx22x29]{\AutobibLink{2010}})} prototype to evaluate its
effectiveness, this paper confirms the idea  with a prototype
for the R programming language\Autobibref{~(\hyperref[t:x28autobib_x22R_Development_Core_TeamR_Language_DefinitionR_Development_Core_Teamx2c_3x2e3x2e12016httpx3ax2fx2fwebx2emitx2eedux2fx7erx2fcurrentx2farchx2famd64x5flinux26x2flibx2fRx2fdocx2fmanualx2fRx2dlangx2epdfx22x29]{\AutobibLink{R Development Core Team}} \hyperref[t:x28autobib_x22R_Development_Core_TeamR_Language_DefinitionR_Development_Core_Teamx2c_3x2e3x2e12016httpx3ax2fx2fwebx2emitx2eedux2fx7erx2fcurrentx2farchx2famd64x5flinux26x2flibx2fRx2fdocx2fmanualx2fRx2dlangx2epdfx22x29]{\AutobibLink{2016}})}. The
creation of this second prototype confirms the validity of
feature{-}specific profiling beyond Racket. It also enlarges
the body of features for which programmers may benefit from
a feature{-}specific profiler.

In summary, this expansion of the original conference paper
into an archival one provides a definition for language
features, feature instances, and feature{-}specific profiling,
explains the components that make up a feature{-}specific
profiler, describes two ingredients to make the idea truly
practical, and evaluates prototypes for the actionability of
its results, implementation effort, and run{-}time performance
in the Racket and R contexts.

\sectionNewpage

\Ssection{Linguistic Features and their profiles}{Linguistic Features and their profiles}\label{t:x28part_x22featurex22x29}

An FSP attributes execution costs to instances of linguistic
features, that is, any construct that has both a syntactic presence
in code and a run{-}time cost that can be detected by inspecting
the language{'}s call stack.
Because the computation
associated with a particular instance of a feature can be
dispersed throughout a program, this view can provide
actionable information when a traditional profiler falls
short. To collect this information an
FSP comes with a slightly different architecture than a
traditional profiler.  This section gives an overview of our approach.

\Ssubsection{Linguistic Features}{Linguistic Features}\label{t:x28part_x22featurex2ddescx22x29}

We consider a language feature to be any syntactic construct
that has an operational stack{-}based cost, such as a function calling protocol,
looping constructs, or dynamic dispatch for objects. The
features that a program uses are orthogonal to the actual
algorithm it implements. For example, a program that
implements a list traversal algorithm may use loops,
comprehensions, or recursive functions. While the algorithms
and resulting values are the same in all three cases, their implementation
may have different performance costs.

The goal of feature{-}specific profiling is to find uses of features that are
expensive and not expensive algorithms.  Knowing which features are
expensive in a program is not sufficient for programmers to know how to
speed up their code. An expensive feature may appear in many places, some
innocuous to performance, and may be difficult to remove from a program
entirely. More precisely, a feature may not generally be expensive, but some
uses may be inappropriate. For example, dynamic dispatch is not usually a
critical cost component, but might be when used in a hot loop for a
mega{-}morphic method. An FSP therefore points programmers to individual
feature instances. As a concrete example, while all dynamic dispatch calls
make up a single feature, every single use of dynamic dispatch is a unique
feature instance, and one of them may come with a significant performance
cost.

The cost of feature instances does not necessarily have a
direct one{-}to{-}one mapping to their location in source code.
One way this happens is when the cost centers of one feature
may intersect with the cost centers of another feature. For
example, a concurrent program may wish to attribute program
costs in terms of its individual threads rather than the
functions run by the threads. A traditional profiler
correctly identifies the functions being run, but it fails
to properly attribute them to their underlying threads. We
call these \textit{conflated costs}. An FSP properly attaches
such costs to their appropriate threads.

In additional to having conflated costs, linguistic features
may also come with non{-}local, \textit{dispersed costs}, that
is, costs that manifest themselves at a different point than
their syntactic location in code. Continuing the previous
example, dynamic dispatch is a language construct with
non{-}local costs. One useful way to measure dynamic dispatch
is to attribute its costs to a specific method, rather than
just its call sites. Accounting costs this way disambiguates
time spent in the program{'}s algorithm versus time spent
dispatching. Traditional profilers attribute the
dispatch cost only to the call site, which is
misleading and suggests to programmers that the algorithm
itself is costly, rather than the dispatch mechanism. An FSP
solves this problem by attributing the cost of method calls
to their declarations. Programmers may be able to use this information
to avoid costly uses of dynamic dispatch, without
having to change their underlying algorithm.

\Ssubsection{An Example Feature Profile}{An Example Feature Profile}\label{t:x28part_x22userx2dprofx22x29}

\begin{Figure}\begin{Centerfigure}\begin{FigureInside}\begin{SInsetFlow}\begin{bigtabular}{@{\bigtableleftpad}l@{}l@{}l@{}l@{}l@{}l@{}l@{}l@{}l@{}}
\hbox{ } &
\hbox{ } &
\hbox{ } &
\begin{RktBlk}\begin{tabular}[c]{@{}l@{}}
\hbox{\mbox{\hphantom{\Scribtexttt{x}}}\Smaller{\Scribtexttt{1}}} \\
\hbox{\mbox{\hphantom{\Scribtexttt{x}}}\Smaller{\Scribtexttt{2}}} \\
\hbox{\mbox{\hphantom{\Scribtexttt{x}}}\Smaller{\Scribtexttt{3}}} \\
\hbox{\mbox{\hphantom{\Scribtexttt{x}}}\Smaller{\Scribtexttt{4}}} \\
\hbox{\mbox{\hphantom{\Scribtexttt{x}}}\Smaller{\Scribtexttt{5}}} \\
\hbox{\mbox{\hphantom{\Scribtexttt{x}}}\Smaller{\Scribtexttt{6}}} \\
\hbox{\mbox{\hphantom{\Scribtexttt{x}}}\Smaller{\Scribtexttt{7}}} \\
\hbox{\mbox{\hphantom{\Scribtexttt{x}}}\Smaller{\Scribtexttt{8}}} \\
\hbox{\mbox{\hphantom{\Scribtexttt{x}}}\Smaller{\Scribtexttt{9}}} \\
\hbox{\Smaller{\Scribtexttt{10}}} \\
\hbox{\Smaller{\Scribtexttt{11}}}\end{tabular}\end{RktBlk} &
\hbox{ } &
\hbox{ } &
\hbox{ } &
\hbox{ } &
\begin{RktBlk}\begin{tabular}[c]{@{}l@{}}
\hbox{\RktModLink{\RktMod{\#lang}}\mbox{\hphantom{\Scribtexttt{x}}}\RktSym{racket}} \\
\hbox{\RktPn{(}\RktSym{define}\mbox{\hphantom{\Scribtexttt{x}}}\RktPn{(}\RktSym{fizzbuzz}\mbox{\hphantom{\Scribtexttt{x}}}\RktSym{n}\RktPn{)}} \\
\hbox{\mbox{\hphantom{\Scribtexttt{xx}}}\RktPn{(}\RktSym{for}\mbox{\hphantom{\Scribtexttt{x}}}\RktPn{(}\RktPn{[}\RktSym{i}\mbox{\hphantom{\Scribtexttt{x}}}\RktPn{(}\RktSym{range}\mbox{\hphantom{\Scribtexttt{x}}}\RktSym{n}\RktPn{)}\RktPn{]}\RktPn{)}} \\
\hbox{\mbox{\hphantom{\Scribtexttt{xxxx}}}\RktPn{(}\RktSym{cond}} \\
\hbox{\mbox{\hphantom{\Scribtexttt{xxxxxx}}}\RktPn{[}\RktPn{(}\RktSym{divisible}\mbox{\hphantom{\Scribtexttt{x}}}\RktSym{i}\mbox{\hphantom{\Scribtexttt{x}}}\RktVal{15}\RktPn{)}\mbox{\hphantom{\Scribtexttt{x}}}\RktPn{(}\RktSym{printf}\mbox{\hphantom{\Scribtexttt{x}}}\RktVal{"FizzBuzz{\char`\\}n"}\RktPn{)}\RktPn{]}} \\
\hbox{\mbox{\hphantom{\Scribtexttt{xxxxxx}}}\RktPn{[}\RktPn{(}\RktSym{divisible}\mbox{\hphantom{\Scribtexttt{x}}}\RktSym{i}\mbox{\hphantom{\Scribtexttt{x}}}\RktVal{5}\RktPn{)}\mbox{\hphantom{\Scribtexttt{xx}}}\RktPn{(}\RktSym{printf}\mbox{\hphantom{\Scribtexttt{x}}}\RktVal{"Buzz{\char`\\}n"}\RktPn{)}\RktPn{]}} \\
\hbox{\mbox{\hphantom{\Scribtexttt{xxxxxx}}}\RktPn{[}\RktPn{(}\RktSym{divisible}\mbox{\hphantom{\Scribtexttt{x}}}\RktSym{i}\mbox{\hphantom{\Scribtexttt{x}}}\RktVal{3}\RktPn{)}\mbox{\hphantom{\Scribtexttt{xx}}}\RktPn{(}\RktSym{printf}\mbox{\hphantom{\Scribtexttt{x}}}\RktVal{"Fizz{\char`\\}n"}\RktPn{)}\RktPn{]}} \\
\hbox{\mbox{\hphantom{\Scribtexttt{xxxxxx}}}\RktPn{[}\RktSym{else}\mbox{\hphantom{\Scribtexttt{xxxxxxxxxxxxx}}}\RktPn{(}\RktSym{printf}\mbox{\hphantom{\Scribtexttt{x}}}\RktVal{"$\sim$a{\char`\\}n"}\mbox{\hphantom{\Scribtexttt{x}}}\RktSym{i}\RktPn{)}\RktPn{]}\RktPn{)}\RktPn{)}\RktPn{)}} \\
\hbox{\mbox{\hphantom{\Scribtexttt{x}}}} \\
\hbox{\RktPn{(}\RktSym{feature{-}profile}} \\
\hbox{\mbox{\hphantom{\Scribtexttt{x}}}\RktPn{(}\RktSym{fizzbuzz}\mbox{\hphantom{\Scribtexttt{x}}}\RktVal{10000000}\RktPn{)}\RktPn{)}}\end{tabular}\end{RktBlk}\end{bigtabular}\end{SInsetFlow}

\noindent \identity{\vspace{1em} \hrule \vspace{1em}\begin{small}}

\noindent \begin{SingleColumn}\Scribtexttt{Feature Report}

\Scribtexttt{(Feature times may sum to more or less than 100\% of the total running time)}

\Scribtexttt{}\mbox{\hphantom{\Scribtexttt{x}}}

\Scribtexttt{Output accounts for 68{\hbox{\texttt{.}}}22\% of running time}

\Scribtexttt{}\mbox{\hphantom{\Scribtexttt{xxxxxx}}}\Scribtexttt{(5580 / 8180 ms)}

\Scribtexttt{}\mbox{\hphantom{\Scribtexttt{xx}}}\Scribtexttt{4628 ms {\hbox{\texttt{:}}} fizzbuzz{\hbox{\texttt{.}}}rkt{\hbox{\texttt{:}}}8{\hbox{\texttt{:}}}24}

\Scribtexttt{}\mbox{\hphantom{\Scribtexttt{xx}}}\Scribtexttt{564 ms {\hbox{\texttt{:}}} fizzbuzz{\hbox{\texttt{.}}}rkt{\hbox{\texttt{:}}}7{\hbox{\texttt{:}}}24}

\Scribtexttt{}\mbox{\hphantom{\Scribtexttt{xx}}}\Scribtexttt{232 ms {\hbox{\texttt{:}}} fizzbuzz{\hbox{\texttt{.}}}rkt{\hbox{\texttt{:}}}6{\hbox{\texttt{:}}}24}

\Scribtexttt{}\mbox{\hphantom{\Scribtexttt{xx}}}\Scribtexttt{156 ms {\hbox{\texttt{:}}} fizzbuzz{\hbox{\texttt{.}}}rkt{\hbox{\texttt{:}}}5{\hbox{\texttt{:}}}24}

\Scribtexttt{}\mbox{\hphantom{\Scribtexttt{x}}}

\Scribtexttt{Generic sequences account for 11{\hbox{\texttt{.}}}78\% of running time}

\Scribtexttt{}\mbox{\hphantom{\Scribtexttt{xxxxxx}}}\Scribtexttt{(964 / 8180 ms)}

\Scribtexttt{}\mbox{\hphantom{\Scribtexttt{xx}}}\Scribtexttt{964 ms {\hbox{\texttt{:}}} fizzbuzz{\hbox{\texttt{.}}}rkt{\hbox{\texttt{:}}}3{\hbox{\texttt{:}}}11}\end{SingleColumn}

\noindent \identity{\end{small}}\end{FigureInside}\end{Centerfigure}

\Centertext{\Legend{\FigureTarget{\label{t:x28counter_x28x22figurex22_x22fizzbuzzx22x29x29}\textsf{Fig.}~\textsf{1}. }{t:x28counter_x28x22figurex22_x22fizzbuzzx22x29x29}\textsf{Feature profile for FizzBuzz}}}\end{Figure}

To illustrate the workings of an FSP, figure~\hyperref[t:x28counter_x28x22figurex22_x22fizzbuzzx22x29x29]{\FigureRef{1}{t:x28counter_x28x22figurex22_x22fizzbuzzx22x29x29}}
presents a concrete example, the Fizzbuzz\NoteBox{\NoteContent{\href{https://immanent.com/2007/01/24/using-fizzbuzz-to-find-developers-who-grok-coding/}{\Snolinkurl{https://immanent.com/2007/01/24/using-fizzbuzz-to-find-developers-who-grok-coding/}}}}program in Racket, and shows the report from the
FSP for a call to the function with an input value of 10,000,000.
The profiler report notes the use of two Racket features
with a large impact on performance: output and
iterations over generic sequences. Five seconds were spent
on output. Most of this time is spent on printing numbers
not divisible by either 3 or 5 (line 16), which includes
most numbers. Unfortunately output is core to
Fizzbuzz and it cannot be avoided. On the
other hand, the \RktSym{for}{-}loop spends about one second in
generic sequence dispatch. Specifically, while the
\RktSym{range} function produces a list, the \RktSym{for}
construct iterates over all types of sequences and must
therefore process its input generically. In Racket,
this is actionable advice. A programmer
can reduce this cost by using \RktSym{in{-}range}, rather
than \RktSym{range}, thus informing the compiler that the
\RktSym{for} loop iterates over a range sequence.

\Ssubsection{A Four Part Profiler}{A Four Part Profiler}\label{t:x28part_x22profx2ddescx22x29}

Feature{-}specific profiling relies on one optional and three
required ingredients. First, the language{'}s run{-}time system
must support a way to keep track of dynamic extents. Second,
the language must also support statistical or sampling
profiling. Third, the author of features must be able to modify the
code of their features so that they mark their dynamic
extent following an FSP{-}specific protocol. Finally,
optional feature{-}specific plugins augment the
protocol by turning the FSP{'}s collected data into useful information.

\Ssubsubsectionstarx{Dynamic Extent}{Dynamic Extent}\label{t:x28part_x22Dynamicx5fExtentx22x29}

An FSP relies on a language{'}s ability to track the dynamic
extent of features. Our approach
is to place annotations on the call stack. A feature{'}s
implementation adds a mark to the stack at the begining of
its extent. The mark carries information that identifies both
the feature and its specific instance. When an instance{'}s
execution ends, the annotation is removed from the stack.
Many features contain {``}callbacks{''} to user code, such as the
\RktSym{for}{-}loop located at line 11 of the Fizzbuzz example
in figure~\hyperref[t:x28counter_x28x22figurex22_x22fizzbuzzx22x29x29]{\FigureRef{1}{t:x28counter_x28x22figurex22_x22fizzbuzzx22x29x29}}. The cost of running these
callbacks should not be accounted as part of the feature{'}s
cost. Our way to handle this situation is to add an
additional annotation to the stack. When the callback
finishes, this annotation is popped off the stack, which
indicates that the program has gone back to executing feature code.
Some languages such as Racket directly support stack annotations.
Racket refers to these as continuation marks
\Autobibref{~(\hyperref[t:x28autobib_x22John_Clementsx2c_Matthew_Flattx2c_and_Matthias_FelleisenModeling_an_algebraic_stepperIn_Procx2e_European_Symposium_on_Programmingx2c_ppx2e_320x2dx2d3342001x22x29]{\AutobibLink{Clements et al\Sendabbrev{.}}} \hyperref[t:x28autobib_x22John_Clementsx2c_Matthew_Flattx2c_and_Matthias_FelleisenModeling_an_algebraic_stepperIn_Procx2e_European_Symposium_on_Programmingx2c_ppx2e_320x2dx2d3342001x22x29]{\AutobibLink{2001}})}, which are similar to stack annotations.
Others, such as R, do not, but we show that adding stack
annotations is straightforward (\ChapRef{\SectionNumberLink{t:x28part_x22otherlangsx22x29}{8}}{Broader applicability: Profiling R}).

\Ssubsubsectionstarx{Sampling Profiler}{Sampling Profiler}\label{t:x28part_x22Samplingx5fProfilerx22x29}

An FSP additionally requires its host language to support
sampling profiling. Such a profiler collects samples of the
stack and its annotations at fixed intervals during
program execution. It uses these samples to determine what features,
if any, are being executed. After the program has finished, these
collected samples are analyzed and presented, as in
figure~\hyperref[t:x28counter_x28x22figurex22_x22fizzbuzzx22x29x29]{\FigureRef{1}{t:x28counter_x28x22figurex22_x22fizzbuzzx22x29x29}}. The total time spent in features
tends to differ from the program{'}s total execution time.
These differences stem from the distribution of annotations
in the collected samples. Any individual sample may contain
the cost of multiple features, meaning a sample with
multiple annotations is associated with multiple features.
Likewise, in the case of an annotation{-}free stack, a sample
is not associated with any features. The cost of a feature
is composed entirely of all of its specific instances. That
is, a feature is only executing when exactly one of
its instances are running.

\Ssubsubsectionstarx{Feature Annotations}{Feature Annotations}\label{t:x28part_x22Featurex5fAnnotationsx22x29}

Every feature comes with a different notion about what
costs are related to that feature, and which dynamic extent the
profiler should track. Features also have different notions
about what code is not related to the feature, and thus the
profiler should not track. For example, the \RktSym{for}{-}loop
in figure~\hyperref[t:x28counter_x28x22figurex22_x22fizzbuzzx22x29x29]{\FigureRef{1}{t:x28counter_x28x22figurex22_x22fizzbuzzx22x29x29}} must account for the time spent
generating and iterating over the list as a part of its
feature, but it is not responsible for the time spent in its
body. Because every feature has a unique notion of cost, its
authors are responsible for modifying their libraries to add
annotating indicating feature code. While modifying a
feature{'}s implemenation code puts some burden on authors,
we show that adding these annotations is manageable.

\Ssubsubsectionstarx{Feature Plugins}{Feature Plugins}\label{t:x28part_x22Featurex5fPluginsx22x29}

 While
annotations denote a feature{'}s dynamic extent, a
plugin denotes the profile with the interpretation.
Specifically, a plugin enables features to report their cost
centers even when multiple instances have overlapping and
non{-}local cost centers. This plugin is completely optional
and many features rely entirely on the protocol.

\sectionNewpage

\Ssection{Profiling Racket Contracts}{Profiling Racket Contracts}\label{t:x28part_x22contractsx22x29}

The Fizzbuzz example is simplistic and does not necessitate
a new type of profiling. To motivate a feature{-}centric
reporting of behavioral costs, this section illustrates the
profiling of contracts\Autobibref{~(\hyperref[t:x28autobib_x22Robert_Bruce_Findler_and_Matthias_FelleisenContracts_for_Higherx2dorder_FunctionsIn_Procx2e_International_Conference_on_Functional_Programming2002httpsx3ax2fx2fdoix2eorgx2f10x2e1145x2f581478x2e581484x22x29]{\AutobibLink{Findler and Felleisen}} \hyperref[t:x28autobib_x22Robert_Bruce_Findler_and_Matthias_FelleisenContracts_for_Higherx2dorder_FunctionsIn_Procx2e_International_Conference_on_Functional_Programming2002httpsx3ax2fx2fdoix2eorgx2f10x2e1145x2f581478x2e581484x22x29]{\AutobibLink{2002}})}, a feature with
dispersed costs.

In
Racket, contracts are used to
monitor the flow of values across module boundaries.
One common use case is to ensure that statically typed
modules interact safely with untyped modules.
The left half of figure~\hyperref[t:x28counter_x28x22figurex22_x22typedx2duntypedx22x29x29]{\FigureRef{2}{t:x28counter_x28x22figurex22_x22typedx2duntypedx22x29x29}} shows an untyped module
\Scribtexttt{"const{\hbox{\texttt{.}}}rkt"} and a typed module \Scribtexttt{"utils{\hbox{\texttt{.}}}rkt"}. The
untyped module defines and exports \raisebox{-2.640625bp}{\makebox[8.8bp][l]{\includegraphics[trim=2.4000000000000004 2.4000000000000004 2.4000000000000004 2.4000000000000004]{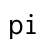}}} as
\RktVal{3{\hbox{\texttt{.}}}14}. That value is used in a test for \RktSym{arc{-}area} to
convert the radius of an arc to its area. The
value \raisebox{-2.640625bp}{\makebox[8.8bp][l]{\includegraphics[trim=2.4000000000000004 2.4000000000000004 2.4000000000000004 2.4000000000000004]{pict.pdf}}} passes through a contract (represented
by the gray box), as it passes to the typed module. If
\raisebox{-2.640625bp}{\makebox[8.8bp][l]{\includegraphics[trim=2.4000000000000004 2.4000000000000004 2.4000000000000004 2.4000000000000004]{pict.pdf}}} is not a number, the contract prevents the
value from passing through. Likewise, if \raisebox{-2.640625bp}{\makebox[8.8bp][l]{\includegraphics[trim=2.4000000000000004 2.4000000000000004 2.4000000000000004 2.4000000000000004]{pict.pdf}}} is a
number, the computation of \Scribtexttt{"utils{\hbox{\texttt{.}}}rkt"} may safely rely
on the fact that \raisebox{-2.640625bp}{\makebox[8.8bp][l]{\includegraphics[trim=2.4000000000000004 2.4000000000000004 2.4000000000000004 2.4000000000000004]{pict.pdf}}} is a number and can compile
accordingly.
Not all contracts can be checked immediately when values
cross boundaries, especially contracts for higher{-}order
functions or first{-}class objects. These contracts, shown in
the right half of figure~\hyperref[t:x28counter_x28x22figurex22_x22typedx2duntypedx22x29x29]{\FigureRef{2}{t:x28counter_x28x22figurex22_x22typedx2duntypedx22x29x29}}, are
implemented as wrappers that check the arguments and results
for every function or method call.
Here, the module defines a function \RktSym{rads{-}{\Stttextmore}dgrs},
which converts a function that operates on radians into one
that operates on degrees. The \RktSym{arc{-}area} function is
used in a higher{-}order manner. As such, the contract
boundary must wrap the function, represented as a gray box
surrounding \RktSym{arc{-}area}, to ensure that the function meets the
type it is given.

\begin{Figure}\begin{Centerfigure}\begin{FigureInside}\raisebox{-0.7093749999999943bp}{\makebox[272.00000000000006bp][l]{\includegraphics[trim=2.4000000000000004 2.4000000000000004 2.4000000000000004 2.4000000000000004]{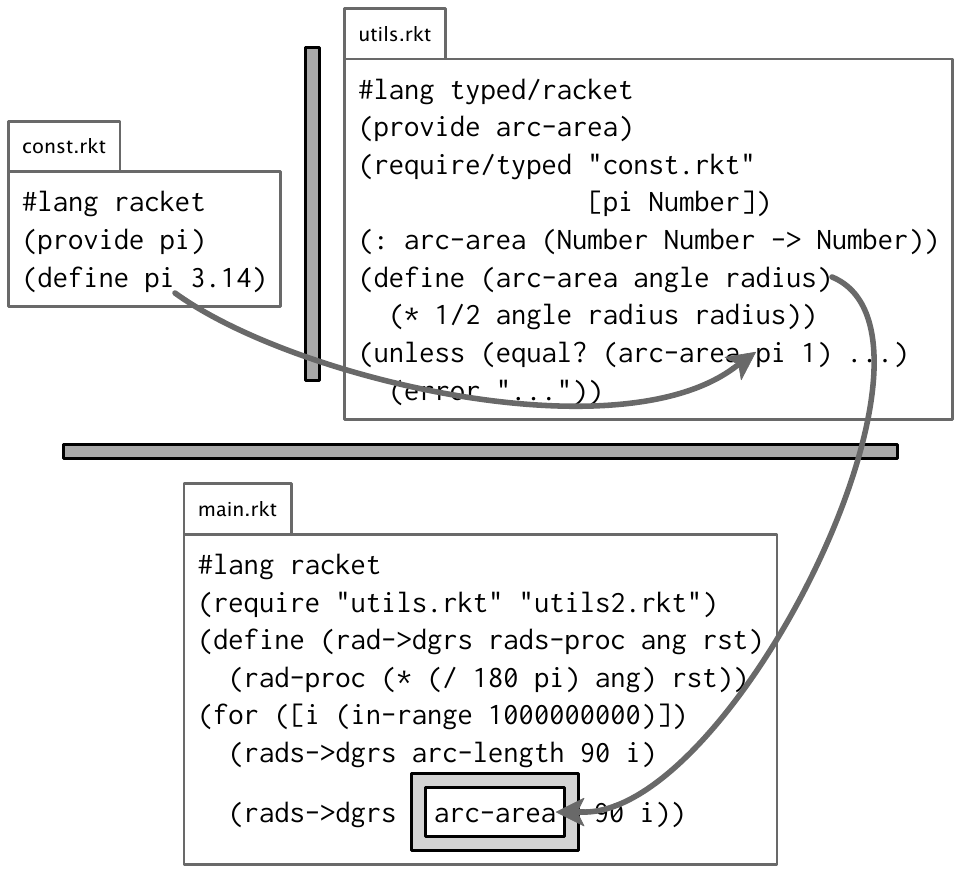}}}\end{FigureInside}\end{Centerfigure}

\Centertext{\Legend{\FigureTarget{\label{t:x28counter_x28x22figurex22_x22typedx2duntypedx22x29x29}\textsf{Fig.}~\textsf{2}. }{t:x28counter_x28x22figurex22_x22typedx2duntypedx22x29x29}\textsf{Flat (top) and higher{-}order (bottom) contracts for typed and untyped modules}}}\end{Figure}

Traditional profilers properly track the costs of flat
contracts but fail to properly track the delayed checking of
higher{-}order contracts. The left side of figure~\hyperref[t:x28counter_x28x22figurex22_x22profx2dcomparex22x29x29]{\FigureRef{3}{t:x28counter_x28x22figurex22_x22profx2dcomparex22x29x29}} shows the results when profiling the program
in figure~\hyperref[t:x28counter_x28x22figurex22_x22typedx2duntypedx22x29x29]{\FigureRef{2}{t:x28counter_x28x22figurex22_x22typedx2duntypedx22x29x29}} with a traditional profiler.
This profiler is able to detect that the program spends
roughly 10\% of execution time checking contracts, but it is
unable to determine the time spent in individual contract
instances. Worse still, the profiler associates the costs of
checking contracts with the for loop rather than where the
contracts are actually introduced, at the typed{-}untyped
boundaries. This behavior does not help programmers solve
performance problems with their code.

An FSP properly attributes the run{-}time
costs of contracts. The right side of
figure~\hyperref[t:x28counter_x28x22figurex22_x22profx2dcomparex22x29x29]{\FigureRef{3}{t:x28counter_x28x22figurex22_x22profx2dcomparex22x29x29}} shows the result when running the
same program in a feature{-}specific profiler. The profiler
determines that contracts account for roughly 25\% of execution
time. Additionally, the profiler determines that the
\Scribtexttt{arc{-}area} and \Scribtexttt{arc{-}length} contracts take
comparable time to check.

\begin{Figure}\begin{Centerfigure}\begin{FigureInside}\raisebox{-0.2874999999999943bp}{\makebox[353.20000000000005bp][l]{\includegraphics[trim=2.4000000000000004 2.4000000000000004 2.4000000000000004 2.4000000000000004]{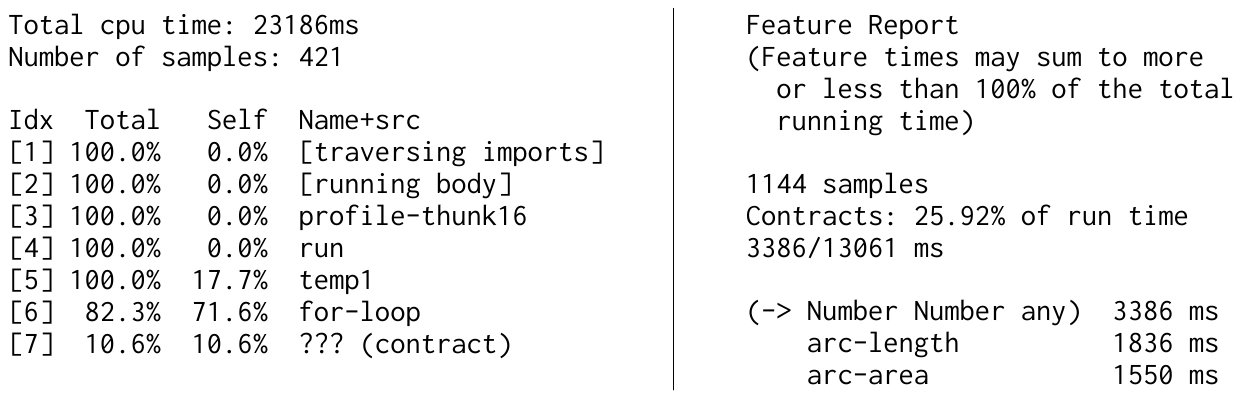}}}\end{FigureInside}\end{Centerfigure}

\Centertext{\Legend{\FigureTarget{\label{t:x28counter_x28x22figurex22_x22profx2dcomparex22x29x29}\textsf{Fig.}~\textsf{3}. }{t:x28counter_x28x22figurex22_x22profx2dcomparex22x29x29}\textsf{Output Traditional Profiler (left) and Feature{-}Specific Profiler (right)}}}\end{Figure}

The FSP{'}s output is broken down into distinct features and
instances of features. In the case of figure~\hyperref[t:x28counter_x28x22figurex22_x22profx2dcomparex22x29x29]{\FigureRef{3}{t:x28counter_x28x22figurex22_x22profx2dcomparex22x29x29}}, only
one feature takes a noticeable amount of time: contracts. It
additionally notices two particular instances of contracts and reports
the amount of time each spent.

Many features run simultaneously, such as pattern matching
and function calls. In these cases, the profiler collects
information for all running features or none in cases where
no features are running. As a result, not all of the
features put together may not add up to 100\% of the
execution time. In this case, contracts are the only feature
the profile tracked, and they account for roughly 26\% of the
run time. In contrast, a feature{'}s total cost is the sum of
all instances. As such, all instances for a particular
feature will make up 100\% of that feature{'}s total cost.

\sectionNewpage

\Ssection{Profiler Architecture}{Profiler Architecture}\label{t:x28part_x22architecturex22x29}

An FSP consists of four parts (shown in
figure~\hyperref[t:x28counter_x28x22figurex22_x22architecturex22x29x29]{\FigureRef{4}{t:x28counter_x28x22figurex22_x22architecturex22x29x29}}): a sampling profiler, an
analysis to process the raw samples, a protocol for features
to mark the extent of feature execution, and optional
analysis plug{-}ins for generating reports on individual
features. The architecture allows programmers to add
profiler support for features on an incremental basis. In
this section, we describe our implementation of an FSP for
Racket\NoteBox{\NoteContent{\href{https://github.com/stamourv/feature-profile}{\Snolinkurl{https://github.com/stamourv/feature-profile}}}}
in detail. We illustrate it with features that do not
require custom analysis plug{-}ins, such as output, type
casts, and optional function arguments. In the next section
we discuss the optional analysis plug{-}ins and features that
benefit from them.

\begin{Figure}\begin{Centerfigure}\begin{FigureInside}\raisebox{-0.19999999999998863bp}{\makebox[360.00000000000006bp][l]{\includegraphics[trim=2.4000000000000004 2.4000000000000004 2.4000000000000004 2.4000000000000004]{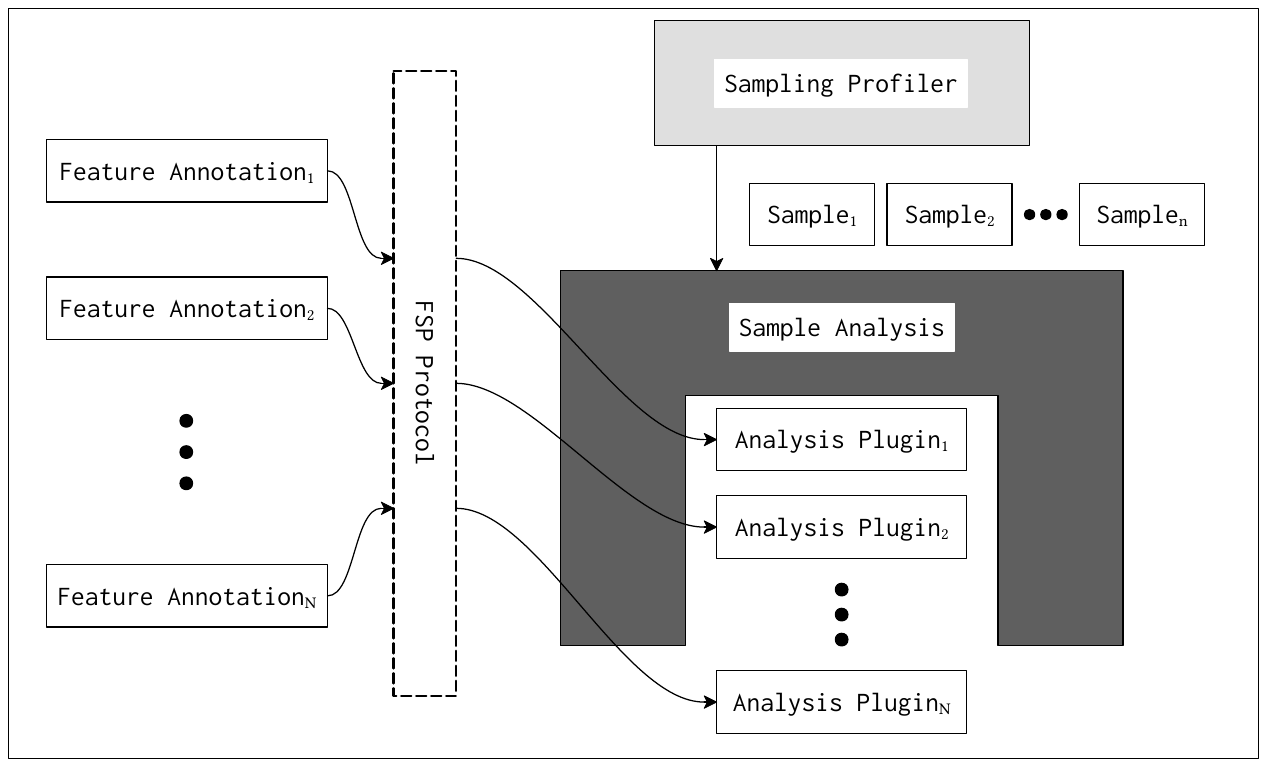}}}\end{FigureInside}\end{Centerfigure}

\Centertext{\Legend{\FigureTarget{\label{t:x28counter_x28x22figurex22_x22architecturex22x29x29}\textsf{Fig.}~\textsf{4}. }{t:x28counter_x28x22figurex22_x22architecturex22x29x29}\textsf{Architecture for an FSP}}}\end{Figure}

The profiler employs a sampling{-}thread architecture to
detect when programs execute certain pieces of code. When a
programmer turns on the profiler, a run of the program
spawns a separate sampling thread, which inspects the main
thread{'}s stack at regular intervals on the order of one
sample per 50 milliseconds. Once the program terminates, an
offline analysis deals with the collected samples and
produces programmer{-}facing reports.

The sample analysis relies on a protocol between itself and
the feature implementations. The protocol is articulated in
terms of markers on the control stack. Each marker indicates
when a feature executes its specific code. The offline
analysis can thus use these markers to attribute specific
slices of time consumption to a feature.

For our Racket{-}based prototype, the protocol heavily relies
on Racket{'}s continuation marks, an API for stack
inspection\Autobibref{~(\hyperref[t:x28autobib_x22John_Clementsx2c_Matthew_Flattx2c_and_Matthias_FelleisenModeling_an_algebraic_stepperIn_Procx2e_European_Symposium_on_Programmingx2c_ppx2e_320x2dx2d3342001x22x29]{\AutobibLink{Clements et al\Sendabbrev{.}}} \hyperref[t:x28autobib_x22John_Clementsx2c_Matthew_Flattx2c_and_Matthias_FelleisenModeling_an_algebraic_stepperIn_Procx2e_European_Symposium_on_Programmingx2c_ppx2e_320x2dx2d3342001x22x29]{\AutobibLink{2001}})}. Since this API differs from stack
inspection protocols in other languages, the first part of
this section provides some background information on
continuation marks. The second part explains how the
implementer of a feature uses continuation marks to interact
with the profiler framework. The last subsection presents
the offline analysis.

\Ssubsection{Inspecting the Stack with Continuation Marks}{Inspecting the Stack with Continuation Marks}\label{t:x28part_x22contx2dmarksx22x29}

Any program may use continuation marks to attach key{-}value pairs to frames on the control stack and
retrieve them later. Racket{'}s API provides two operations critical to FSPs:

\noindent \begin{itemize}\atItemizeStart

\item \RktPn{(}\RktSym{with{-}continuation{-}mark}\Scribtexttt{ }\RktVar{key}\Scribtexttt{ }\RktVar{value}\Scribtexttt{ }\RktVar{expr}\RktPn{)},
which attaches a (\RktVar{key}, \RktVar{value}) pair to the
current stack frame and then evaluates \RktVar{expr}. The
markers automatically disappear when the evaluation of
\RktVar{expr} terminates.

\item \RktPn{(}\RktSym{current{-}continuation{-}marks}\Scribtexttt{ }\RktVar{thread}\RktPn{)},
which walks the stack and retrieves all key{-}value pairs from the stack of a
specified thread.\end{itemize}

\noindent Programs can also filter marks with
\RktPn{(}\RktSym{continuation{-}mark{-}set{-}{\Stttextmore}list}\Scribtexttt{ }\RktVar{marks}\Scribtexttt{ }\RktVar{key}\RktPn{)}.
This operation returns a filtered list of \RktVar{marks} whose keys match \RktVar{key}.
 Outside of these operations, continuation marks do not
affect a program{'}s behavior.\NoteBox{\NoteContent{Continuation marks also
preserve the proper implementation of tail calls.}}

\begin{Figure}\begin{Centerfigure}\begin{FigureInside}\begin{SInsetFlow}\begin{bigtabular}{@{\bigtableleftpad}l@{}l@{}l@{}l@{}l@{}l@{}l@{}l@{}l@{}}
\hbox{ } &
\hbox{ } &
\hbox{ } &
\begin{RktBlk}\begin{tabular}[c]{@{}l@{}}
\hbox{\mbox{\hphantom{\Scribtexttt{x}}}\Smaller{\Scribtexttt{1}}} \\
\hbox{\mbox{\hphantom{\Scribtexttt{x}}}\Smaller{\Scribtexttt{2}}} \\
\hbox{\mbox{\hphantom{\Scribtexttt{x}}}\Smaller{\Scribtexttt{3}}} \\
\hbox{\mbox{\hphantom{\Scribtexttt{x}}}\Smaller{\Scribtexttt{4}}} \\
\hbox{\mbox{\hphantom{\Scribtexttt{x}}}\Smaller{\Scribtexttt{5}}} \\
\hbox{\mbox{\hphantom{\Scribtexttt{x}}}\Smaller{\Scribtexttt{6}}} \\
\hbox{\mbox{\hphantom{\Scribtexttt{x}}}\Smaller{\Scribtexttt{7}}} \\
\hbox{\mbox{\hphantom{\Scribtexttt{x}}}\Smaller{\Scribtexttt{8}}} \\
\hbox{\mbox{\hphantom{\Scribtexttt{x}}}\Smaller{\Scribtexttt{9}}} \\
\hbox{\Smaller{\Scribtexttt{10}}} \\
\hbox{\Smaller{\Scribtexttt{11}}} \\
\hbox{\Smaller{\Scribtexttt{12}}} \\
\hbox{\Smaller{\Scribtexttt{13}}} \\
\hbox{\Smaller{\Scribtexttt{14}}} \\
\hbox{\Smaller{\Scribtexttt{15}}} \\
\hbox{\Smaller{\Scribtexttt{16}}} \\
\hbox{\Smaller{\Scribtexttt{17}}} \\
\hbox{\Smaller{\Scribtexttt{18}}}\end{tabular}\end{RktBlk} &
\hbox{ } &
\hbox{ } &
\hbox{ } &
\hbox{ } &
\begin{RktBlk}\begin{tabular}[c]{@{}l@{}}
\hbox{\RktPn{(}\RktSym{struct}\mbox{\hphantom{\Scribtexttt{x}}}\RktSym{tree}\mbox{\hphantom{\Scribtexttt{x}}}\RktPn{(}\RktPn{)}\RktPn{)}} \\
\hbox{\RktPn{(}\RktSym{struct}\mbox{\hphantom{\Scribtexttt{x}}}\RktSym{leaf}\mbox{\hphantom{\Scribtexttt{x}}}\RktSym{tree}\mbox{\hphantom{\Scribtexttt{x}}}\RktPn{(}\RktSym{n}\RktPn{)}\RktPn{)}} \\
\hbox{\RktPn{(}\RktSym{struct}\mbox{\hphantom{\Scribtexttt{x}}}\RktSym{node}\mbox{\hphantom{\Scribtexttt{x}}}\RktSym{tree}\mbox{\hphantom{\Scribtexttt{x}}}\RktPn{(}\RktSym{l}\mbox{\hphantom{\Scribtexttt{x}}}\RktSym{n}\mbox{\hphantom{\Scribtexttt{x}}}\RktSym{r}\RktPn{)}\RktPn{)}} \\
\hbox{\mbox{\hphantom{\Scribtexttt{x}}}} \\
\hbox{\RktCmt{;}\RktCmt{~}\RktCmt{paths {\hbox{\texttt{:}}} Tree {-}{\Stttextmore} [Listof [Listof Number]]}} \\
\hbox{\RktPn{(}\RktSym{define}\mbox{\hphantom{\Scribtexttt{x}}}\RktPn{(}\RktSym{paths}\mbox{\hphantom{\Scribtexttt{x}}}\RktSym{t}\RktPn{)}} \\
\hbox{\mbox{\hphantom{\Scribtexttt{xx}}}\RktPn{(}\RktSym{cond}} \\
\hbox{\mbox{\hphantom{\Scribtexttt{xxxx}}}\RktPn{[}\RktPn{(}\RktSym{leaf{\hbox{\texttt{?}}}}\mbox{\hphantom{\Scribtexttt{x}}}\RktSym{t}\RktPn{)}} \\
\hbox{\mbox{\hphantom{\Scribtexttt{xxxxx}}}\RktPn{(}\RktSym{list}\mbox{\hphantom{\Scribtexttt{x}}}\RktPn{(}\RktSym{cons}\mbox{\hphantom{\Scribtexttt{x}}}\RktPn{(}\RktSym{leaf{-}n}\mbox{\hphantom{\Scribtexttt{x}}}\RktSym{t}\RktPn{)}} \\
\hbox{\mbox{\hphantom{\Scribtexttt{xxxxxxxxxxxxxxxxx}}}\RktPn{(}\RktSym{continuation{-}mark{-}set{-}{\Stttextmore}list}} \\
\hbox{\mbox{\hphantom{\Scribtexttt{xxxxxxxxxxxxxxxxxx}}}\RktPn{(}\RktSym{current{-}continuation{-}marks}\RktPn{)}} \\
\hbox{\mbox{\hphantom{\Scribtexttt{xxxxxxxxxxxxxxxxxx}}}\RktVal{{\textquotesingle}}\RktVal{paths}\RktPn{)}\RktPn{)}\RktPn{)}\RktPn{]}} \\
\hbox{\mbox{\hphantom{\Scribtexttt{xxxx}}}\RktPn{[}\RktPn{(}\RktSym{node{\hbox{\texttt{?}}}}\mbox{\hphantom{\Scribtexttt{x}}}\RktSym{t}\RktPn{)}} \\
\hbox{\mbox{\hphantom{\Scribtexttt{xxxxx}}}\RktPn{(}\RktSym{with{-}continuation{-}mark}\mbox{\hphantom{\Scribtexttt{x}}}\RktVal{{\textquotesingle}}\RktVal{paths}\mbox{\hphantom{\Scribtexttt{x}}}\RktPn{(}\RktSym{node{-}l}\mbox{\hphantom{\Scribtexttt{x}}}\RktSym{t}\RktPn{)}} \\
\hbox{\mbox{\hphantom{\Scribtexttt{xxxxxxx}}}\RktPn{(}\RktSym{append}\mbox{\hphantom{\Scribtexttt{x}}}\RktPn{(}\RktSym{paths}\mbox{\hphantom{\Scribtexttt{x}}}\RktPn{(}\RktSym{node{-}n}\mbox{\hphantom{\Scribtexttt{x}}}\RktSym{t}\RktPn{)}\RktPn{)}\mbox{\hphantom{\Scribtexttt{x}}}\RktPn{(}\RktSym{paths}\mbox{\hphantom{\Scribtexttt{x}}}\RktPn{(}\RktSym{node{-}r}\mbox{\hphantom{\Scribtexttt{x}}}\RktSym{t}\RktPn{)}\RktPn{)}\RktPn{)}\RktPn{)}\RktPn{]}\RktPn{)}\RktPn{)}} \\
\hbox{\mbox{\hphantom{\Scribtexttt{x}}}} \\
\hbox{\RktPn{(}\RktSym{check{-}equal{\hbox{\texttt{?}}}}\mbox{\hphantom{\Scribtexttt{x}}}\RktPn{(}\RktSym{paths}\mbox{\hphantom{\Scribtexttt{x}}}\RktPn{(}\RktSym{node}\mbox{\hphantom{\Scribtexttt{x}}}\RktVal{1}\mbox{\hphantom{\Scribtexttt{x}}}\RktPn{(}\RktSym{node}\mbox{\hphantom{\Scribtexttt{x}}}\RktVal{2}\mbox{\hphantom{\Scribtexttt{x}}}\RktPn{(}\RktSym{leaf}\mbox{\hphantom{\Scribtexttt{x}}}\RktVal{3}\RktPn{)}\mbox{\hphantom{\Scribtexttt{x}}}\RktPn{(}\RktSym{leaf}\mbox{\hphantom{\Scribtexttt{x}}}\RktVal{4}\RktPn{)}\RktPn{)}\mbox{\hphantom{\Scribtexttt{x}}}\RktPn{(}\RktSym{leaf}\mbox{\hphantom{\Scribtexttt{x}}}\RktVal{5}\RktPn{)}\RktPn{)}\RktPn{)}} \\
\hbox{\mbox{\hphantom{\Scribtexttt{xxxxxxxxxxxxxx}}}\RktVal{{\textquotesingle}}\RktVal{(}\RktVal{(}\RktVal{3}\mbox{\hphantom{\Scribtexttt{x}}}\RktVal{2}\mbox{\hphantom{\Scribtexttt{x}}}\RktVal{1}\RktVal{)}\mbox{\hphantom{\Scribtexttt{x}}}\RktVal{(}\RktVal{4}\mbox{\hphantom{\Scribtexttt{x}}}\RktVal{2}\mbox{\hphantom{\Scribtexttt{x}}}\RktVal{1}\RktVal{)}\mbox{\hphantom{\Scribtexttt{x}}}\RktVal{(}\RktVal{5}\mbox{\hphantom{\Scribtexttt{x}}}\RktVal{1}\RktVal{)}\RktVal{)}\RktPn{)}}\end{tabular}\end{RktBlk}\end{bigtabular}\end{SInsetFlow}

\noindent \raisebox{-0.5999999999999943bp}{\makebox[396.00000000000006bp][l]{\includegraphics[trim=2.4000000000000004 2.4000000000000004 2.4000000000000004 2.4000000000000004]{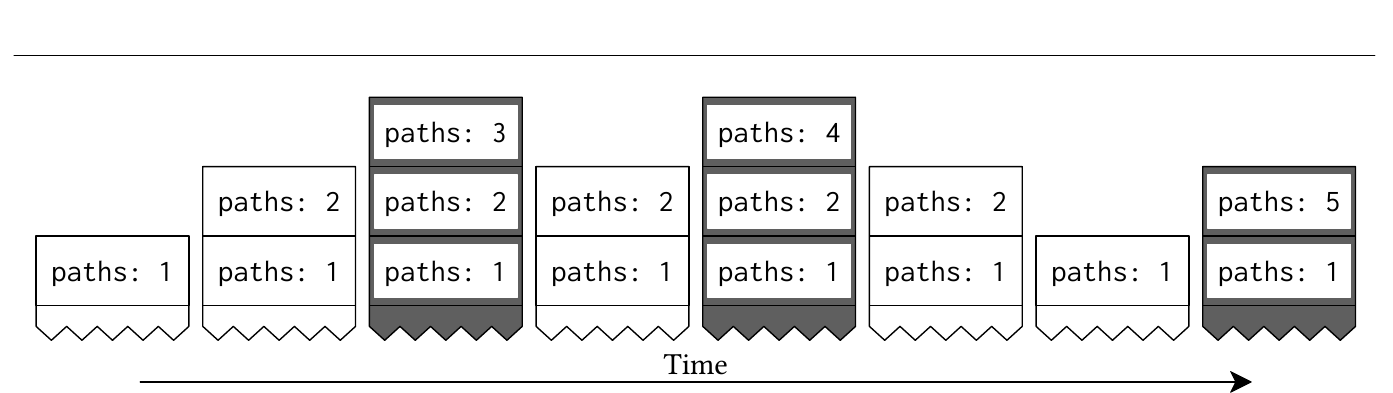}}}\end{FigureInside}\end{Centerfigure}

\Centertext{\Legend{\FigureTarget{\label{t:x28counter_x28x22figurex22_x22markx2dpathsx22x29x29}\textsf{Fig.}~\textsf{5}. }{t:x28counter_x28x22figurex22_x22markx2dpathsx22x29x29}\textsf{Recording paths in a tree with continuation marks}}}\end{Figure}

Figure~\hyperref[t:x28counter_x28x22figurex22_x22markx2dpathsx22x29x29]{\FigureRef{5}{t:x28counter_x28x22figurex22_x22markx2dpathsx22x29x29}} illustrates the working of
continuation marks with a function that traverses binary
trees and records paths from roots to leaves. The top half
of the figure shows the code that performs the
traversal. Whenever the function reaches an internal node,
it leaves a continuation mark recording that node{'}s value.
When it reaches a leaf, it collects those marks, adds the
leaf to the path and returns the completed path. A trace of
the continuation mark stack is shown in the bottom half of
the figure. It highlights the execution points where the
stack is reported to the user.

Continuation marks are extensively used in the Racket
ecosystem, e.g., the generation of error messages in
the DrRacket IDE\Autobibref{~(\hyperref[t:x28autobib_x22Robert_Bruce_Findlerx2c_John_Clementsx2c_Cormac_Flanaganx2c_Matthew_Flattx2c_Shriram_Krishnamurthix2c_Paul_Stecklerx2c_and_Matthias_FelleisenDrSchemex3a_a_programming_environment_for_SchemeJornal_of_Functional_Programming_12x282x29x2c_ppx2e_159x2dx2d1822002x22x29]{\AutobibLink{Findler et al\Sendabbrev{.}}} \hyperref[t:x28autobib_x22Robert_Bruce_Findlerx2c_John_Clementsx2c_Cormac_Flanaganx2c_Matthew_Flattx2c_Shriram_Krishnamurthix2c_Paul_Stecklerx2c_and_Matthias_FelleisenDrSchemex3a_a_programming_environment_for_SchemeJornal_of_Functional_Programming_12x282x29x2c_ppx2e_159x2dx2d1822002x22x29]{\AutobibLink{2002}})}, an algebraic stepper
\Autobibref{~(\hyperref[t:x28autobib_x22John_Clementsx2c_Matthew_Flattx2c_and_Matthias_FelleisenModeling_an_algebraic_stepperIn_Procx2e_European_Symposium_on_Programmingx2c_ppx2e_320x2dx2d3342001x22x29]{\AutobibLink{Clements et al\Sendabbrev{.}}} \hyperref[t:x28autobib_x22John_Clementsx2c_Matthew_Flattx2c_and_Matthias_FelleisenModeling_an_algebraic_stepperIn_Procx2e_European_Symposium_on_Programmingx2c_ppx2e_320x2dx2d3342001x22x29]{\AutobibLink{2001}})}, the DrRacket debugger, for thread{-}local
dynamic binding\Autobibref{~(\hyperref[t:x28autobib_x22Rx2e_Kent_DybvigChez_Scheme_Version_8_Userx27s_GuideCadence_Research_Systems2009x22x29]{\AutobibLink{Dybvig}} \hyperref[t:x28autobib_x22Rx2e_Kent_DybvigChez_Scheme_Version_8_Userx27s_GuideCadence_Research_Systems2009x22x29]{\AutobibLink{2009}})},
for exception handling, and even serializable continuations in the
PLT web server\Autobibref{~(\hyperref[t:x28autobib_x22Jay_McCarthyThe_twox2dstate_solutionx3a_native_and_serializable_continuations_accordIn_Procx2e_Objectx2doriented_Programmingx2c_Systemsx2c_Languagesx2c_and_Applicationsx2c_ppx2e_567x2dx2d5822010x22x29]{\AutobibLink{McCarthy}} \hyperref[t:x28autobib_x22Jay_McCarthyThe_twox2dstate_solutionx3a_native_and_serializable_continuations_accordIn_Procx2e_Objectx2doriented_Programmingx2c_Systemsx2c_Languagesx2c_and_Applicationsx2c_ppx2e_567x2dx2d5822010x22x29]{\AutobibLink{2010}})}.

Beyond Racket, continuation marks have also been
added to Microsoft{'}s CLR\Autobibref{~(\hyperref[t:x28autobib_x22Greg_Pettyjohnx2c_John_Clementsx2c_Joe_Marshallx2c_Shriram_Krishnamurthix2c_and_Matthias_FelleisenContinuations_from_generalized_stack_inspectionIn_Procx2e_International_Conference_on_Functional_Programmingx2c_ppx2e_216x2dx2d2272005x22x29]{\AutobibLink{Pettyjohn et al\Sendabbrev{.}}} \hyperref[t:x28autobib_x22Greg_Pettyjohnx2c_John_Clementsx2c_Joe_Marshallx2c_Shriram_Krishnamurthix2c_and_Matthias_FelleisenContinuations_from_generalized_stack_inspectionIn_Procx2e_International_Conference_on_Functional_Programmingx2c_ppx2e_216x2dx2d2272005x22x29]{\AutobibLink{2005}})} and
JavaScript\Autobibref{~(\hyperref[t:x28autobib_x22John_Clementsx2c_Ayswarya_Sundaramx2c_and_David_HermanImplementing_continuation_marks_in_JavaScriptIn_Procx2e_Scheme_and_Functional_Programming_Workshopx2c_ppx2e_1x2dx2d102008x22x29]{\AutobibLink{Clements et al\Sendabbrev{.}}} \hyperref[t:x28autobib_x22John_Clementsx2c_Ayswarya_Sundaramx2c_and_David_HermanImplementing_continuation_marks_in_JavaScriptIn_Procx2e_Scheme_and_Functional_Programming_Workshopx2c_ppx2e_1x2dx2d102008x22x29]{\AutobibLink{2008}})}.
Other languages provide similar mechanisms, such as stack reflection in
Smalltalk and the stack introspection used by the GHCi
debugger\Autobibref{~(\hyperref[t:x28autobib_x22Simon_Marlowx2c_Josxe9_Iborrax2c_Bernard_Popex2c_and_Andy_GillA_lightweight_interactive_debugger_for_HaskellIn_Procx2e_Haskell_Workshopx2c_ppx2e_13x2dx2d242007x22x29]{\AutobibLink{Marlow et al\Sendabbrev{.}}} \hyperref[t:x28autobib_x22Simon_Marlowx2c_Josxe9_Iborrax2c_Bernard_Popex2c_and_Andy_GillA_lightweight_interactive_debugger_for_HaskellIn_Procx2e_Haskell_Workshopx2c_ppx2e_13x2dx2d242007x22x29]{\AutobibLink{2007}})} for Haskell.

\Ssubsection{Feature{-}specific Data Gathering : The Protocol}{Feature{-}specific Data Gathering : The Protocol}\label{t:x28part_x22instrx2dflatx22x29}

The stack{-}sample analysis requires that a feature implementation places a
marker with a certain key on the control stack when it
begins to evaluate feature{-}specific code.

\Ssubsubsectionstarx{Marking}{Marking}\label{t:x28part_x22Markingx22x29}

Feature authors who wish to enable feature{-}specific profiling for
their features must change the implementation of the feature
so that instances mark their dynamic extents with \textit{feature marks}.
It suffices to wrap the relevant code with
\RktSym{with{-}continuation{-}mark}. These marks, added to the call stack, allow the
profiler to observe whether a thread is currently executing
code related to a feature.

Figure~\hyperref[t:x28counter_x28x22figurex22_x22flatx2dinstrx2dcodex22x29x29]{\FigureRef{6}{t:x28counter_x28x22figurex22_x22flatx2dinstrx2dcodex22x29x29}} shows an excerpt from the
instrumentation of type assertions in Typed Racket, a
variant of Racket that is statically type
checked\Autobibref{~(\hyperref[t:x28autobib_x22Sam_Tobinx2dHochstadt_and_Matthias_FelleisenThe_design_and_implementation_of_Typed_SchemeIn_Procx2e_Principles_of_Programming_Languagesx2c_ppx2e_395x2dx2d4062008x22x29]{\AutobibLink{Tobin{-}Hochstadt and Felleisen}} \hyperref[t:x28autobib_x22Sam_Tobinx2dHochstadt_and_Matthias_FelleisenThe_design_and_implementation_of_Typed_SchemeIn_Procx2e_Principles_of_Programming_Languagesx2c_ppx2e_395x2dx2d4062008x22x29]{\AutobibLink{2008}})}. The underlined conditional is
responsible for performing the actual assertion. The mark{'}s
key should uniquely identify the construct. In this case, we
use the symbol \RktVal{{\textquotesingle}}\RktVal{TR{-}assertion} as the key. Unique
choices avoid false reports and interference by distinct
features. In addition, choosing unique keys also permits the
composition of arbitrary features. As a consequence, the
analysis component of the FSP can present a unified report
to users; it also implies that users need not select in
advance the constructs they deem problematic.

The mark value{---}or \textit{payload}{---}can be anything that identifies the
feature instance to which the cost should be assigned.
In figure~\hyperref[t:x28counter_x28x22figurex22_x22flatx2dinstrx2dcodex22x29x29]{\FigureRef{6}{t:x28counter_x28x22figurex22_x22flatx2dinstrx2dcodex22x29x29}}, the payload is the source location of a
specific assertion in the program, which allows the profiler to compute the
cost of individual instances of \RktSym{assert}.

Annotating features is simple and involves only
non{-}instrusive, local code changes, but it does require
access to the implementation for the feature of interest.
Because it does not require any specialized profiling
knowledge, however, it is well within the reach of the
authors of linguistic constructs.

\begin{Figure}\begin{Centerfigure}\begin{FigureInside}\begin{SInsetFlow}\begin{bigtabular}{@{\bigtableleftpad}l@{}l@{}l@{}l@{}l@{}l@{}l@{}l@{}l@{}}
\hbox{ } &
\hbox{ } &
\hbox{ } &
\begin{RktBlk}\begin{tabular}[c]{@{}l@{}}
\hbox{\mbox{\hphantom{\Scribtexttt{x}}}\Smaller{\Scribtexttt{1}}} \\
\hbox{\mbox{\hphantom{\Scribtexttt{x}}}\Smaller{\Scribtexttt{2}}} \\
\hbox{\mbox{\hphantom{\Scribtexttt{x}}}\Smaller{\Scribtexttt{3}}} \\
\hbox{\mbox{\hphantom{\Scribtexttt{x}}}\Smaller{\Scribtexttt{4}}} \\
\hbox{\mbox{\hphantom{\Scribtexttt{x}}}\Smaller{\Scribtexttt{5}}} \\
\hbox{\mbox{\hphantom{\Scribtexttt{x}}}\Smaller{\Scribtexttt{6}}} \\
\hbox{\mbox{\hphantom{\Scribtexttt{x}}}\Smaller{\Scribtexttt{7}}} \\
\hbox{\mbox{\hphantom{\Scribtexttt{x}}}\Smaller{\Scribtexttt{8}}}\end{tabular}\end{RktBlk} &
\hbox{ } &
\hbox{ } &
\hbox{ } &
\hbox{ } &
\begin{RktBlk}\begin{tabular}[c]{@{}l@{}}
\hbox{\RktPn{(}\RktSym{define{-}syntax}\mbox{\hphantom{\Scribtexttt{x}}}\RktPn{(}\RktSym{assert}\mbox{\hphantom{\Scribtexttt{x}}}\RktSym{stx}\RktPn{)}} \\
\hbox{\mbox{\hphantom{\Scribtexttt{xx}}}\RktPn{(}\RktSym{syntax{-}case}\mbox{\hphantom{\Scribtexttt{x}}}\RktSym{stx}\mbox{\hphantom{\Scribtexttt{x}}}\RktPn{(}\RktPn{)}} \\
\hbox{\mbox{\hphantom{\Scribtexttt{xxxx}}}\RktPn{[}\RktPn{(}\RktSym{assert}\mbox{\hphantom{\Scribtexttt{x}}}\RktSym{v}\mbox{\hphantom{\Scribtexttt{x}}}\RktSym{p}\RktPn{)}\mbox{\hphantom{\Scribtexttt{x}}}\RktCmt{;}\RktCmt{~}\RktCmt{the compiler rewrites this to{\hbox{\texttt{:}}}}} \\
\hbox{\mbox{\hphantom{\Scribtexttt{xxxxx}}}\RktPn{(}\RktSym{quasisyntax}} \\
\hbox{\mbox{\hphantom{\Scribtexttt{xxxxxx}}}\RktPn{(}\RktSym{let}\mbox{\hphantom{\Scribtexttt{x}}}\RktPn{(}\RktPn{[}\RktSym{val}\mbox{\hphantom{\Scribtexttt{x}}}\RktSym{v}\RktPn{]}\mbox{\hphantom{\Scribtexttt{x}}}\RktPn{[}\RktSym{pred}\mbox{\hphantom{\Scribtexttt{x}}}\RktSym{p}\RktPn{]}\RktPn{)}} \\
\hbox{\mbox{\hphantom{\Scribtexttt{xxxxxxxx}}}\RktPn{(}\RktSym{with{-}continuation{-}mark}\mbox{\hphantom{\Scribtexttt{x}}}\RktVal{{\textquotesingle}}\RktVal{TR{-}assertion}} \\
\hbox{\mbox{\hphantom{\Scribtexttt{xxxxxxxxxx}}}\RktPn{(}\RktSym{unsyntax}\mbox{\hphantom{\Scribtexttt{x}}}\RktPn{(}\RktSym{source{-}location}\mbox{\hphantom{\Scribtexttt{x}}}\RktSym{stx}\RktPn{)}\RktPn{)}} \\
\hbox{\mbox{\hphantom{\Scribtexttt{xxxxxxxxxx}}}\RktPn{(}\RktSym{if}\mbox{\hphantom{\Scribtexttt{x}}}\RktPn{(}\RktSym{pred}\mbox{\hphantom{\Scribtexttt{x}}}\RktSym{val}\RktPn{)}\mbox{\hphantom{\Scribtexttt{x}}}\RktSym{val}\mbox{\hphantom{\Scribtexttt{x}}}\RktPn{(}\RktSym{error}\mbox{\hphantom{\Scribtexttt{x}}}\RktVal{"Assertion failed{\hbox{\texttt{.}}}"}\RktPn{)}\RktPn{)}\RktPn{)}\RktPn{)}\RktPn{)}\RktPn{]}\RktPn{)}\RktPn{)}}\end{tabular}\end{RktBlk}\end{bigtabular}\end{SInsetFlow}

\noindent \identity{\vspace{0.1em}}\end{FigureInside}\end{Centerfigure}

\Centertext{\Legend{\FigureTarget{\label{t:x28counter_x28x22figurex22_x22flatx2dinstrx2dcodex22x29x29}\textsf{Fig.}~\textsf{6}. }{t:x28counter_x28x22figurex22_x22flatx2dinstrx2dcodex22x29x29}\textsf{Instrumentation of assertions (excerpt)}}}\end{Figure}

\Ssubsubsectionstarx{Antimarking}{Antimarking}\label{t:x28part_x22Antimarkingx22x29}

 Features are
seldom {``}leaves{''} in a program; i.e., they usually run
user code whose execution time may not have to count towards
the time spent in the feature. For example, the profiler
must not count the time spent in function bodies towards the
cost of the language{'}s function call protocol.

\begin{Figure}\begin{Centerfigure}\begin{FigureInside}\begin{SInsetFlow}\begin{bigtabular}{@{\bigtableleftpad}l@{}l@{}l@{}l@{}l@{}l@{}l@{}l@{}l@{}}
\hbox{ } &
\hbox{ } &
\hbox{ } &
\begin{RktBlk}\begin{tabular}[c]{@{}l@{}}
\hbox{\mbox{\hphantom{\Scribtexttt{x}}}\Smaller{\Scribtexttt{1}}} \\
\hbox{\mbox{\hphantom{\Scribtexttt{x}}}\Smaller{\Scribtexttt{2}}} \\
\hbox{\mbox{\hphantom{\Scribtexttt{x}}}\Smaller{\Scribtexttt{3}}} \\
\hbox{\mbox{\hphantom{\Scribtexttt{x}}}\Smaller{\Scribtexttt{4}}} \\
\hbox{\mbox{\hphantom{\Scribtexttt{x}}}\Smaller{\Scribtexttt{5}}} \\
\hbox{\mbox{\hphantom{\Scribtexttt{x}}}\Smaller{\Scribtexttt{6}}} \\
\hbox{\mbox{\hphantom{\Scribtexttt{x}}}\Smaller{\Scribtexttt{7}}} \\
\hbox{\mbox{\hphantom{\Scribtexttt{x}}}\Smaller{\Scribtexttt{8}}} \\
\hbox{\mbox{\hphantom{\Scribtexttt{x}}}\Smaller{\Scribtexttt{9}}} \\
\hbox{\Smaller{\Scribtexttt{10}}}\end{tabular}\end{RktBlk} &
\hbox{ } &
\hbox{ } &
\hbox{ } &
\hbox{ } &
\begin{RktBlk}\begin{tabular}[c]{@{}l@{}}
\hbox{\RktPn{(}\RktSym{define{-}syntax}\mbox{\hphantom{\Scribtexttt{x}}}\RktPn{(}\RktSym{lambda/keyword}\mbox{\hphantom{\Scribtexttt{x}}}\RktSym{stx}\RktPn{)}} \\
\hbox{\mbox{\hphantom{\Scribtexttt{xx}}}\RktPn{(}\RktSym{syntax{-}case}\mbox{\hphantom{\Scribtexttt{x}}}\RktSym{stx}\mbox{\hphantom{\Scribtexttt{x}}}\RktPn{(}\RktPn{)}} \\
\hbox{\mbox{\hphantom{\Scribtexttt{xxxx}}}\RktPn{[}\RktPn{(}\RktSym{lambda/keyword}\mbox{\hphantom{\Scribtexttt{x}}}\RktSym{formals}\mbox{\hphantom{\Scribtexttt{x}}}\RktSym{body}\RktPn{)}\mbox{\hphantom{\Scribtexttt{x}}}\RktCmt{;}\RktCmt{~}\RktCmt{the compiler rewrites this to{\hbox{\texttt{:}}}}} \\
\hbox{\mbox{\hphantom{\Scribtexttt{xxxxx}}}\RktPn{(}\RktSym{quasisyntax}} \\
\hbox{\mbox{\hphantom{\Scribtexttt{xxxxxxx}}}\RktPn{(}\RktSym{lambda}\mbox{\hphantom{\Scribtexttt{x}}}\RktPn{(}\RktSym{unsyntax}\mbox{\hphantom{\Scribtexttt{x}}}\RktPn{(}\RktSym{handle{-}keywords}\mbox{\hphantom{\Scribtexttt{x}}}\RktSym{formals}\RktPn{)}\RktPn{)}} \\
\hbox{\mbox{\hphantom{\Scribtexttt{xxxxxxxxx}}}\RktPn{(}\RktSym{with{-}continuation{-}mark}\mbox{\hphantom{\Scribtexttt{x}}}\RktVal{{\textquotesingle}}\RktVal{kw{-}protocol}} \\
\hbox{\mbox{\hphantom{\Scribtexttt{xxxxxxxxxxxxxxxxxxxxxxxxxxxxxxxxx}}}\RktPn{(}\RktSym{unsyntax}\mbox{\hphantom{\Scribtexttt{x}}}\RktPn{(}\RktSym{source{-}location}\mbox{\hphantom{\Scribtexttt{x}}}\RktSym{stx}\RktPn{)}\RktPn{)}} \\
\hbox{\mbox{\hphantom{\Scribtexttt{xxxxxxxxxxx}}}\Scribtexttt{$\cdots$parse keyword arguments, compute default values$\cdots$}} \\
\hbox{\mbox{\hphantom{\Scribtexttt{xxxxxxxxxxx}}}\RktPn{(}\RktSym{with{-}continuation{-}mark}\mbox{\hphantom{\Scribtexttt{x}}}\RktVal{{\textquotesingle}}\RktVal{kw{-}protocol}\mbox{\hphantom{\Scribtexttt{x}}}\RktVal{{\textquotesingle}}\RktVal{antimark}} \\
\hbox{\mbox{\hphantom{\Scribtexttt{xxxxxxxxxxxxx}}}\RktSym{body}\RktPn{)}\RktPn{)}\RktPn{)}\RktPn{)}\RktPn{]}\RktPn{)}\RktPn{)}\mbox{\hphantom{\Scribtexttt{x}}}\RktCmt{;}\RktCmt{~}\RktCmt{body is use{-}site code}}\end{tabular}\end{RktBlk}\end{bigtabular}\end{SInsetFlow}

\noindent \raisebox{-0.5999999999999943bp}{\makebox[396.00000000000006bp][l]{\includegraphics[trim=2.4000000000000004 2.4000000000000004 2.4000000000000004 2.4000000000000004]{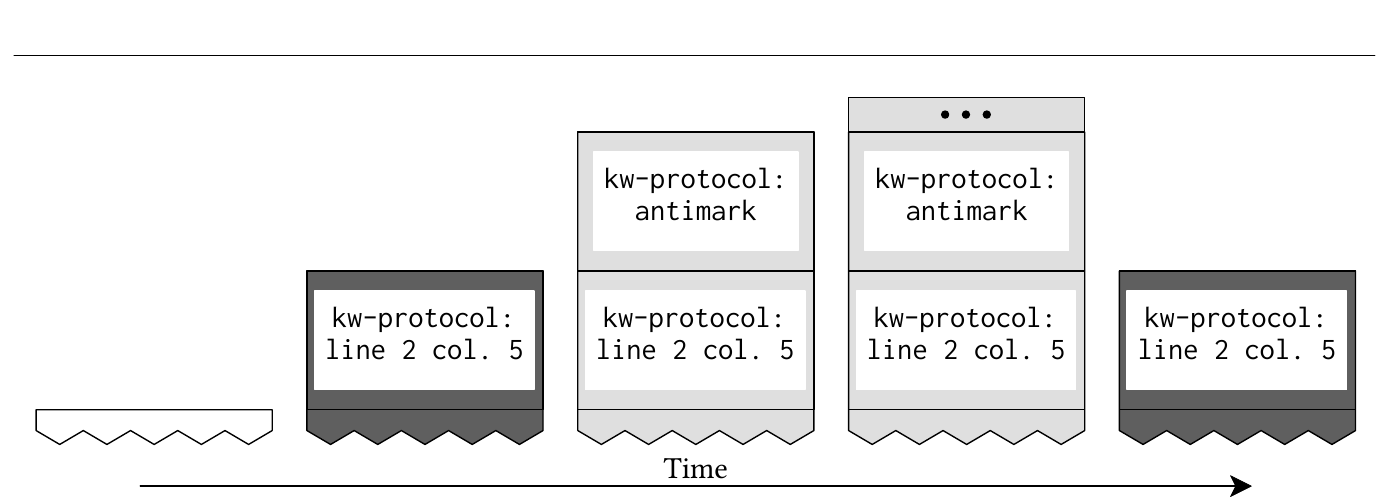}}}\end{FigureInside}\end{Centerfigure}

\Centertext{\Legend{\FigureTarget{\label{t:x28counter_x28x22figurex22_x22kwx2dantimarkx2dcodex22x29x29}\textsf{Fig.}~\textsf{7}. }{t:x28counter_x28x22figurex22_x22kwx2dantimarkx2dcodex22x29x29}\textsf{Use of antimarks in instrumentation}}}\end{Figure}

To account for user code, features place \textit{antimarks} on
the stack. Such antimarks are continuation marks with a
distinguished value, a payload of \RktVal{{\textquotesingle}}\RktVal{antimark}, that
delimit a feature{'}s code. The analysis phase recognizes
antimarks and uses them to cancel out feature marks. Cost is
attributed to a feature only if the most recent mark is a
feature mark. If it is an antimark, the program is currently
executing user code, which should not be counted. An
antimark only cancels marks for its original feature. Marks
and antimarks, for the same or different features can be
nested.

Figure~\hyperref[t:x28counter_x28x22figurex22_x22kwx2dantimarkx2dcodex22x29x29]{\FigureRef{7}{t:x28counter_x28x22figurex22_x22kwx2dantimarkx2dcodex22x29x29}} illustrates the idea with code
that instruments a simplified version of Racket{'}s optional
and keyword argument protocol \Autobibref{~(\hyperref[t:x28autobib_x22Matthew_Flatt_and_Eli_BarzilayKeyword_and_Optional_Arguments_in_PLT_SchemeIn_Procx2e_Workshop_on_Scheme_and_Functional_Programming2009x22x29]{\AutobibLink{Flatt and Barzilay}} \hyperref[t:x28autobib_x22Matthew_Flatt_and_Eli_BarzilayKeyword_and_Optional_Arguments_in_PLT_SchemeIn_Procx2e_Workshop_on_Scheme_and_Functional_Programming2009x22x29]{\AutobibLink{2009}})}. The
simplified implementation appears in the top half of the
figure and a sample trace of a function call using keyword
arguments is displayed in the bottom half. When the function
call begins, a \RktVal{{\textquotesingle}}\RktVal{kw{-}protocol} mark is placed on
the stack (annotated in \intextrgbcolor{0.37254901960784315,0.37254901960784315,0.37254901960784315}{DARK GRAY})
with a source location as its payload. Once evaluation of
the function begins, an antimark is placed on the stack
(annotated in \intextrgbcolor{0.8745098039215686,0.8745098039215686,0.8745098039215686}{LIGHT GRAY}). Once the
antimark has been removed from the stack, cost accounting is
again attributed towards keyword arguments.

In contrast, the assertions from figure~\hyperref[t:x28counter_x28x22figurex22_x22flatx2dinstrx2dcodex22x29x29]{\FigureRef{6}{t:x28counter_x28x22figurex22_x22flatx2dinstrx2dcodex22x29x29}} do
not require antimarks because user code evaluation happens
exclusively outside the marked region (line 8). Another feature that
has this behavior is program output, which also never calls
user code from within the feature.

\Ssubsubsectionstarx{Sampling}{Sampling}\label{t:x28part_x22Samplingx22x29}

 During program
execution, the FSP{'}s sampling thread periodically collects
and stores continuation marks from the main thread. The
sampling thread knows which keys correspond to features it
should track, and collects marks for all features at once.\NoteBox{\NoteContent{In
general, the sampling thread could additionally
collect samples of all marks and sort the marks in the
analysis phase.}}

\Ssubsection{Analyzing Feature{-}specific Data}{Analyzing Feature{-}specific Data}\label{t:x28part_x22analysisx2dflatx22x29}

After the program execution terminates, the analysis
component processes the data collected by the sampling
thread to produce a feature cost report. The tool analyses
each feature separately, then combines the results into a
unified report.

\Ssubsubsectionstarx{Cost assignment}{Cost assignment}\label{t:x28part_x22Costx5fassignmentx22x29}

 The profiler uses a standard sliding window
technique to assign a time cost to each sample based on the elapsed time
between the sample, its predecessor and its successor.
Only samples with a feature mark as the most recent mark contribute time
towards features.

\Ssubsubsectionstarx{Payload grouping}{Payload grouping}\label{t:x28part_x22Payloadx5fgroupingx22x29}

Payloads identify
individual feature instances. Our accounting algorithm
groups samples by payload and adds up the cost of each
sample; the sums correspond to the cost of each feature
instance. Payloads can be grouped in arbitrary equivalence
classes. Our profiler currently groups them based on
equality, but library authors can implement grouping
according to any criteria they desire. The FSP then
generates reports for each feature, using payloads as keys
and time costs as values.

\begin{Figure}\begin{Centerfigure}\begin{FigureInside}\raisebox{-0.7552083333333255bp}{\makebox[392.00000000000006bp][l]{\includegraphics[trim=2.4000000000000004 2.4000000000000004 2.4000000000000004 2.4000000000000004]{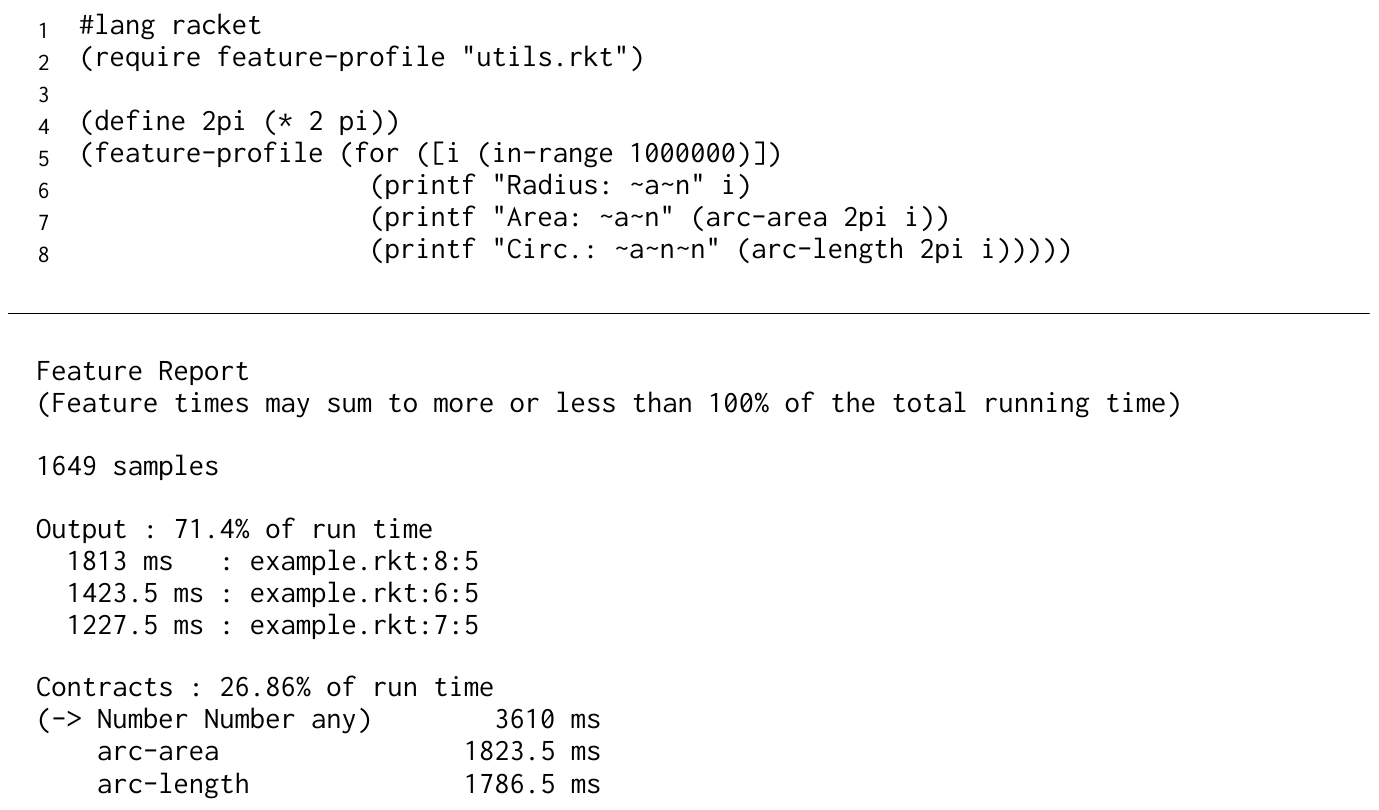}}}\end{FigureInside}\end{Centerfigure}

\Centertext{\Legend{\FigureTarget{\label{t:x28counter_x28x22figurex22_x22circlex2dprofilex22x29x29}\textsf{Fig.}~\textsf{8}. }{t:x28counter_x28x22figurex22_x22circlex2dprofilex22x29x29}\textsf{Feature Profiler Results for Circle Properties}}}\end{Figure}

\Ssubsubsectionstarx{Report composition}{Report composition}\label{t:x28part_x22Reportx5fcompositionx22x29}

Finally, after generating individual feature reports, the
FSP combines them into a unified report. Constructs absent
from the program and those inexpensive enough to never be
sampled are pruned to avoid clutter. The report lists
features in descending order of cost. Likewise, each feature
instance is listed in descending order grouped by their
associated feature.

\notitlesection\Sincsubsubsection\label{t:x28part_x22archx2dhiddenx22x29}Figure~\hyperref[t:x28counter_x28x22figurex22_x22circlex2dprofilex22x29x29]{\FigureRef{8}{t:x28counter_x28x22figurex22_x22circlex2dprofilex22x29x29}} shows a program that uses the
\Scribtexttt{utils{\hbox{\texttt{.}}}rkt} library shown in
figure~\hyperref[t:x28counter_x28x22figurex22_x22typedx2duntypedx22x29x29]{\FigureRef{2}{t:x28counter_x28x22figurex22_x22typedx2duntypedx22x29x29}}. Specifically, the program
prints the radius, area, and circumference for 1,000,000
circles of increasing size. The right half of the figure
also gives a profile report for this program. Most of the
execution time is spent printing the circles{'} properties
(lines 7{-}11), and thus appears first in the feature list.
Specifically, printing the circle{'}s circumference (line 9)
takes the most time (18 s). Finally, the second item,
contract verification, has a relatively small cost compared
to output for this program (4 s).

\sectionNewpage

\Ssection{Profiling Complex Features}{Profiling Complex Features}\label{t:x28part_x22richx22x29}

The feature{-}specific protocol in the preceding section
assumes that there is a one{-}to{-}one correspondence from the
placement of a feature to the location where it incurs a
run{-}time cost. This process, however, does not apply to
features whose instances have costs appear either in
multiple places or in different places than than their
syntactic location suggests. These are features with \textit{non{-}local costs}, because a feature instance and its cost
are separated. Higher{-}order contracts illustrate this idea
particularly well because they are specified in one place
yet incur costs at many others.
In other cases, several different instances of a feature
contribute to a single cost center, such as a concurrent
program that wants to attribute a cost to the program as a
whole as well as the particular thread or actor running
associated with it. These features have \textit{conflated
costs}.

While the creator of features with non{-}local or conflated
costs can use the FSP protocol to measure some aspects of
their costs, adopting a better protocol produces better
results when evaluating such features. This section shows
both how to extend the FSP{'}s analysis component with
feature{-}specific plug{-}ins and how to adapt the communication
protocol appropriately. It is divided into two parts. First,
we discuss custom payloads, values that the authors of
features use to describe their non{-}local or conflated costs
(\SecRef{\SectionNumberLink{t:x28part_x22implx2drichx22x29}{5.1}}{Custom Payloads}). Using custom payloads, an analysis
plug{-}in may convert the information into a form that
programmers can digest and act on (\SecRef{\SectionNumberLink{t:x28part_x22analysisx2drichx22x29}{5.2}}{Analyzing Complex{-}Cost Features}). We use
three running examples to demonstrate non{-}local and
conflated features and their payloads: contracts,
actor{-}based concurrency, and parser backtracking.

\Ssubsection{Custom Payloads}{Custom Payloads}\label{t:x28part_x22implx2drichx22x29}

The instrumentation for features with complex{-}cost
accounting, non{-}local or conflated, makes use of arbitrary
values to mark payloads instead of source locations. These
payloads must contain enough information to identify a
feature{'}s cost center and to distinguish specific instances.
Contracts, actor{-}based concurrency and parser backtracking
are three cases where features benefit from having such
custom payloads.

Although storing precise and detailed data in payloads is
attractive, developers must also avoid excessive computation
or allocation when constructing their payloads. After all,
payloads are constructed every time feature code is
executed, whether or not the sampler observes it.

\Ssubsubsectionstarx{Contracts}{Contracts}\label{t:x28part_x22contracts2x22x29}

As discussed in \ChapRef{\SectionNumberLink{t:x28part_x22contractsx22x29}{3}}{Profiling Racket Contracts}, higher{-}order behavioral
contracts have non{-}local costs. Rather than using source
locations as cost{-}centers, a contract uses \textit{blame
objects}. The latter tracks the parties to a contract so
that its possible to poinpoint the faulty party in case of a
violation. Every time an object traverses a higher{-}order
contract boundary, the contract system attaches a blame
object. This blame object holds enough
information to reconstruct a complete picture of contract
checking events{---}the contract to check, the name of the
contracted value, and the names of the components that
agreed to the contract.

\Ssubsubsectionstarx{Actor{-}Based Concurrency}{Actor{-}Based Concurrency}\label{t:x28part_x22Actorx2dBasedx5fConcurrencyx22x29}

Marketplace is a DSL for writing programs in terms of actor{-}based\Autobibref{~(\hyperref[t:x28autobib_x22Carl_Hewittx2c_Peter_Bishopx2c_and_Richard_SteigerA_Universal_Modular_ACTOR_Formalism_for_Artificial_IntelligenceIn_Procx2e_International_Joint_Conference_on_Artificial_Intelligence1973x22x29]{\AutobibLink{Hewitt et al\Sendabbrev{.}}} \hyperref[t:x28autobib_x22Carl_Hewittx2c_Peter_Bishopx2c_and_Richard_SteigerA_Universal_Modular_ACTOR_Formalism_for_Artificial_IntelligenceIn_Procx2e_International_Joint_Conference_on_Artificial_Intelligence1973x22x29]{\AutobibLink{1973}})}
concurrency\Autobibref{~(\hyperref[t:x28autobib_x22Tony_Garnockx2dJonesx2c_Sam_Tobinx2dHochstadtx2c_and_Matthias_FelleisenThe_network_as_a_language_constructIn_Procx2e_European_Symposium_on_Programming_Languagesx2c_ppx2e_473x2dx2d4922014x22x29]{\AutobibLink{Garnock{-}Jones et al\Sendabbrev{.}}} \hyperref[t:x28autobib_x22Tony_Garnockx2dJonesx2c_Sam_Tobinx2dHochstadtx2c_and_Matthias_FelleisenThe_network_as_a_language_constructIn_Procx2e_European_Symposium_on_Programming_Languagesx2c_ppx2e_473x2dx2d4922014x22x29]{\AutobibLink{2014}})}. Programs that use
Marketplace features have conflated costs. The cost{-}centers
of these programs are attributed in terms of the processes
the language uses, rather than the functions that an
individual process runs. To handle this, Marketplace uses
process identifiers as payloads. Since
\RktSym{current{-}continuation{-}marks} gathers all the marks
currently on the stack, the sampling thread can gather
\textit{core samples}.\NoteBox{\NoteContent{In analogy to geology, a core
sample includes marks from the entire stack, rather than
the top most mark.}} Because Marketplace VMs are spawned and
transfer control using function calls, these core samples
include not only the current process but also all its
ancestors{---}its parent VM, its grandparent, etc.

\begin{Figure}\begin{Centerfigure}\begin{FigureInside}\raisebox{-0.22812499999998836bp}{\makebox[394.4000000000001bp][l]{\includegraphics[trim=2.4000000000000004 2.4000000000000004 2.4000000000000004 2.4000000000000004]{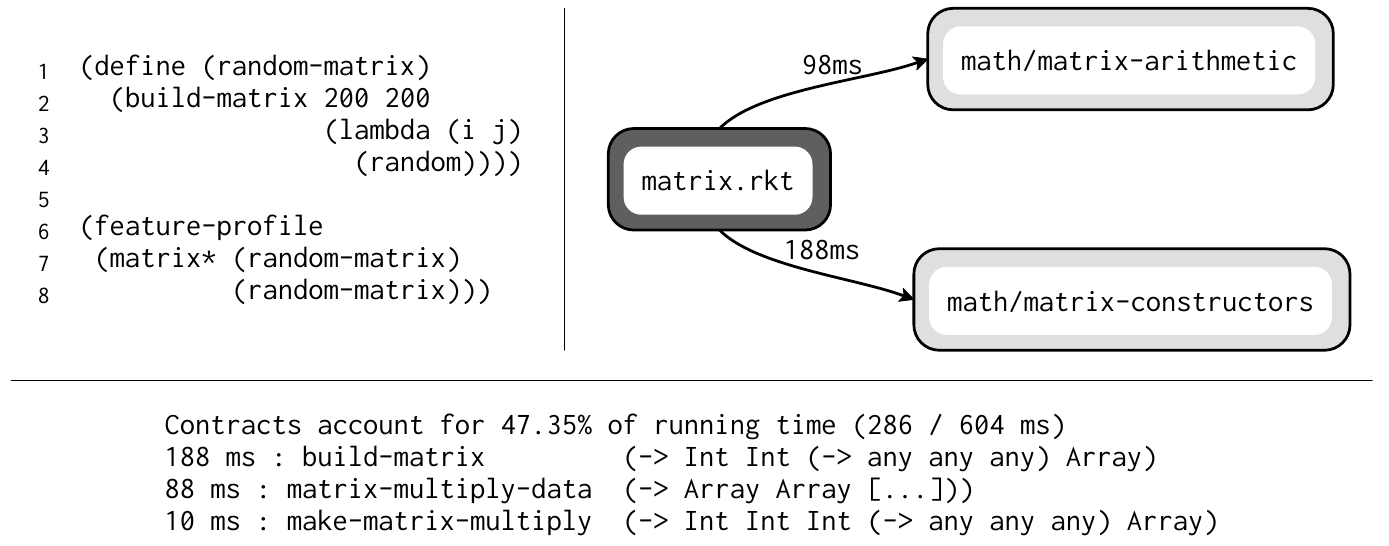}}}\end{FigureInside}\end{Centerfigure}

\Centertext{\Legend{\FigureTarget{\label{t:x28counter_x28x22figurex22_x22matrixx2dmodulex2dviewx22x29x29}\textsf{Fig.}~\textsf{9}. }{t:x28counter_x28x22figurex22_x22matrixx2dmodulex2dviewx22x29x29}\textsf{Module graph and by{-}value views of a contract boundary}}}\end{Figure}

\Ssubsubsectionstarx{Parser Backtracking}{Parser Backtracking}\label{t:x28part_x22parserx2drich1x22x29}

The Racket ecosystem includes a parser generator named
Parsack. A parser{'}s cost{-}centers are the particular parse
path that it follows, rather than any particular production
rule that the parser happens to be using. In particular, a
feature{-}specific approach shines when determining on which
paths the parser eventually backtracks. This allows a
programmer to improve a program{'}s performance by
reordering production rules when possible. To accommodate
this, payloads for Parsack combine three values into a
payload: the source location of the current production rule
disjunction, the index of the active branch within the
disjunction, and the offset in the input where the parser is
currently matching. Because parsing a term may require
recursively parsing sub{-}terms, a Parsack payload includes
core samples that allow the plugin to to attribute time to
all active non{-}terminals.

\Ssubsection{Analyzing Complex{-}Cost Features}{Analyzing Complex{-}Cost Features}\label{t:x28part_x22analysisx2drichx22x29}

Even if payloads contain enough information to uniquely
identify a feature instance{'}s cost{-}center, programmers
usually cannot directly digest the complex information in
the corresponding payloads. When a feature uses such
payloads, its creator is encouraged to implement an analysis plug{-}in that
generates user{-}facing reports.

\Ssubsubsectionstarx{Contracts}{Contracts}\label{t:x28part_x22contracts1x22x29}

The goal of the contract plug{-}in is to report which pairs of
parties impose contract checking and how much this checking
costs. A programmer can act only after identifying the
relevant components. Hence, the analysis aims to provide an
at{-}a{-}glance overview of the cost of each contract and
boundary.

To this end, the contract analysis generates a \textit{module graph} view of
contract boundaries. This graph shows modules as nodes, contract boundaries as
edges and contract costs as labels on edges.
Because typed{-}untyped boundaries are an important source of contracts,
the module graph distinguishes typed modules (in \intextrgbcolor{0.37254901960784315,0.37254901960784315,0.37254901960784315}{DARK GRAY}) from untyped modules
(in \intextrgbcolor{0.8745098039215686,0.8745098039215686,0.8745098039215686}{LIGHT GRAY}).
To generate this view, the analysis extracts component names from blame objects.
It then groups payloads that share pairs of parties and computes costs as
discussed in \SecRef{\SectionNumberLink{t:x28part_x22analysisx2dflatx22x29}{4.3}}{Analyzing Feature{-}specific Data}.
The top{-}right part of figure~\hyperref[t:x28counter_x28x22figurex22_x22matrixx2dmodulex2dviewx22x29x29]{\FigureRef{9}{t:x28counter_x28x22figurex22_x22matrixx2dmodulex2dviewx22x29x29}} shows the module graph
for a program that constructs two random matrices and multiplies them.
This latter code resides in an untyped module, but the matrix functions of
the \Scribtexttt{math} library reside in a typed module.
Hence linking the client and the library introduces a contract boundary between
them.

In addition to the module graph, an FSP can provides other
views as well. For example, the bottom portion of
figure~\hyperref[t:x28counter_x28x22figurex22_x22matrixx2dmodulex2dviewx22x29x29]{\FigureRef{9}{t:x28counter_x28x22figurex22_x22matrixx2dmodulex2dviewx22x29x29}} shows the \textit{by{-}value}
view, which provides fine{-}grained information about the cost
of individual contracted values.

\begin{Figure}\begin{Centerfigure}\begin{FigureInside}\identity{\begin{minipage}[c]{0.68\textwidth}} \raisebox{-0.8718749999999997bp}{\makebox[272.80000000000007bp][l]{\includegraphics[trim=2.4000000000000004 2.4000000000000004 2.4000000000000004 2.4000000000000004]{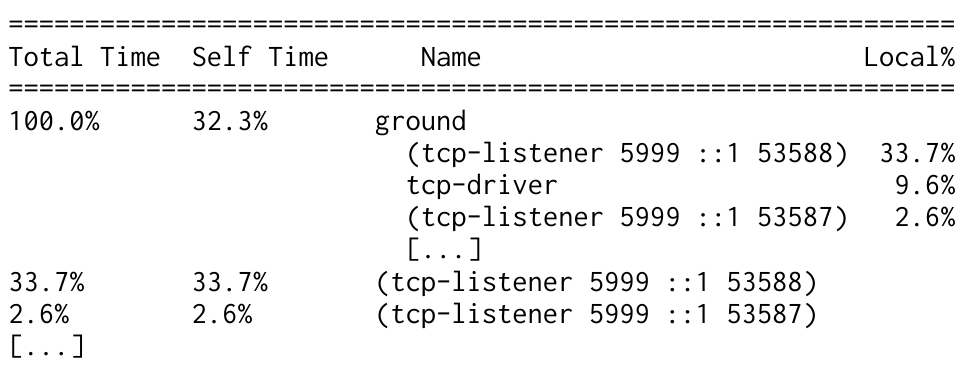}}}
\identity{\vspace{0.1em}\end{minipage}}\end{FigureInside}\end{Centerfigure}

\Centertext{\Legend{\FigureTarget{\label{t:x28counter_x28x22figurex22_x22echox2dprocessx2daccountingx22x29x29}\textsf{Fig.}~\textsf{10}. }{t:x28counter_x28x22figurex22_x22echox2dprocessx2daccountingx22x29x29}\textsf{Marketplace process accounting (excerpt)}}}\end{Figure}

\Ssubsubsectionstarx{Actor{-}Based Concurrency}{Actor{-}Based Concurrency}\label{t:x28part_x22actorx2dbasedx2dconcurrency1x22x29}

The goal of the
Marketplace analysis plug{-}in is to assign costs to individual Marketplace processes
and VMs, as opposed to the code they execute.
Marketplace feature marks use the names of processes and VMs as payloads, which
allows the plug{-}in to distinguish separate processes executing the same functions.

The plug{-}in uses full core samples to attribute costs to VMs
based on the costs of their children. These core samples
record the entire ancestry of processes in the same way the
call stack records the function calls that led to a certain
point in the execution. We exploit that similarity and reuse
standard edge profiling techniques\NoteBox{\NoteContent{VM cost assignment
is simpler than edge profiling because VM/process graphs are
in fact trees. Edge profiling techniques still apply,
though, which allows us to reuse part of the Racket edge
profiler{'}s implementation.}} to attribute costs to the entire
ancestry of a process. To disambiguate between similar
processes in its reports, the plug{-}in uses a process{'}s full
ancestry as an identity.

Figure~\hyperref[t:x28counter_x28x22figurex22_x22echox2dprocessx2daccountingx22x29x29]{\FigureRef{10}{t:x28counter_x28x22figurex22_x22echox2dprocessx2daccountingx22x29x29}} shows the accounting from a
Marketplace{-}based echo server. The first entry of the profile shows the ground
VM, which spawns all other VMs and processes.
The rightmost column shows how execution time is split across the ground VM{'}s
children.
Of note are the processes handling requests from two clients.
As reflected in the profile, the client on port 53588 is sending ten times as
much input as the one on port 53587.

The plug{-}in also reports the overhead of the Marketplace library itself.
Any time attributed directly to a VM; i.e., not to any of its children{---}is
overhead from the library. In our echo server example, 32.3\% of the total
execution time is reported as the ground VM{'}s \textit{self time}, which
corresponds to the library{'}s overhead.\NoteBox{\NoteContent{The echo server
performs no actual work which, by comparison, increases the
library{'}s relative overhead.}}

\begin{Figure}\begin{Centerfigure}\begin{FigureInside}\raisebox{-0.23645833333331412bp}{\makebox[392.00000000000006bp][l]{\includegraphics[trim=2.4000000000000004 2.4000000000000004 2.4000000000000004 2.4000000000000004]{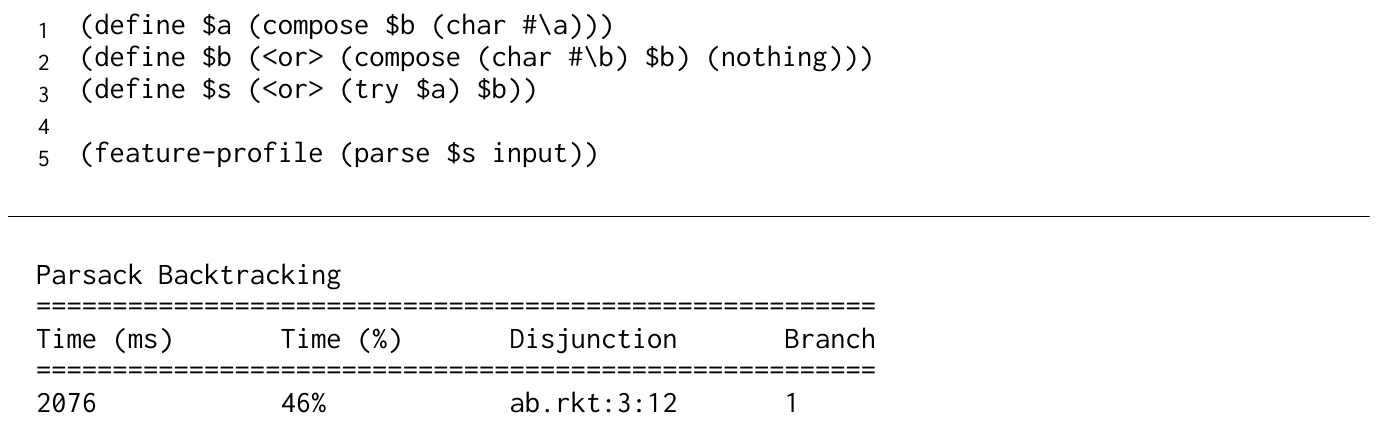}}}\end{FigureInside}\end{Centerfigure}

\Centertext{\Legend{\FigureTarget{\label{t:x28counter_x28x22figurex22_x22backtrackx22x29x29}\textsf{Fig.}~\textsf{11}. }{t:x28counter_x28x22figurex22_x22backtrackx22x29x29}\textsf{An example Parsack{-}based parser and its backtracking profile}}}\end{Figure}

\Ssubsubsectionstarx{Parser backtracking}{Parser backtracking}\label{t:x28part_x22parserx2drich2x22x29}

The feature{-}specific analysis for Parsack determines how much time is spent
backtracking for each branch of each production rule disjunction.
The source locations and input offsets in the payload allows the plug{-}in
to identify each unique visit that the parser makes to each disjunction during
parsing.

The plug{-}in detects backtracking as follows. Because disjunctions are
ordered, the parser must backtrack from early branches
in the disjuction before it reaches a production rule that
parses. Therefore, whenever the analysis observes a sample
from the matching branch at a given input
location, it attributes backtracking cost to the preceding
branches. It computes that cost from the samples taken in
these branches at the same input location. As with the
Marketplace plug{-}in, the Parsack plug{-}in uses core samples
and edge profiling to handle the recursive structure of the
process.

Figure~\hyperref[t:x28counter_x28x22figurex22_x22backtrackx22x29x29]{\FigureRef{11}{t:x28counter_x28x22figurex22_x22backtrackx22x29x29}} shows a simple parser that first attempts to parse a
sequence of \Scribtexttt{b}s followed by an \Scribtexttt{a}, and in case of failure, backtracks
in order to parse a sequence of \Scribtexttt{b}s.
The right portion of figure~\hyperref[t:x28counter_x28x22figurex22_x22backtrackx22x29x29]{\FigureRef{11}{t:x28counter_x28x22figurex22_x22backtrackx22x29x29}} shows the output of the FSP when
running the parser on a sequence of 9,000,000 \Scribtexttt{b}s. It confirms that the
parser had to backtrack from the first branch after spending almost half of the
program{'}s execution attempting it.
Swapping the \RktSym{\$a} and \RktSym{\$b} branches in the disjunction eliminates
this backtracking.

\sectionNewpage

\Ssection{Controlling Profiler Costs}{Controlling Profiler Costs}\label{t:x28part_x22costsx22x29}

Features that implement the feature{-}specific protocol insert continuation
marks regardless of whether a programmer wishes to profile the program. For
features where individual instances perform a significant amount of work,
such as contracts, the overhead of marks is usually not observable as shown
in \SecRef{\SectionNumberLink{t:x28part_x22resultsx2doverheadx22x29}{7.3}}{Overhead}. For other features, such as fine{-}grained
console output, where the aggregate cost of individually inexpensive
instance annotations are significant, the overhead of marks can be
problematic. In such cases, programmers want to choose when marks are
applied on a by{-}execution basis.

In addition, programmers may also want to control when
mark insertions take place to avoid reporting costs in code
that they wish to ignore or cannot modify. For instance,
reporting that the plot library heavily relies on
pattern{-}matching in its implementation is useless to most
programmers; they cannot fix it.
It makes sense only if they are prepared to
replace the plotting library altogether.

To establish control over when and where continuation marks are added, a
profiler must support two kinds of marks: active and latent.  We refer to
the marks described in the previous sections as active marks A latent mark
is an annotation that can be turned into an active mark as needed. An
implementation may employ a preprocessor for this purpose. We distinguish
between \textit{syntactically latent marks} for use with compile{-}time
meta{-}programming and \textit{functional latent marks} for use with library or
run{-}time functions.

\Ssubsection{Syntactically Latent Marks}{Syntactically Latent Marks}\label{t:x28part_x22Syntacticallyx5fLatentx5fMarksx22x29}

Syntactically latent marks exist as annotations on the intermediate
representation (IR) of a program.  To add a latent mark, the feature
implementation leaves tags\NoteBox{\NoteContent{Many compilers have means to attach
information to nodes in the IR. Our Racket prototype uses syntax properties
\Autobibref{~(\hyperref[t:x28autobib_x22Rx2e_Kent_Dybvigx2c_Robert_Hiebx2c_and_Carl_BruggemanSyntax_Abstracton_in_SchemeIn_Procx2e_Lisp_and_Symbolic_Computation1993x22x29]{\AutobibLink{Dybvig et al\Sendabbrev{.}}} \hyperref[t:x28autobib_x22Rx2e_Kent_Dybvigx2c_Robert_Hiebx2c_and_Carl_BruggemanSyntax_Abstracton_in_SchemeIn_Procx2e_Lisp_and_Symbolic_Computation1993x22x29]{\AutobibLink{1993}})}.}}  on the residual program{'}s IR instead of directly
inserting feature marks and antimarks. These tags are discarded after
compilation and thus have no run{-}time effect on the program execution.
Other meta{-}programs or the compiler can observe latent marks and turn them
into active marks.

A feature{-}specific profiler can rely on a dedicated compiler pass to convert
syntactic latent marks into active ones. Many compilers have some mechanism
to modify a program{'}s pre{-}compiled source.  Racket, for example, uses the
language{'}s \textit{compilation handler} mechanism to interpose this activation
pass. The pass traverses the input program, replacing every relevant
syntactic latent mark it finds with an active mark.  As this mechanism
relies on the compiler, a programmer using latent marks must recompile the
user{'}s code. The library code, however, does not need to be re{-}compiled,
which make syntactic latent marks practical for large environments.

This implementation method applies only to features implemented using
meta{-}programming such as the sntactic extensions used in many Racket or R
programs.  Thus many of these features use syntactically latent
marks. Languages without any meta{-}programming facilities can still support
latent marks with external tools that emulate meta{-}programming.

\Ssubsection{Functional Latent Marks}{Functional Latent Marks}\label{t:x28part_x22Functionalx5fLatentx5fMarksx22x29}

Functional latent marks offer an alternative to syntactically latent
marks. Instead of tagging the programmer{'}s code, a preprocessor recognizes
calls to feature{-}related functions and rewrites the program{'}s code to wrap
such calls with active marks.  Like syntactic latent marks, functional
latent marks require recompilation of code that uses the relevant
functions.  Also like syntactic latent marks, they do not require
recompiling libraries that \textit{provide} feature{-}related functions, which
makes them appropriate for functions provided as runtime primitives.

As an example, Racket{'}s output feature uses functional
latent marks instead of active marks. Functional latent
marks are appropriate here because a program may contain
many instances of the output feature, each having little
overhead. The output feature includes a list of runtime and
standard library functions that emit output and adds feature
marks around all calls to those functions, as well as
antimarks around their arguments to avoid measuring their
evaluation.

\sectionNewpage

\Ssection{Evaluation: Profiler Results}{Evaluation: Profiler Results}\label{t:x28part_x22resultsx22x29}

Our evaluation of the Racket feature{-}specific profiler addresses three
promises: that measuring in a feature{-}specific way supplies useful insights
into performance problems; that it is easy to add support for new features;
and that the run{-}time overhead of profiling manageable. This section first
presents case studies that demonstrate how feature{-}specific profiling
improves the performance of programs. Then it reports on the effort required
to mark features and implement plug{-}ins. Finally, it discusses the run{-}time
overhead imposed by the profiler.

\begin{Figure}\begin{Leftfigure}\begin{FigureInside}\begin{SCentered}\raisebox{-0.19999999999998863bp}{\makebox[367.42968750000006bp][l]{\includegraphics[trim=2.4000000000000004 2.4000000000000004 2.4000000000000004 2.4000000000000004]{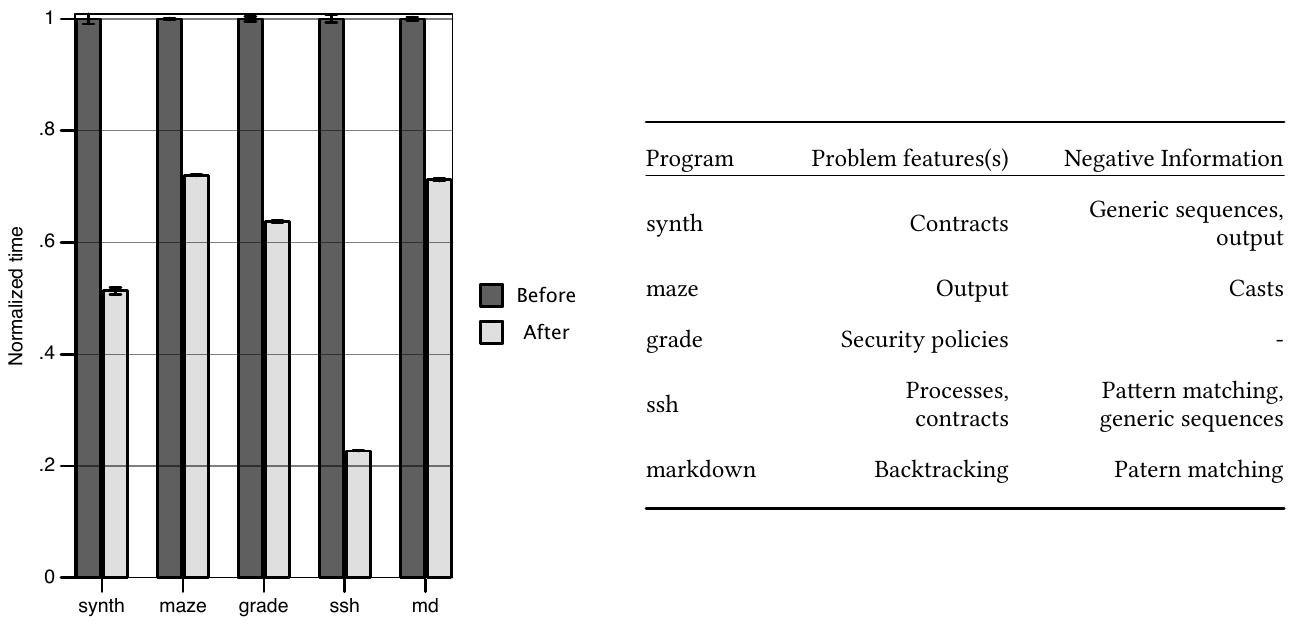}}}\end{SCentered}

\raisebox{-0.1999999999999993bp}{\makebox[8.0bp][l]{\includegraphics[trim=2.4000000000000004 2.4000000000000004 2.4000000000000004 2.4000000000000004]{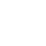}}}

\begin{smaller}Results are the mean of 30 executions on a 6{-}core 64{-}bit Debian GNU/Linux
system with 12GB of RAM.

\noindent Because Shill supports only FreeBSD, results for \textit{grade} are from a 6{-}core
FreeBSD system with 6GB of RAM.

\noindent Error bars are one standard deviation on either side.\end{smaller}\end{FigureInside}\end{Leftfigure}

\Centertext{\Legend{\FigureTarget{\label{t:x28counter_x28x22figurex22_x22casex2dstudiesx2dplotx22x29x29}\textsf{Fig.}~\textsf{12}. }{t:x28counter_x28x22figurex22_x22casex2dstudiesx2dplotx22x29x29}\textsf{Execution time after profiling and improvements (lower is better)}}}\end{Figure}

\Ssubsection{Case Studies}{Case Studies}\label{t:x28part_x22casex2dstudiesx22x29}

To be useful, a profiler must accurately identify feature use
costs and provide \textit{actionable} information to
programmers. Ideally, it identifies specific feature uses
that are responsible for significant performance costs in a
given program. When it finds such instances, the profiler
must point programmers towards solutions. Additionally, it
must also provide \textit{negative} information, i.e., confirm
that some uses of language constructs need not be
investigated.

Here we present five case studies. Each one describes a
program, summarizes the profiler{'}s feedback, and explains
the changes that directly follow from the report.
Figure~\hyperref[t:x28counter_x28x22figurex22_x22casex2dstudiesx2dplotx22x29x29]{\FigureRef{12}{t:x28counter_x28x22figurex22_x22casex2dstudiesx2dplotx22x29x29}} displays a concise overview
of the performance after incorporating this feedback. These
case{-}studies range in size from 1 to 15 modules, the
difference in size did not affect the effectiveness of the
project.

\notitlesection\label{t:x28part_x22synthx22x29}\textit{Sound Synthesis Engine}
This case study concerns a sound synthesis engine written by
St{-}Amour. The engine uses the \Scribtexttt{math} library{'}s
arrays to represent sound signals. It consists of a \Scribtexttt{mixer} module that handles most of the interaction with the
\Scribtexttt{math} library as well as a number of specialized
synthesis modules that interface with the mixer, such as
function generators, sequencers, and a drum machine. Unlike
the engine, the \Scribtexttt{math} library is written in
Typed Racket. To ensure a sound interaction between the
languages, a contract boundary separates it from the untyped
synthesis engine. For scale, the synthesis engine spans 452
lines of code, and we profile it with ten seconds of music.\NoteBox{\NoteContent{The synthesized song is {``}Funky Town{''}, by Lipps Inc.}}

\begin{Figure}\begin{Centerfigure}\begin{FigureInside}\raisebox{-0.59375bp}{\makebox[397.6000000000001bp][l]{\includegraphics[trim=2.4000000000000004 2.4000000000000004 2.4000000000000004 2.4000000000000004]{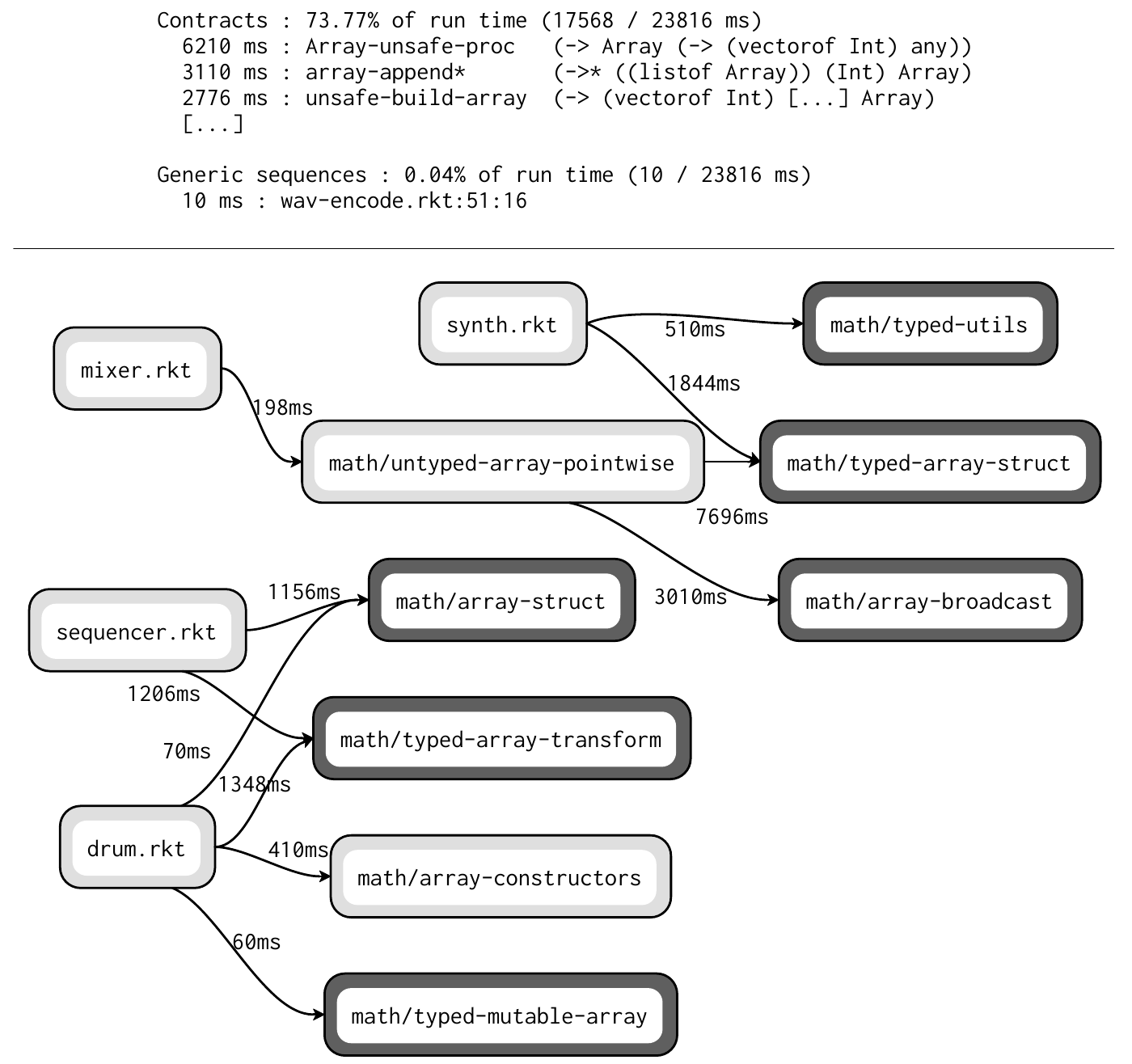}}}\end{FigureInside}\end{Centerfigure}

\Centertext{\Legend{\FigureTarget{\label{t:x28counter_x28x22figurex22_x22funkyx2dtownx2dmultix2dprofilex22x29x29}\textsf{Fig.}~\textsf{13}. }{t:x28counter_x28x22figurex22_x22funkyx2dtownx2dmultix2dprofilex22x29x29}\textsf{Feature profile (excerpt) and module
graph view for the synthesizer}}}\end{Figure}

Racket{'}s traditional statistical profiler reports that around
40\% of total execution time is spent in two functions from
the \Scribtexttt{math} library:

\noindent \begin{SCentered}\raisebox{-0.3343749999999912bp}{\makebox[233.2bp][l]{\includegraphics[trim=2.4000000000000004 2.4000000000000004 2.4000000000000004 2.4000000000000004]{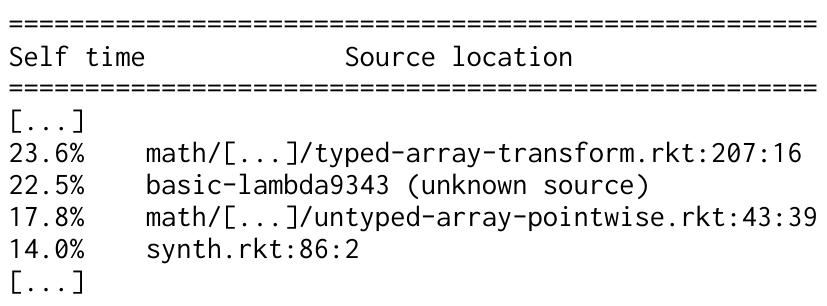}}}\end{SCentered}

\noindent Such profiling results suggest a problem with
the \Scribtexttt{math} library. Rewriting or avoiding
it altogether would be a significant undertaking.

Figure~\hyperref[t:x28counter_x28x22figurex22_x22funkyx2dtownx2dmultix2dprofilex22x29x29]{\FigureRef{13}{t:x28counter_x28x22figurex22_x22funkyx2dtownx2dmultix2dprofilex22x29x29}} shows the FSP{'}s take
of the same program. According to its report, almost three
quarters of the program{'}s execution time is spent checking
contracts, the most expensive being attached to the \Scribtexttt{math} library{'}s array functions. Consequently, any
significant performance improvements must come from
those contracts. Since the \Scribtexttt{math}
library{'}s contracts are automatically generated by Typed
Racket, improving their performance directly is not
practical. Reducing the use of contracts is more
likely to be profitable. Because contract generation happens
only at the boundary of typed and untyped code, modifying a
few modules that create this boundary may lower the imposed
cost.
In order to determine how to move a boundary, the programmer
turns to the module graph view in the lower portion of
figure~\hyperref[t:x28counter_x28x22figurex22_x22funkyx2dtownx2dmultix2dprofilex22x29x29]{\FigureRef{13}{t:x28counter_x28x22figurex22_x22funkyx2dtownx2dmultix2dprofilex22x29x29}}. This graph is provided
by our feature{-}specific analysis for contracts.
Almost half the total execution time lies between the
untyped interface to the \Scribtexttt{math} library used by the \Scribtexttt{mixer} module (in
\intextrgbcolor{0.8745098039215686,0.8745098039215686,0.8745098039215686}{LIGHT GRAY}) and the typed portions of the library (in \intextrgbcolor{0.37254901960784315,0.37254901960784315,0.37254901960784315}{DARK GRAY}).
This suggests converting the \Scribtexttt{mixer} module to Typed Racket;
a 15{-}minute effort that improves performance by$\sim$48\%.

Figure~\hyperref[t:x28counter_x28x22figurex22_x22funkyx2dtownx2dmultix2dprofilex22x29x29]{\FigureRef{13}{t:x28counter_x28x22figurex22_x22funkyx2dtownx2dmultix2dprofilex22x29x29}} also shows that generic sequence
operations, while often expensive, do not impose a significant cost in this
program, despite their pervasive use.
Manually specializing sequences would be a waste of time.
Similarly, since the report does not feature file output costs, optimizing how
the generated signal is emitted as a WAVE file would also be a waste of time.

\begin{Figure}\begin{Centerfigure}\begin{FigureInside}\raisebox{-1.7010416666666686bp}{\makebox[392.00000000000006bp][l]{\includegraphics[trim=2.4000000000000004 2.4000000000000004 2.4000000000000004 2.4000000000000004]{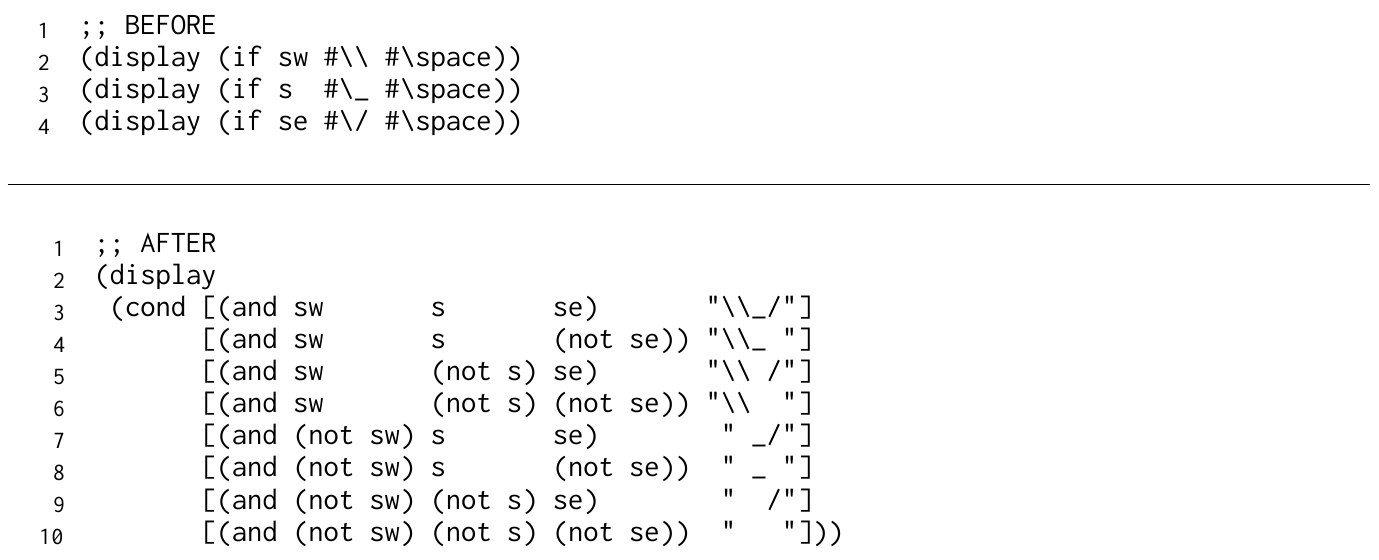}}}\end{FigureInside}\end{Centerfigure}

\Centertext{\Legend{\FigureTarget{\label{t:x28counter_x28x22figurex22_x22mazex2dbeforex2dafterx22x29x29}\textsf{Fig.}~\textsf{14}. }{t:x28counter_x28x22figurex22_x22mazex2dbeforex2dafterx22x29x29}\textsf{Fusing output operations in the maze generator}}}\end{Figure}

\notitlesection\label{t:x28part_x22mazex22x29}\textit{Maze Generator} The second case study
employs a version of a maze generator written by Olin Shivers. The program
is 758 lines of Racket; it generates a maze on a hexagonal grid, ensures
that it is solvable, and prints it.

The top portion of the output of an FSP shows 55\% of
the execution time is spent on output:

\noindent \begin{SCentered}\raisebox{-0.25937499999999547bp}{\makebox[259.6000000000001bp][l]{\includegraphics[trim=2.4000000000000004 2.4000000000000004 2.4000000000000004 2.4000000000000004]{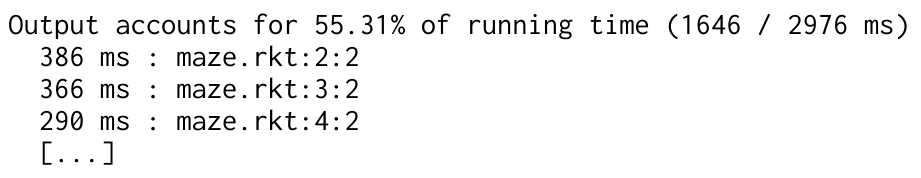}}}\end{SCentered}

\noindent Three calls to \RktSym{display}, each responsible for
printing part of the bottom of hexagons, stand out as
especially expensive. Printing each part separately results
in a large number of single{-}character output operations.
This report suggests fusing all three output operations into
one. The result of this reorganization is shown in
figure~\hyperref[t:x28counter_x28x22figurex22_x22mazex2dbeforex2dafterx22x29x29]{\FigureRef{14}{t:x28counter_x28x22figurex22_x22mazex2dbeforex2dafterx22x29x29}}. Following this advice
results in a \identity{1.39$\times$} speedup.

The profiler reports that a dynamic cast inside an inner loop has no effect
on performance. This result deviates from the more intuitive thought that
such a cast would be costly. Programmers can use this information to keep the
benefits of the cast.

\notitlesection\label{t:x28part_x22shillx22x29}\textit{Shill{-}Based Grading Script} Our third
case study involves a grading script, written by Scott Moore, that tests
students{'} OCaml code. The script is 330 lines of Shill\Autobibref{~(\hyperref[t:x28autobib_x22Scott_Moorex2c_Christos_Dimoulasx2c_Dan_Kingx2c_and_Stephen_ChongSHILLx3a_a_secure_shell_scripting_languageIn_Procx2e_USENIX_Symposium_on_Operating_Systems_Design_and_Implementation2014httpsx3ax2fx2fwwwx2eusenixx2eorgx2fconferencex2fosdi14x2ftechnicalx2dsessionsx2fpresentationx2fmoorex22x29]{\AutobibLink{Moore et al\Sendabbrev{.}}} \hyperref[t:x28autobib_x22Scott_Moorex2c_Christos_Dimoulasx2c_Dan_Kingx2c_and_Stephen_ChongSHILLx3a_a_secure_shell_scripting_languageIn_Procx2e_USENIX_Symposium_on_Operating_Systems_Design_and_Implementation2014httpsx3ax2fx2fwwwx2eusenixx2eorgx2fconferencex2fosdi14x2ftechnicalx2dsessionsx2fpresentationx2fmoorex22x29]{\AutobibLink{2014}})} code; Shill is a
least{-}privilege shell scripting language written in Racket.

According to the FSP, contracts for security permissions
account for more than 66\% of execution time:

\noindent \begin{SCentered}\raisebox{-0.534374999999994bp}{\makebox[343.20000000000005bp][l]{\includegraphics[trim=2.4000000000000004 2.4000000000000004 2.4000000000000004 2.4000000000000004]{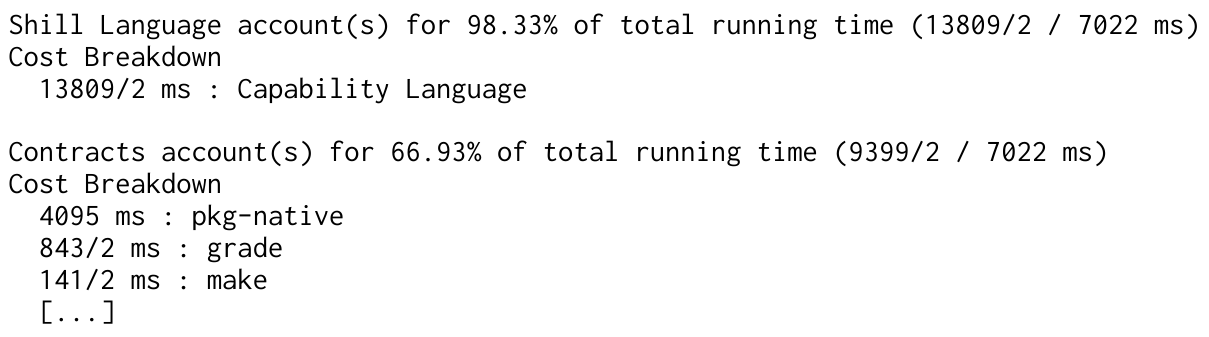}}}\end{SCentered}

\noindent  Overhead from calling external programs causes the
most slowdown. Unlike the sound synthesis example, Shill
uses contracts and a kernel extension to ensure external
programs do not violate Shill{'}s security properties. The
script contains three external programs, one being OCaml and
the other two being text manipulation utilities.
Reimplementing the two text manipulation utilities in Shill
reduces the time spent in permission checking, resulting in
a 32\% improvement in the script{'}s performance.

The results of this profile also contain useful negative information. Shill
uses an ambient language to interface between traditional operating system
permission models and Shill{'}s capability language. The FSP shows that
capability code accounts for 98\% of the time spent inside of the Racket
environment. This demonstrates that the transition layer imposed by the
ambient language has little overhead.

\begin{Figure}\begin{Centerfigure}\begin{FigureInside}\raisebox{-0.7562499999999659bp}{\makebox[392.00000000000006bp][l]{\includegraphics[trim=2.4000000000000004 2.4000000000000004 2.4000000000000004 2.4000000000000004]{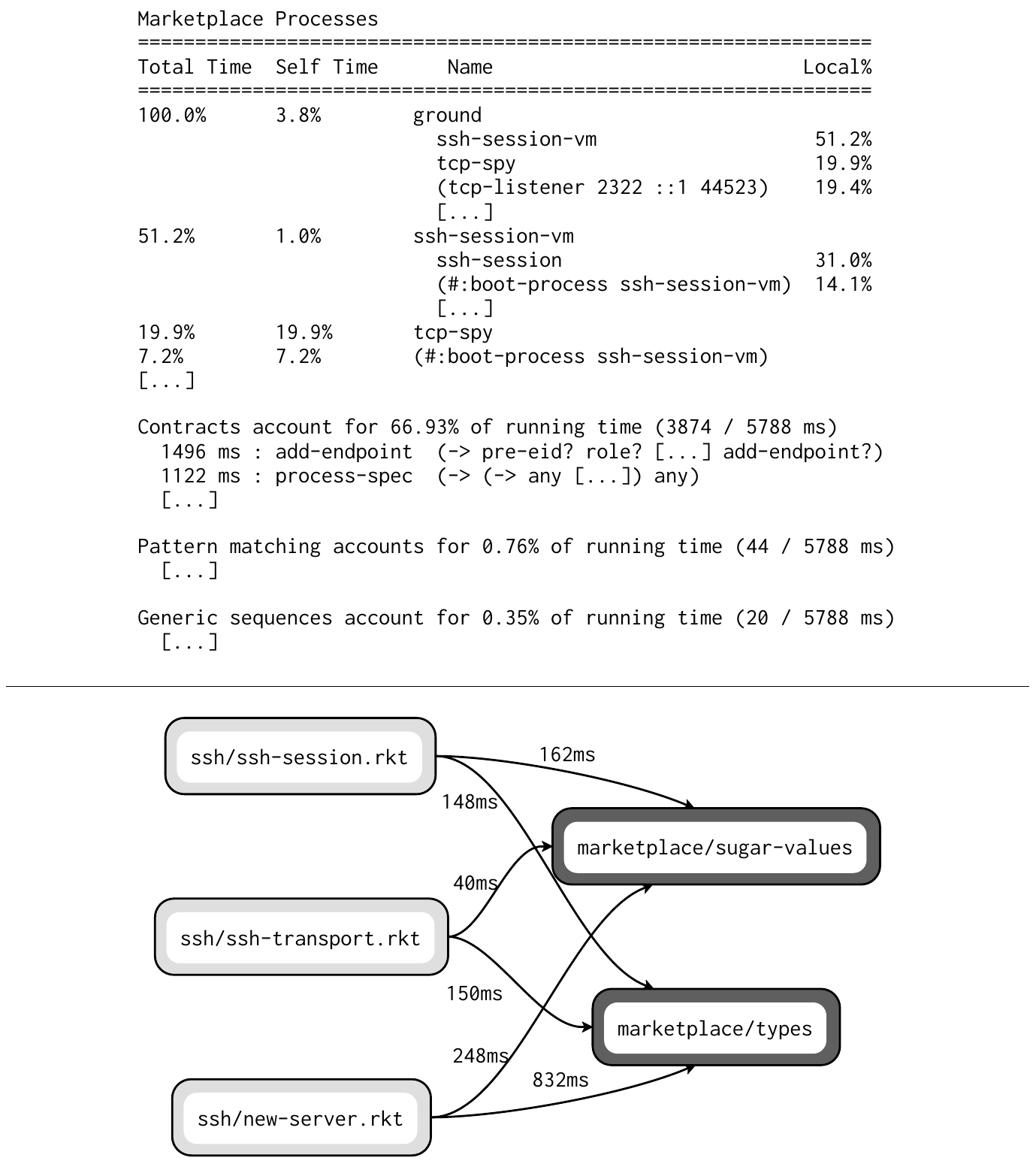}}}\end{FigureInside}\end{Centerfigure}

\Centertext{\Legend{\FigureTarget{\label{t:x28counter_x28x22figurex22_x22sshx2dmultix2dprofilex22x29x29}\textsf{Fig.}~\textsf{15}. }{t:x28counter_x28x22figurex22_x22sshx2dmultix2dprofilex22x29x29}\textsf{Profiling results for the SSH server (excerpt, top) module graph view of SSH server (bottom)}}}\end{Figure}

\notitlesection\label{t:x28part_x22sshx22x29}\textit{Marketplace{-}Based SSH Server}
The fourth case study involves an SSH
server\NoteBox{\NoteContent{\href{https://github.com/tonyg/marketplace-ssh}{\Snolinkurl{https://github.com/tonyg/marketplace-ssh}}}} in Marketplace.
The SSH server is 3,762 lines of untyped Marketplace
code and Marketplace itself is 4,801 lines of Typed Racket code.
To exercise it, a driver script starts the server, connects to it, launches
a Racket read{-}eval{-}print{-}loop on the local host, evaluates the expression
\RktPn{(}\RktSym{+}\Scribtexttt{ }\RktVal{1}\Scribtexttt{ }\RktVal{2}\Scribtexttt{ }\RktVal{3}\Scribtexttt{ }\RktVal{4}\Scribtexttt{ }\RktVal{5}\Scribtexttt{ }\RktVal{6}\RktPn{)}, disconnects and terminates the server.

As figure~\hyperref[t:x28counter_x28x22figurex22_x22sshx2dmultix2dprofilex22x29x29]{\FigureRef{15}{t:x28counter_x28x22figurex22_x22sshx2dmultix2dprofilex22x29x29}} shows, the profiler brings out two useful
facts.  First, two \textit{spy} processes{---}the \Scribtexttt{tcp{-}spy} process and the
boot process of the \Scribtexttt{ssh{-}session} VM{---}account for 25\% of execution time.
In Marketplace, spies are processes that observe other processes for logging
purposes.  The SSH server spawns these spy processes even when logging is
ignored, resulting in unnecessary overhead.  Second, contracts account for
close to 67\% of the running time.  The module view, shown in
figure~\hyperref[t:x28counter_x28x22figurex22_x22sshx2dmultix2dprofilex22x29x29]{\FigureRef{15}{t:x28counter_x28x22figurex22_x22sshx2dmultix2dprofilex22x29x29}}, shows that the majority of these contracts
lie at the boundary between the typed Marketplace library and the untyped
SSH server.  We can selectively remove these contracts in one of two ways:
by adding types to the SSH server or by disabling typechecking in
Marketplace.  Disabling spy processes and type{-}induced contracts results in
a speedup of around \identity{4.41$\times$}. In addition, the report provides
negative information. First, pattern matching again shows to have little
cost despite its pervasive use.  Additionally, Racket data structures can be
implicitly coerced to a sequence that a program is capable of iterating
over. This coercion has a runtime cost, but we show it is small.

\notitlesection\label{t:x28part_x22markdownx22x29}\textit{Markdown Parser}
Our last case study involves a Parsack{-}based Markdown
parser\NoteBox{\NoteContent{\href{https://github.com/greghendershott/markdown}{\Snolinkurl{https://github.com/greghendershott/markdown}}}}
written by Greg Hendershott.
The Markdown parser is 4,058 lines of Racket code that
we run on 1,000 lines of sample text.\NoteBox{\NoteContent{The sample text
is {``}The Time Machine{''}, by H. G. Wells. \href{http://www.gutenberg.org/ebooks/35}{\Snolinkurl{http://www.gutenberg.org/ebooks/35}}}}

The FSP{'}s feedback shows one interesting result.
Specifically, backtracking from three branches takes noticeable
time and accounts for 34\%, 2\%, and 2\% of total execution
time, respectively:

\noindent \begin{SCentered}\raisebox{-0.3031249999999912bp}{\makebox[290.40000000000003bp][l]{\includegraphics[trim=2.4000000000000004 2.4000000000000004 2.4000000000000004 2.4000000000000004]{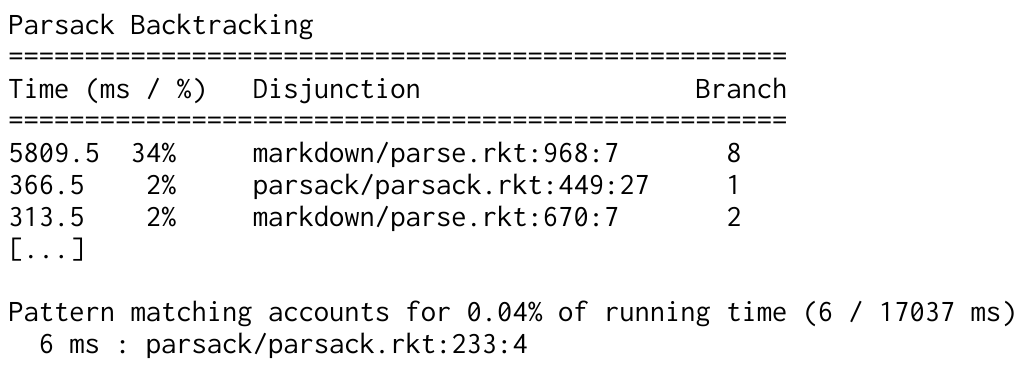}}}\end{SCentered}

\noindent Based on the tool{'}s report, moving the problematic
branches further down in their enclosing disjunction is the
appropriate action. Making this change leads to a speedup of \identity{1.40$\times$}.

For comparison, Parsack{'}s author, Stephen Chang, manually optimized the same
version of the Markdown parser using ad{-}hoc, low{-}level, and hand{-}written,
instrumentation.  His application specific instrumentation leads to a speed
up of \identity{1.37$\times$}. With no knowledge of the parser{'}s internals, we
were able to achieve a similar speedup in only a few minutes of work.

\Ssubsection{Plug{-}in Implementation Effort}{Plug{-}in Implementation Effort}\label{t:x28part_x22effortx22x29}

\begin{Figure}\begin{Centerfigure}\begin{FigureInside}\begin{SInsetFlow}\begin{bigtabular}{@{\bigtableleftpad}l@{}l@{}l@{}l@{}l@{}l@{}l@{}l@{}l@{}}
\hbox{ } &
\hbox{ } &
\hbox{ } &
\begin{RktBlk}\begin{tabular}[c]{@{}l@{}}
\hbox{\mbox{\hphantom{\Scribtexttt{x}}}\Smaller{\Scribtexttt{1}}} \\
\hbox{\mbox{\hphantom{\Scribtexttt{x}}}\Smaller{\Scribtexttt{2}}} \\
\hbox{\mbox{\hphantom{\Scribtexttt{x}}}\Smaller{\Scribtexttt{3}}} \\
\hbox{\mbox{\hphantom{\Scribtexttt{x}}}\Smaller{\Scribtexttt{4}}} \\
\hbox{\mbox{\hphantom{\Scribtexttt{x}}}\Smaller{\Scribtexttt{5}}} \\
\hbox{\mbox{\hphantom{\Scribtexttt{x}}}\Smaller{\Scribtexttt{6}}} \\
\hbox{\mbox{\hphantom{\Scribtexttt{x}}}\Smaller{\Scribtexttt{7}}} \\
\hbox{\mbox{\hphantom{\Scribtexttt{x}}}\Smaller{\Scribtexttt{8}}} \\
\hbox{\mbox{\hphantom{\Scribtexttt{x}}}\Smaller{\Scribtexttt{9}}} \\
\hbox{\Smaller{\Scribtexttt{10}}}\end{tabular}\end{RktBlk} &
\hbox{ } &
\hbox{ } &
\hbox{ } &
\hbox{ } &
\begin{RktBlk}\begin{tabular}[c]{@{}l@{}}
\hbox{\RktPn{(}\inrgbcolorbox{0.9019607843137255,0.9019607843137255,0.9019607843137255}{\RktSym{define}\Scribtexttt{ }\RktSym{marketplace{-}continuation{-}mark{-}key}}} \\
\hbox{\mbox{\hphantom{\Scribtexttt{xx}}}\inrgbcolorbox{0.9019607843137255,0.9019607843137255,0.9019607843137255}{\RktPn{(}\RktSym{make{-}continuation{-}mark{-}key}\Scribtexttt{ }\RktVal{{\textquotesingle}}\RktVal{marketplace}\RktPn{)}}\RktPn{)}} \\
\hbox{\mbox{\hphantom{\Scribtexttt{x}}}} \\
\hbox{\RktPn{[}\RktSym{{\hbox{\texttt{.}}}{\hbox{\texttt{.}}}{\hbox{\texttt{.}}}}\RktPn{]}} \\
\hbox{\mbox{\hphantom{\Scribtexttt{x}}}} \\
\hbox{\RktPn{(}\RktSym{marketplace{-}log}\mbox{\hphantom{\Scribtexttt{x}}}\RktVal{{\textquotesingle}}\RktVal{debug}\mbox{\hphantom{\Scribtexttt{x}}}\RktVal{"Entering process $\sim$v($\sim$v)"}\mbox{\hphantom{\Scribtexttt{x}}}\RktSym{debug{-}name}\mbox{\hphantom{\Scribtexttt{x}}}\RktSym{pid}\RktPn{)}} \\
\hbox{\RktPn{(}\RktSym{define}\mbox{\hphantom{\Scribtexttt{x}}}\RktSym{result}\mbox{\hphantom{\Scribtexttt{x}}}\RktPn{(}\inrgbcolorbox{0.9019607843137255,0.9019607843137255,0.9019607843137255}{\RktSym{with{-}continuation{-}mark}}} \\
\hbox{\mbox{\hphantom{\Scribtexttt{xxxxxxxxxxxxxxxxx}}}\inrgbcolorbox{0.9019607843137255,0.9019607843137255,0.9019607843137255}{\RktSym{marketplace{-}continuation{-}mark{-}key}\Scribtexttt{ }\RktPn{(}\RktSym{or}\Scribtexttt{ }\RktSym{debug{-}name}\Scribtexttt{ }\RktSym{pid}\RktPn{)}}} \\
\hbox{\mbox{\hphantom{\Scribtexttt{xxxxxxxxxxxxxxxxx}}}\RktSym{enclosed{-}expr}\RktPn{)}\RktPn{)}} \\
\hbox{\RktPn{(}\RktSym{marketplace{-}log}\mbox{\hphantom{\Scribtexttt{x}}}\RktVal{{\textquotesingle}}\RktVal{debug}\mbox{\hphantom{\Scribtexttt{x}}}\RktVal{"Leaving}\mbox{\hphantom{\Scribtexttt{xx}}}\RktVal{process $\sim$v($\sim$v)"}\mbox{\hphantom{\Scribtexttt{x}}}\RktSym{debug{-}name}\mbox{\hphantom{\Scribtexttt{x}}}\RktSym{pid}\RktPn{)}}\end{tabular}\end{RktBlk}\end{bigtabular}\end{SInsetFlow}\end{FigureInside}\end{Centerfigure}

\Centertext{\Legend{\FigureTarget{\label{t:x28counter_x28x22figurex22_x22marketplacex2dinstrx22x29x29}\textsf{Fig.}~\textsf{16}. }{t:x28counter_x28x22figurex22_x22marketplacex2dinstrx22x29x29}\textsf{Instrumentation for Marketplace (excerpt)}}}\end{Figure}

Getting a Racket library ready for feature{-}specific profiling requires
little effort, both in terms of the profilier{'}s protocol and the creation of an
optional analysis plug{-}in. It is easily within reach for library authors,
especially because it does not require advanced profiling knowledge. To
support this claim, we report anecdotal evidence and the lines of code for
adding marks to other features, as well as their plug{-}ins.

For illustrative purposes, the instrumentation for Marketplace is shown in
figure~\hyperref[t:x28counter_x28x22figurex22_x22marketplacex2dinstrx22x29x29]{\FigureRef{16}{t:x28counter_x28x22figurex22_x22marketplacex2dinstrx22x29x29}} with the added code highlighted.
Unlike other examples, which use symbols as continuation mark keys,
this code creates a fresh key using \RktSym{make{-}continuation{-}mark{-}key} to
avoid key collisions.

We report the number of lines of code for each remaining
features{'} plug{-}in in figure~\hyperref[t:x28counter_x28x22figurex22_x22profilersx2dlocx22x29x29]{\FigureRef{17}{t:x28counter_x28x22figurex22_x22profilersx2dlocx22x29x29}}. The second
column reports the number of lines that are required to
instrument the feature with marks. The third column reports
the number of lines of plug{-}in analysis code. Finally, the
fourth column reports the feature{'}s implementation size in
lines of code. The line counts for Marketplace and Parsack
do not include the roughly 500 lines of Racket{'}s edge
profiler, which are re{-}linked into the plug{-}ins. With the
exception of contract instrumentation{---}which covers
multiple kinds of contracts and is spread across about
16,000 lines of the contract system{---}instrumentation is
local and non{-}intrusive.

\begin{Figure}\begin{Centerfigure}\begin{FigureInside}\raisebox{-0.09687499999999716bp}{\makebox[353.6742187500001bp][l]{\includegraphics[trim=2.4000000000000004 2.4000000000000004 2.4000000000000004 2.4000000000000004]{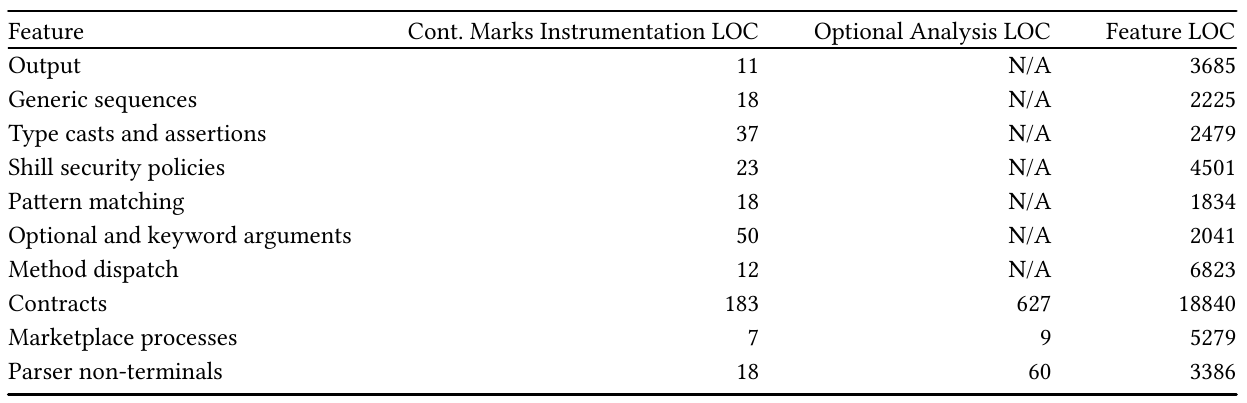}}}\end{FigureInside}\end{Centerfigure}

\Centertext{\Legend{\FigureTarget{\label{t:x28counter_x28x22figurex22_x22profilersx2dlocx22x29x29}\textsf{Fig.}~\textsf{17}. }{t:x28counter_x28x22figurex22_x22profilersx2dlocx22x29x29}\textsf{Instrumentation and analysis LOC per feature}}}\end{Figure}

\Ssubsection{Overhead}{Overhead}\label{t:x28part_x22resultsx2doverheadx22x29}

Our prototype imposes an acceptable overhead on program execution.
figure~\hyperref[t:x28counter_x28x22figurex22_x22overheadx2dplotx22x29x29]{\FigureRef{18}{t:x28counter_x28x22figurex22_x22overheadx2dplotx22x29x29}} summarizes our measurements. The results are
the mean of 30 executions with 95\% confidence error bars. The machine for
these tests is a 64{-}bit Debian GNU/Linux system with 12 core Intel Xeon CPU
clocked at 2.4 GHz and 11 GB of 1333 MHz DDR3 ram.

\begin{Figure}\begin{Centerfigure}\begin{FigureInside}\raisebox{-0.27812499999998863bp}{\makebox[392.00000000000006bp][l]{\includegraphics[trim=2.4000000000000004 2.4000000000000004 2.4000000000000004 2.4000000000000004]{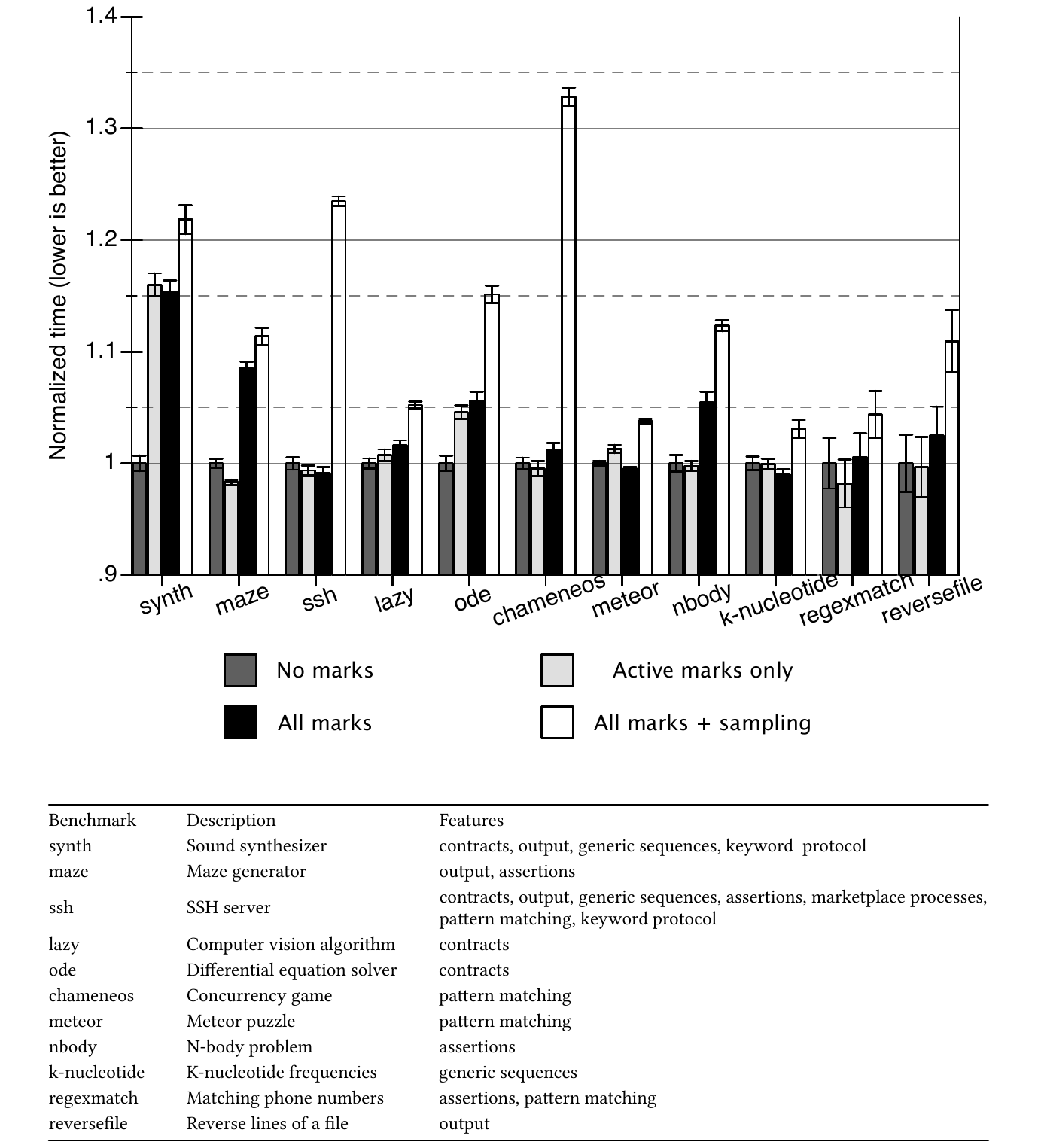}}}\end{FigureInside}\end{Centerfigure}

\Centertext{\Legend{\FigureTarget{\label{t:x28counter_x28x22figurex22_x22overheadx2dplotx22x29x29}\textsf{Fig.}~\textsf{18}. }{t:x28counter_x28x22figurex22_x22overheadx2dplotx22x29x29}\textsf{Instrumentation and sampling overhead}}}\end{Figure}

We use the programs listed in figure~\hyperref[t:x28counter_x28x22figurex22_x22overheadx2dplotx22x29x29]{\FigureRef{18}{t:x28counter_x28x22figurex22_x22overheadx2dplotx22x29x29}} as benchmarks.
They include three of the case studies from \SecRef{\SectionNumberLink{t:x28part_x22casex2dstudiesx22x29}{7.1}}{Case Studies}, two programs that
make heavy use of contracts (lazy and ode), and six programs from the Computer
Language Benchmarks Game\NoteBox{\NoteContent{\href{http://benchmarksgame.alioth.debian.org}{\Snolinkurl{http://benchmarksgame.alioth.debian.org}}}}
that use the features supported by our prototype.
The first column of figure~\hyperref[t:x28counter_x28x22figurex22_x22overheadx2dplotx22x29x29]{\FigureRef{18}{t:x28counter_x28x22figurex22_x22overheadx2dplotx22x29x29}} corresponds to programs
executing without any feature marks and serves as our baseline.
The second column reports results for programs that include only marks that are
active by default: contract marks and Marketplace marks. This bar represents
the default mode for executing programs without profiling.
The third column reports results for a program that is run with all marks activated.
The fourth column includes all of the above as well as the overhead from the
sampling thread; it is closest to the user experience when profiling.

With all marks activated, the overhead is lower than 6\% for all but two
programs, synth and maze, where it accounts for 16\% and 8.5\% respectively.
The overhead for marks that are active by default is only noticeable for two of
the four programs that include such marks, synth and ode, and account for 16\%
and 4.5\% respectively.
Total overhead, including sampling, ranges from 3\% to 33\%.

Based on this experiment, we conclude that instrumentation overhead is
reasonable in general.  The one exception, the synth benchmark, involves a
large quantity of contract checking for cheap contracts, which is the worst
case scenario for contract instrumentation.  Further engineering effort
could lower this overhead.  The overhead from sampling is similar to that of
state{-}of{-}the{-}art sampling profilers \Autobibref{~(\hyperref[t:x28autobib_x22Todd_Mytkowiczx2c_Amer_Diwanx2c_Matthias_Hauswirthx2c_and_Peter_Fx2e_SweeneyEvaluating_the_accuracy_of_Java_profilersIn_Procx2e_Programming_Langauges_Design_and_Implementationx2c_ppx2e_187x2dx2d1972010x22x29]{\AutobibLink{Mytkowicz et al\Sendabbrev{.}}} \hyperref[t:x28autobib_x22Todd_Mytkowiczx2c_Amer_Diwanx2c_Matthias_Hauswirthx2c_and_Peter_Fx2e_SweeneyEvaluating_the_accuracy_of_Java_profilersIn_Procx2e_Programming_Langauges_Design_and_Implementationx2c_ppx2e_187x2dx2d1972010x22x29]{\AutobibLink{2010}})}.

This evaluation has one threat to validity. Because instrumentation is
localized to feature code, its overhead is also localized. That is to say,
the act of profiling a feature makes that feature slightly slower compared
to the rest of the program. This may cause feature execution time to be
overestimated. However, we conjecture that this is not a  problem in
practice because these overheads are low in general. In contrast, sampling
overhead is uniformily\NoteBox{\NoteContent{Assuming random sampling, which we did not
verify.}} distributed across a program{'}s execution and should not introduce
such biases.

\sectionNewpage

\Ssection{Broader applicability: Profiling R}{Broader applicability: Profiling R}\label{t:x28part_x22otherlangsx22x29}

The applicability of feature{-}specific profiling is not limited to a
particular language.  Clearly linguistic features with complex costs are not
unique to Racket, and many languages support some sort of user{-}defined
features.  Specifically, languages with first{-}class functions,
macros, or facilities for embedding DSLs tend to come with complex{-}cost
features and can therefore benefit from our idea.

This section demonstrates the feasibility of implementing a feature{-}specific
profiler for the R programming language.  For a straightforward adaptation
of the Racket prototype, a language must have a sampling profiler and a
stack annotation mechanism. While sampling profilers have been implemented
for many languages, stack annotations are less commonly
supported. In particular, R lacks them. Fortunately, adding continuation marks to a language such as R
takes only a few lines of code.

\Ssubsection{A Sample Feature in R}{A Sample Feature in R}\label{t:x28part_x22rx2dfeaturex22x29}

\begin{Figure}\begin{Centerfigure}\begin{FigureInside}\raisebox{-0.28020833333333117bp}{\makebox[392.00000000000006bp][l]{\includegraphics[trim=2.4000000000000004 2.4000000000000004 2.4000000000000004 2.4000000000000004]{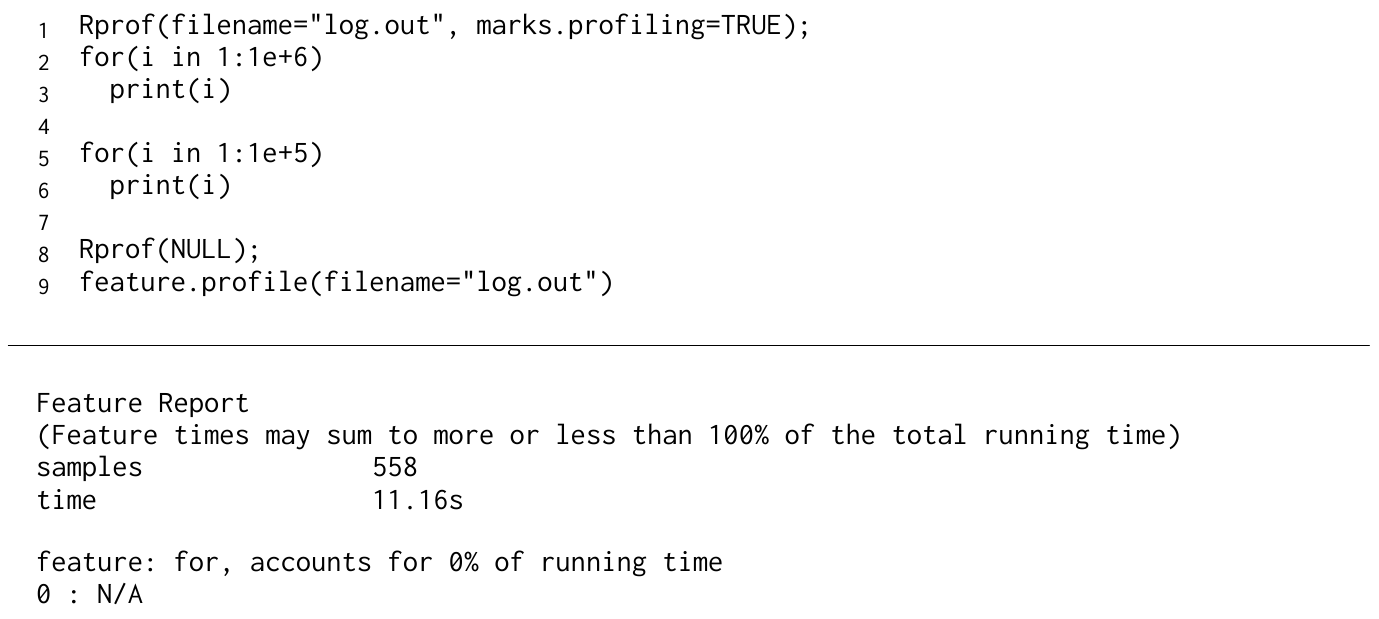}}}\end{FigureInside}\end{Centerfigure}

\Centertext{\Legend{\FigureTarget{\label{t:x28counter_x28x22figurex22_x22loopingx2dconstructsx2dsamplex22x29x29}\textsf{Fig.}~\textsf{19}. }{t:x28counter_x28x22figurex22_x22loopingx2dconstructsx2dsamplex22x29x29}\textsf{Looping Constructs}}}\end{Figure}

Like most programming languages, R provides
looping and mapping constructs such as \RktMeta{}\RktMeta{for}\RktMeta{}, \RktMeta{}\RktMeta{while}\RktMeta{},
and \RktMeta{lapply}.\NoteBox{\NoteContent{\RktMeta{lapply} is similar to \Scribtexttt{map} in
functional languages.}} Unfortunately, R implementers and
users have different opinions on the performance of loops. Folklore
in the R community suggests that looping constructs
are slow and should be avoided in favor of
vectorized operations. By contrast, R implementers claim that
loops run reasonably fast and are slow only
because of \textit{secondary effects}. That is, loops are slow
because of effects that are a by{-}product of using a feature
but are not caused by using the feature directly.
A profiler can help decide which
of the common beliefs matters.

The left{-}hand side of figure~\hyperref[t:x28counter_x28x22figurex22_x22loopingx2dconstructsx2dsamplex22x29x29]{\FigureRef{19}{t:x28counter_x28x22figurex22_x22loopingx2dconstructsx2dsamplex22x29x29}}
shows two \RktMeta{}\RktMeta{for}\RktMeta{} loop instances, the first on line
2 and the second on line 5. These loops have an
accumulator whose costs must be attributed to the feature
and a body of user code whose costs must \textit{not} be attributed
to the feature.

The right{-}hand side of figure~\hyperref[t:x28counter_x28x22figurex22_x22loopingx2dconstructsx2dsamplex22x29x29]{\FigureRef{19}{t:x28counter_x28x22figurex22_x22loopingx2dconstructsx2dsamplex22x29x29}} shows a run
of these loops with a feature{-}specific profiler. As with the Racket
prototype, a sampling profiler collects marks and antimarks, and an analyzer
converts the data into information for programmers. The resulting display
shows that no time is spent on the looping constructs.  That is, the output
(figure~\hyperref[t:x28counter_x28x22figurex22_x22loopingx2dconstructsx2dsamplex22x29x29]{\FigureRef{19}{t:x28counter_x28x22figurex22_x22loopingx2dconstructsx2dsamplex22x29x29}}) shows no samples collected during
code associated with looping constructs. While this one run is not
conclusive evidence, it supports the R implementers{'} claim that the direct
overhead of looping constructs is not significant. R code that uses loops
may still be slow, but the slowdown is not directly caused by the loop
construct.

\Ssubsection{Implementation}{Implementation}\label{t:x28part_x22rx2dcontinuationx2dmarksx22x29}

Only a few modifications to R{'}s implementation were required to support
feature{-}specific profiling.  We implemented continuation marks in 134 lines
of C. The extension to Rprof to inspect the new continuation marks accounted
for 105 lines of code.  Finally, we created a library to implement the
analysis tool in 136 lines of R code.  The implementation was created over a
week with no prior experience with the R language or its internals. These
results suggest that implementing feature{-}specific profiling may be possible
even when the host language does support continuation marks or stack
annotations.

\begin{Figure}\begin{Centerfigure}\begin{FigureInside}\begin{SCodeFlow}\begin{RktBlk}\begin{SingleColumn}\Smaller{\Smaller{\mbox{\hphantom{\Scribtexttt{x}}}\Scribtexttt{1}\mbox{\hphantom{\Scribtexttt{x}}}}}\mbox{\hphantom{\Scribtexttt{x}}}\RktMeta{SEXP}\mbox{\hphantom{\Scribtexttt{x}}}\RktMeta{attribute{\char`\_}hidden}\mbox{\hphantom{\Scribtexttt{x}}}\RktMeta{do{\char`\_}for}\RktPn{(}\RktMeta{SEXP}\mbox{\hphantom{\Scribtexttt{x}}}\RktMeta{call}\RktPn{,}\RktMeta{}\mbox{\hphantom{\Scribtexttt{x}}}\RktMeta{SEXP}\mbox{\hphantom{\Scribtexttt{x}}}\RktMeta{op}\RktPn{,}\RktMeta{}\mbox{\hphantom{\Scribtexttt{x}}}\RktMeta{SEXP}\mbox{\hphantom{\Scribtexttt{x}}}\RktMeta{args}\RktPn{,}\RktMeta{}\mbox{\hphantom{\Scribtexttt{x}}}\RktMeta{SEXP}\mbox{\hphantom{\Scribtexttt{x}}}\RktMeta{rho}\RktPn{)}\RktMeta{}\mbox{\hphantom{\Scribtexttt{x}}}\RktMeta{}\RktPn{{\char`\{}}\RktMeta{}

\Smaller{\Smaller{\mbox{\hphantom{\Scribtexttt{x}}}\Scribtexttt{2}\mbox{\hphantom{\Scribtexttt{x}}}}}\mbox{\hphantom{\Scribtexttt{x}}}\RktMeta{}\mbox{\hphantom{\Scribtexttt{xxxx}}}\RktMeta{}\RktPn{[}\RktPn{{\hbox{\texttt{.}}}}\RktPn{{\hbox{\texttt{.}}}}\RktPn{{\hbox{\texttt{.}}}}\RktPn{]}\RktMeta{}

\Smaller{\Smaller{\mbox{\hphantom{\Scribtexttt{x}}}\Scribtexttt{3}\mbox{\hphantom{\Scribtexttt{x}}}}}\mbox{\hphantom{\Scribtexttt{x}}}\RktMeta{}\mbox{\hphantom{\Scribtexttt{xxxx}}}\RktMeta{R{\char`\_}AddMark}\RktPn{(}\RktMeta{FOR}\RktPn{,}\RktMeta{}\mbox{\hphantom{\Scribtexttt{x}}}\RktMeta{call}\RktPn{,}\RktMeta{}\mbox{\hphantom{\Scribtexttt{x}}}\RktMeta{}\RktVal{TRUE}\RktPn{)}\RktPn{;}\RktMeta{}

\Smaller{\Smaller{\mbox{\hphantom{\Scribtexttt{x}}}\Scribtexttt{4}\mbox{\hphantom{\Scribtexttt{x}}}}}\mbox{\hphantom{\Scribtexttt{x}}}\RktMeta{}\mbox{\hphantom{\Scribtexttt{xxxx}}}\RktMeta{}\RktMeta{for}\RktMeta{}\mbox{\hphantom{\Scribtexttt{x}}}\RktMeta{}\RktPn{(}\RktMeta{i}\mbox{\hphantom{\Scribtexttt{x}}}\RktMeta{}\RktPn{=}\RktMeta{}\mbox{\hphantom{\Scribtexttt{x}}}\RktMeta{}\RktVal{0}\RktPn{;}\RktMeta{}\mbox{\hphantom{\Scribtexttt{x}}}\RktMeta{i}\mbox{\hphantom{\Scribtexttt{x}}}\RktMeta{}\RktPn{{\Stttextless}}\RktMeta{}\mbox{\hphantom{\Scribtexttt{x}}}\RktMeta{n}\RktPn{;}\RktMeta{}\mbox{\hphantom{\Scribtexttt{x}}}\RktMeta{i}\RktPn{+}\RktPn{+}\RktPn{)}\RktMeta{}\mbox{\hphantom{\Scribtexttt{x}}}\RktMeta{}\RktPn{{\char`\{}}\RktMeta{}

\Smaller{\Smaller{\mbox{\hphantom{\Scribtexttt{x}}}\Scribtexttt{5}\mbox{\hphantom{\Scribtexttt{x}}}}}\mbox{\hphantom{\Scribtexttt{x}}}\RktMeta{}\mbox{\hphantom{\Scribtexttt{xxxxxxxx}}}\RktMeta{switch}\mbox{\hphantom{\Scribtexttt{x}}}\RktMeta{}\RktPn{(}\RktMeta{val{\char`\_}type}\RktPn{)}\RktMeta{}\mbox{\hphantom{\Scribtexttt{x}}}\RktMeta{}\RktPn{{\char`\{}}\RktMeta{}\mbox{\hphantom{\Scribtexttt{xx}}}\RktMeta{}\RktPn{{\hbox{\texttt{.}}}}\RktPn{{\hbox{\texttt{.}}}}\RktPn{{\hbox{\texttt{.}}}}\RktMeta{}\mbox{\hphantom{\Scribtexttt{x}}}\RktMeta{}\RktPn{{\char`\}}}\RktMeta{}

\Smaller{\Smaller{\mbox{\hphantom{\Scribtexttt{x}}}\Scribtexttt{6}\mbox{\hphantom{\Scribtexttt{x}}}}}\mbox{\hphantom{\Scribtexttt{x}}}\RktMeta{}\mbox{\hphantom{\Scribtexttt{xxxxxxxx}}}\RktMeta{}\RktPn{[}\RktPn{{\hbox{\texttt{.}}}}\RktPn{{\hbox{\texttt{.}}}}\RktPn{{\hbox{\texttt{.}}}}\RktPn{]}\RktMeta{}

\Smaller{\Smaller{\mbox{\hphantom{\Scribtexttt{x}}}\Scribtexttt{7}\mbox{\hphantom{\Scribtexttt{x}}}}}\mbox{\hphantom{\Scribtexttt{x}}}\RktMeta{}\mbox{\hphantom{\Scribtexttt{xxxxxxxx}}}\RktMeta{R{\char`\_}AddMark}\RktPn{(}\RktMeta{FOR}\RktPn{,}\RktMeta{}\mbox{\hphantom{\Scribtexttt{x}}}\RktMeta{ANTIMARK}\RktPn{,}\RktMeta{}\mbox{\hphantom{\Scribtexttt{x}}}\RktMeta{}\RktVal{TRUE}\RktPn{)}\RktPn{;}\RktMeta{}

\Smaller{\Smaller{\mbox{\hphantom{\Scribtexttt{x}}}\Scribtexttt{8}\mbox{\hphantom{\Scribtexttt{x}}}}}\mbox{\hphantom{\Scribtexttt{x}}}\RktMeta{}\mbox{\hphantom{\Scribtexttt{xxxxxxxx}}}\RktMeta{eval}\RktPn{(}\RktMeta{body}\RktPn{,}\RktMeta{}\mbox{\hphantom{\Scribtexttt{x}}}\RktMeta{rho}\RktPn{)}\RktPn{;}\RktMeta{}

\Smaller{\Smaller{\mbox{\hphantom{\Scribtexttt{x}}}\Scribtexttt{9}\mbox{\hphantom{\Scribtexttt{x}}}}}\mbox{\hphantom{\Scribtexttt{x}}}\RktMeta{}\mbox{\hphantom{\Scribtexttt{xxxxxxxx}}}\RktMeta{R{\char`\_}AddMark}\RktPn{(}\RktMeta{FOR}\RktPn{,}\RktMeta{}\mbox{\hphantom{\Scribtexttt{x}}}\RktMeta{call}\RktPn{,}\RktMeta{}\mbox{\hphantom{\Scribtexttt{x}}}\RktMeta{}\RktVal{TRUE}\RktPn{)}\RktPn{;}\RktMeta{}

\Smaller{\Smaller{\Scribtexttt{10}\mbox{\hphantom{\Scribtexttt{x}}}}}\mbox{\hphantom{\Scribtexttt{x}}}\RktMeta{}\mbox{\hphantom{\Scribtexttt{xxxx}}}\RktMeta{}\RktPn{{\char`\}}}\RktMeta{}

\Smaller{\Smaller{\Scribtexttt{11}\mbox{\hphantom{\Scribtexttt{x}}}}}\mbox{\hphantom{\Scribtexttt{x}}}\RktMeta{}\mbox{\hphantom{\Scribtexttt{xxxx}}}\RktMeta{}\RktPn{[}\RktPn{{\hbox{\texttt{.}}}}\RktPn{{\hbox{\texttt{.}}}}\RktPn{{\hbox{\texttt{.}}}}\RktPn{]}\RktMeta{}

\Smaller{\Smaller{\Scribtexttt{12}\mbox{\hphantom{\Scribtexttt{x}}}}}\mbox{\hphantom{\Scribtexttt{x}}}\RktMeta{}\mbox{\hphantom{\Scribtexttt{xxxx}}}\RktMeta{return}\mbox{\hphantom{\Scribtexttt{x}}}\RktMeta{R{\char`\_}NilValue}\RktPn{;}\RktMeta{}

\Smaller{\Smaller{\Scribtexttt{13}\mbox{\hphantom{\Scribtexttt{x}}}}}\mbox{\hphantom{\Scribtexttt{x}}}\RktMeta{}\RktPn{{\char`\}}}\RktMeta{}\end{SingleColumn}\end{RktBlk}\end{SCodeFlow}\end{FigureInside}\end{Centerfigure}

\Centertext{\Legend{\FigureTarget{\label{t:x28counter_x28x22figurex22_x22forx2dloopx2dimplementationx22x29x29}\textsf{Fig.}~\textsf{20}. }{t:x28counter_x28x22figurex22_x22forx2dloopx2dimplementationx22x29x29}\textsf{For{-}loop implementation with marks (excerpt)}}}\end{Figure}

\Ssubsubsectionstarx{Continuation Marks}{Continuation Marks}\label{t:x28part_x22Continuationx5fMarksx22x29}

Although R does not support continuation marking directly, R programs can
inspect and manipulate the call stack. It is possible to extend the frames
in the call stack to support continuation marks with modifications to the
R{'}s engine, namely, by extending frames to store marks in a hash map with
unique keys and multiple payloads; by teaching the garbage collector how to
track these maps; and by adding primitives to add and inspect continuation
marks.

The capability to add marks to the stack must be accessible from both R and
C, as R features are written in both languages. While supporting
continuation marks does add to the complexity of the R code base, that
complexity is localized. Marks also do not affect the performance of
programs when they are disabled.\NoteBox{\NoteContent{With our modifications, R can be
compiled with and without continuation marks. While this may seem like a
questionable design, it is actually a standard practice for many R
tools\Autobibref{~(\hyperref[t:x28autobib_x22Florxe9al_Morandatx2c_Brandon_Hillx2c_Leo_Osvaldx2c_and_Jan_VitekEvaluating_the_Design_of_the_R_LanguageIn_Procx2e_European_Conference_on_Objectx2dOriented_Programming2012httpsx3ax2fx2fdoix2eorgx2f10x2e1007x2f978x2d3x2d642x2d31057x2d7x5f6x22x29]{\AutobibLink{Morandat et al\Sendabbrev{.}}} \hyperref[t:x28autobib_x22Florxe9al_Morandatx2c_Brandon_Hillx2c_Leo_Osvaldx2c_and_Jan_VitekEvaluating_the_Design_of_the_R_LanguageIn_Procx2e_European_Conference_on_Objectx2dOriented_Programming2012httpsx3ax2fx2fdoix2eorgx2f10x2e1007x2f978x2d3x2d642x2d31057x2d7x5f6x22x29]{\AutobibLink{2012}})}.}}

The API for continuation marks in R is similar to its Racket variant:

\noindent \begin{itemize}\atItemizeStart

\item \RktMeta{add}\RktPn{{\hbox{\texttt{.}}}}\RktMeta{mark}\RktPn{(}\RktMeta{key}\RktPn{,}\RktMeta{}\mbox{\hphantom{\Scribtexttt{x}}}\RktMeta{value}\RktPn{)}\RktMeta{}, which imperatively adds
(\RktMeta{key},\RktMeta{value}) to the call stack.

\item \RktMeta{marks}\RktPn{(}\RktMeta{key}\RktPn{)}\RktMeta{}, which walks the call stack and
retrieves all marks that match \RktMeta{key}.\end{itemize}

\noindent The API for Racket and R differ in primarily one
aspect. The function to add a mark in Racket takes an
expression, which is missing in the R variant. Unlike in
Racket, \RktMeta{add}\RktPn{{\hbox{\texttt{.}}}}\RktMeta{mark} places the continuation mark on
the stack; the mark is implicitly removed when the current stack frame
is popped.

R features that are implemented in C use the \RktMeta{R{\char`\_}AddMark}
and \RktMeta{R{\char`\_}Marks} functions to manipulate continuation marks.
These functions behave identically to their R equivalents.
As an example, figure~\hyperref[t:x28counter_x28x22figurex22_x22forx2dloopx2dimplementationx22x29x29]{\FigureRef{20}{t:x28counter_x28x22figurex22_x22forx2dloopx2dimplementationx22x29x29}} shows
the marks in R{'}s implementation of \RktMeta{}\RktMeta{for}\RktMeta{}. The modified implementation
places a mark at the beginning of the loop and replaces it
with an antimark when the call to \RktMeta{eval} begins executing
the loop{'}s body. Once finished, the run{-}time removes the
frame for \RktMeta{do}\RktPn{{-}}\RktMeta{for}\RktMeta{} from the call stack, which also removes the mark.

\Ssubsubsectionstarx{Sampling Profiler}{Sampling Profiler}\label{t:x28part_x22rx2dsamplingx2dprofx22x29}

Our prototype profiler uses Rprof, which is R{'}s built{-}in sampling profiler.
This profiler uses Unix interrupts to sample the call stack during
execution. These samples are written to a file for post{-}processing.  We
modified Rprof to capture marks in addition to local variables To enables
continuation marks, one must set \RktMeta{marks}\RktPn{{\hbox{\texttt{.}}}}\RktMeta{profiling}, as shown in
figure~\hyperref[t:x28counter_x28x22figurex22_x22loopingx2dconstructsx2dsamplex22x29x29]{\FigureRef{19}{t:x28counter_x28x22figurex22_x22loopingx2dconstructsx2dsamplex22x29x29}}. Modifying Rprof to track
continuation marks rather than using R{'}s native stack inspection mechanism
allows programmers to use other Rprof features, such as disabling the
profiler during portions of the computation.

\begin{Figure}\begin{Centerfigure}\begin{FigureInside}\raisebox{-1.0489583333333425bp}{\makebox[392.00000000000006bp][l]{\includegraphics[trim=2.4000000000000004 2.4000000000000004 2.4000000000000004 2.4000000000000004]{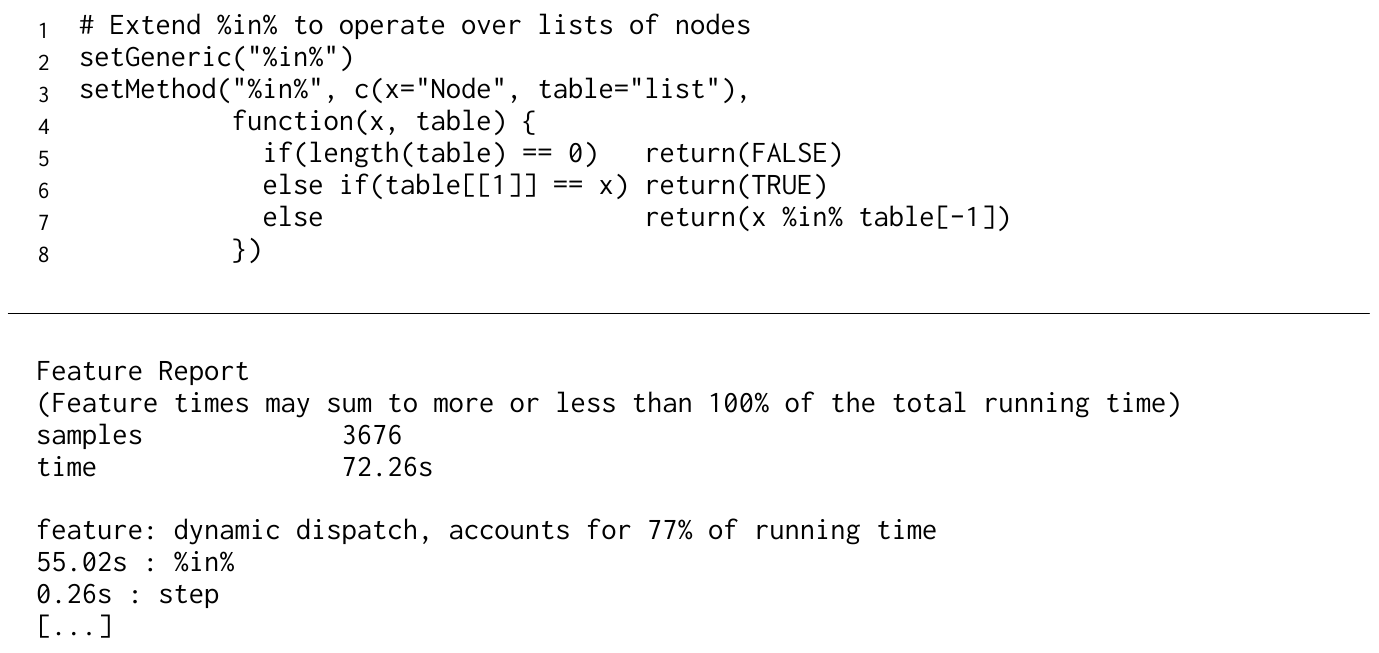}}}\end{FigureInside}\end{Centerfigure}

\Centertext{\Legend{\FigureTarget{\label{t:x28counter_x28x22figurex22_x22dynamicx2ddispatchx2dsamplex22x29x29}\textsf{Fig.}~\textsf{21}. }{t:x28counter_x28x22figurex22_x22dynamicx2ddispatchx2dsamplex22x29x29}\textsf{Dynamic Dispatch (top) and profile output (excerpt, bottom)}}}\end{Figure}

\Ssubsubsectionstarx{Analysis Pass}{Analysis Pass}\label{t:x28part_x22Analysisx5fPassx22x29}

Similar to the analysis pass in Racket, the R analysis pass shows four
pieces of information: (1) the execution time; (2) number of samples
collected; (3) a detailed list of every feature under analysis; (4) as well
as the time spent in that feature and its instances.  Programmers run the
analysis pass by giving the Rprof trace to the \RktMeta{feature}\RktPn{{\hbox{\texttt{.}}}}\RktMeta{profile} function,
as shown in figure~\hyperref[t:x28counter_x28x22figurex22_x22loopingx2dconstructsx2dsamplex22x29x29]{\FigureRef{19}{t:x28counter_x28x22figurex22_x22loopingx2dconstructsx2dsamplex22x29x29}} line 9. Processing each
feature happens again in the same three steps that the Racket analysis
performs.
Figure~\hyperref[t:x28counter_x28x22figurex22_x22dynamicx2ddispatchx2dsamplex22x29x29]{\FigureRef{21}{t:x28counter_x28x22figurex22_x22dynamicx2ddispatchx2dsamplex22x29x29}} shows a report.  It presents the cost
dynamic dispatch for one of R{'}s object systems. The analysis lists feature
instances by method name rather than the source location.  The data is
particularly interesting because, like behavioral contracts, dynamic
dispatch has dispersed costs. The source of dynamic dispatch is where the
method definition is, but the cost manifests itself at the method{'}s call
sites. Because the continuation mark payloads store the name of the method,
we can attribute the cost of dynamic dispatch to the proper
source.

\Ssubsection{Use Cases}{Use Cases}\label{t:x28part_x22rx2dcorpusx22x29}

Next we present four small case studies of features that demonstrate how our
profiler can help programmers. The case studies range over a wide spectrum
of features: dynamic dispatch, parameter{-}naming function applications,
copy{-}on{-}write parameter passing, and vector
subsetting\Autobibref{~(\hyperref[t:x28autobib_x22Hadley_WickhamAdvanced_RFirst_editionx2e_Chapman_and_Hallx2fCRC2014httpx3ax2fx2fadvx2drx2ehadx2ecox2enzx2fx22x29]{\AutobibLink{Wickham}} \hyperref[t:x28autobib_x22Hadley_WickhamAdvanced_RFirst_editionx2e_Chapman_and_Hallx2fCRC2014httpx3ax2fx2fadvx2drx2ehadx2ecox2enzx2fx22x29]{\AutobibLink{2014}})}.\NoteBox{\NoteContent{Called slicing in other languages.}}

\begin{Figure}\begin{Centerfigure}\begin{FigureInside}\raisebox{-0.5864583333333369bp}{\makebox[392.00000000000006bp][l]{\includegraphics[trim=2.4000000000000004 2.4000000000000004 2.4000000000000004 2.4000000000000004]{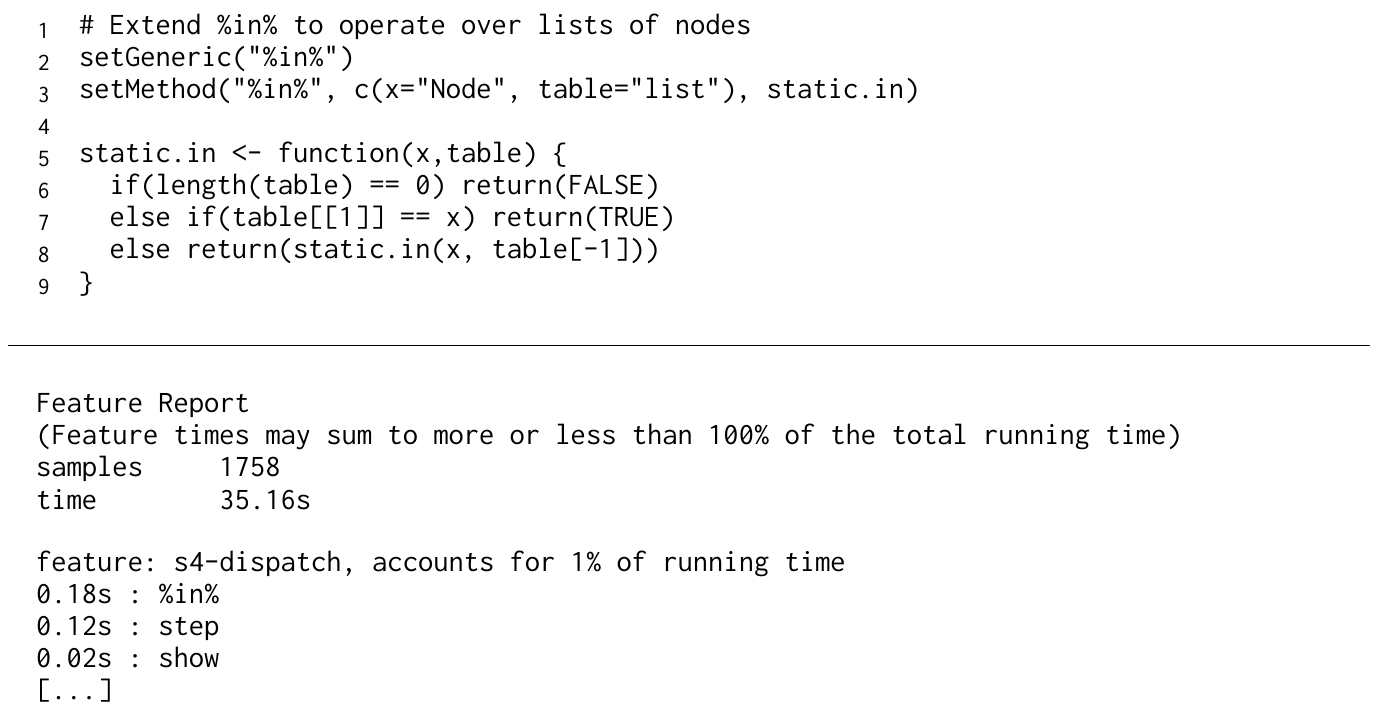}}}\end{FigureInside}\end{Centerfigure}

\Centertext{\Legend{\FigureTarget{\label{t:x28counter_x28x22figurex22_x22dynamicx2ddispatchx2dfixedx2dsamplex22x29x29}\textsf{Fig.}~\textsf{22}. }{t:x28counter_x28x22figurex22_x22dynamicx2ddispatchx2dfixedx2dsamplex22x29x29}\textsf{Dynamic Dispatch (fixed, top) and profile output (excerpt, bottom)}}}\end{Figure}

\Ssubsubsectionstarx{Dynamic Dispatch}{Dynamic Dispatch}\label{t:x28part_x22Dynamicx5fDispatchx22x29}

R{'}s S4 object system supports multiple dispatch.  Any R function, including
primitives, can be transformed into the default implementation of an S4
method. When a method is called, it executes the implementation whose
arguments best match the parameter types. The run{-}time system calls the
default version of the function if no arguments match the required input
types.

Figure~\hyperref[t:x28counter_x28x22figurex22_x22dynamicx2ddispatchx2dsamplex22x29x29]{\FigureRef{21}{t:x28counter_x28x22figurex22_x22dynamicx2ddispatchx2dsamplex22x29x29}} depicts the method \RktMeta{}\RktPn{\%in\%}\RktMeta{}, used here
as a part of Kruskal{'}s algorithm to find a minimum spanning tree of a
graph. This version uses dynamic dispatch recursively until it finds the
desired node or the list is empty. The variant of this code in
figure~\hyperref[t:x28counter_x28x22figurex22_x22dynamicx2ddispatchx2dfixedx2dsamplex22x29x29]{\FigureRef{22}{t:x28counter_x28x22figurex22_x22dynamicx2ddispatchx2dfixedx2dsamplex22x29x29}} uses dynamic dispatch \textit{once}
and thereafter calls a static function. Both variants of this method have
equivalent behavior when the list is a homogeneous list of nodes.  The
recursive use of dynamic dispatch causes the first definition to be slower
than the second.  Conventional profilers identify the use of dynamic
dispatch as having a major performance impact in the program.
Unfortunately, they cannot identify which specific use of dynamic dispatch
is causing the performance problems, as they point to the S4 implementation
but do not trace the costs back to calls.  A feature{-}specific profile, as
shown in figure~\hyperref[t:x28counter_x28x22figurex22_x22dynamicx2ddispatchx2dsamplex22x29x29]{\FigureRef{21}{t:x28counter_x28x22figurex22_x22dynamicx2ddispatchx2dsamplex22x29x29}}, not only identifies dynamic
dispatch as a major problem in the program, but it also points to the
\RktMeta{}\RktPn{\%in\%}\RktMeta{} method as the culprit

\begin{Figure}\begin{Centerfigure}\begin{FigureInside}\raisebox{-0.26770833333337096bp}{\makebox[398.00000000000006bp][l]{\includegraphics[trim=2.4000000000000004 2.4000000000000004 2.4000000000000004 2.4000000000000004]{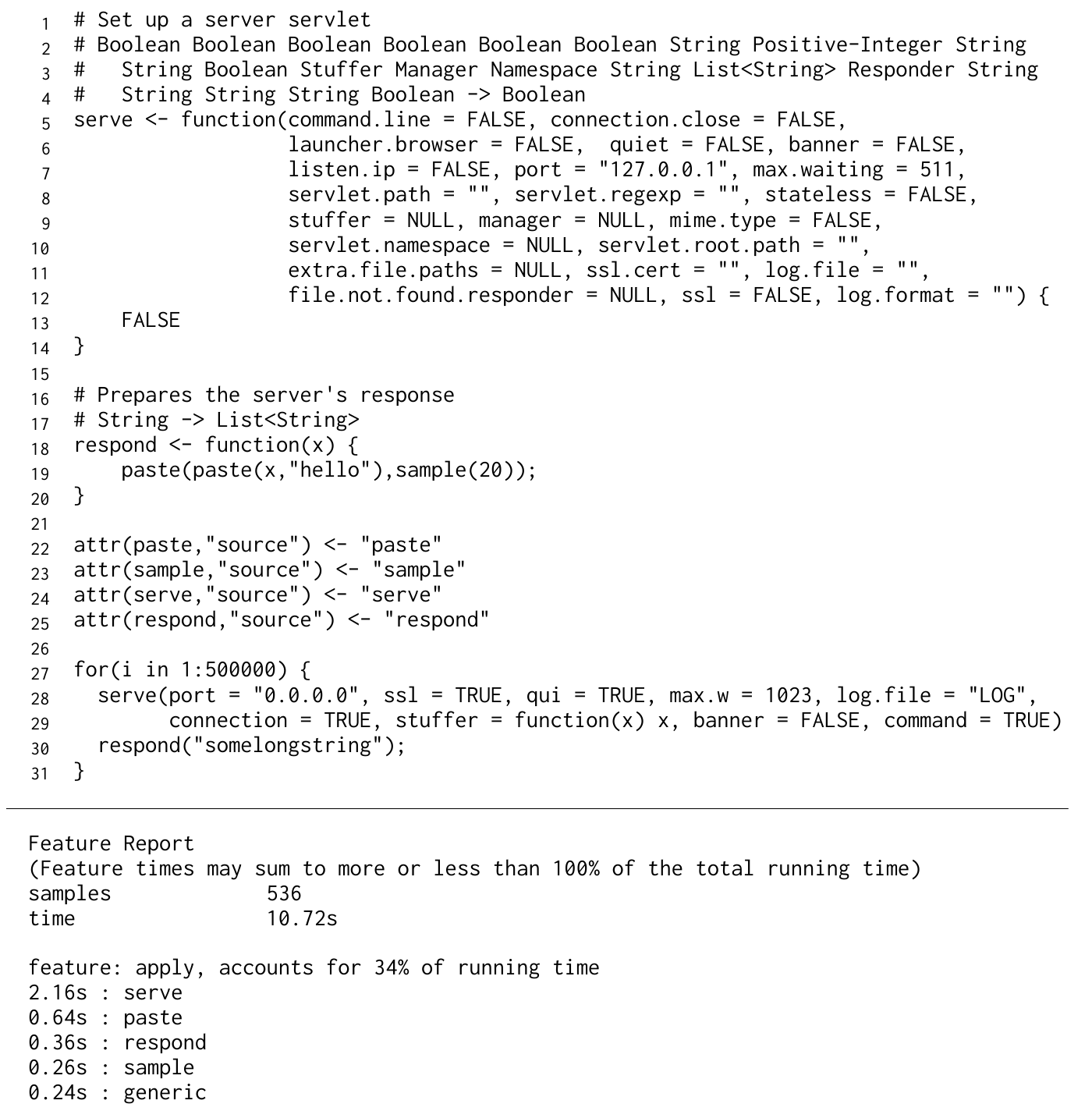}}}\end{FigureInside}\end{Centerfigure}

\Centertext{\Legend{\FigureTarget{\label{t:x28counter_x28x22figurex22_x22functionx2dapplicationx2dsamplex22x29x29}\textsf{Fig.}~\textsf{23}. }{t:x28counter_x28x22figurex22_x22functionx2dapplicationx2dsamplex22x29x29}\textsf{Function Application (top) and Profile Output (bottom)}}}\end{Figure}

\Ssubsubsectionstarx{Function Application}{Function Application}\label{t:x28part_x22Functionx5fApplicationx22x29}

Function calls in R may use named arguments in addition to traditional
positional arguments. Named arguments at call sites are matched with named
parameters. When a function is called and an argument is passed with a name,
the argument is bound to the parameter whose name has the longest matching
prefix of the name given for the argument. Thus, every function used with
named arguments must perform run{-}time string comparisons. Additionally, such
a function application succeeds even if the number of arguments does not
coincide with the number of parameters. Execution halts only when a
parameter without a value is evaluated.  As a result, function calls are
difficult to optimize, and thus programmers consider them to be slow. An
profiler can help identify which function calls cause the most runtime
overhead and which are not cause for concern.

Figure~\hyperref[t:x28counter_x28x22figurex22_x22functionx2dapplicationx2dsamplex22x29x29]{\FigureRef{23}{t:x28counter_x28x22figurex22_x22functionx2dapplicationx2dsamplex22x29x29}} shows the skeleton of two
functions: \RktMeta{serve} and \RktMeta{respond}. The former has a computationally
simple and fast function body compared with a complicated slow calling
interface. The latter has a complicated and slow function body but fast and
simple calling interface. Traditional profilers find similar execution times
for each function, because the combined running time of both the function
body and calling interface are the same. While both timings are similar, \RktMeta{serve} spends more time in the calling interface than required. As shown in
figure~\hyperref[t:x28counter_x28x22figurex22_x22functionx2dapplicationx2dsamplex22x29x29]{\FigureRef{23}{t:x28counter_x28x22figurex22_x22functionx2dapplicationx2dsamplex22x29x29}}, our profiler identifies the
primary bottleneck for \RktMeta{serve}{'}s calling interface. Thus, the program{'}s
performance can be improved by inlining \RktMeta{serve} or simplifiying its
interface, which programmers can do in response to the FSP{'}s actionable
report.

\Ssubsubsectionstarx{Copy{-}on{-}Write}{Copy{-}on{-}Write}\label{t:x28part_x22Copyx2donx2dWritex22x29}

Conceptually, the semantics of R requires a deep copy of every argument
passed into a function. In reality, the implementation only duplicates
objects when absolutely necessary.  Operations such as mutation force the
duplication, creating copies. If no such operation occurs, then objects
are never duplicated. This so{-}called copy{-}on{-}write policy can lead
to unpredictable performance effects.

\begin{Figure}\begin{Centerfigure}\begin{FigureInside}\raisebox{-1.1177083333333369bp}{\makebox[392.00000000000006bp][l]{\includegraphics[trim=2.4000000000000004 2.4000000000000004 2.4000000000000004 2.4000000000000004]{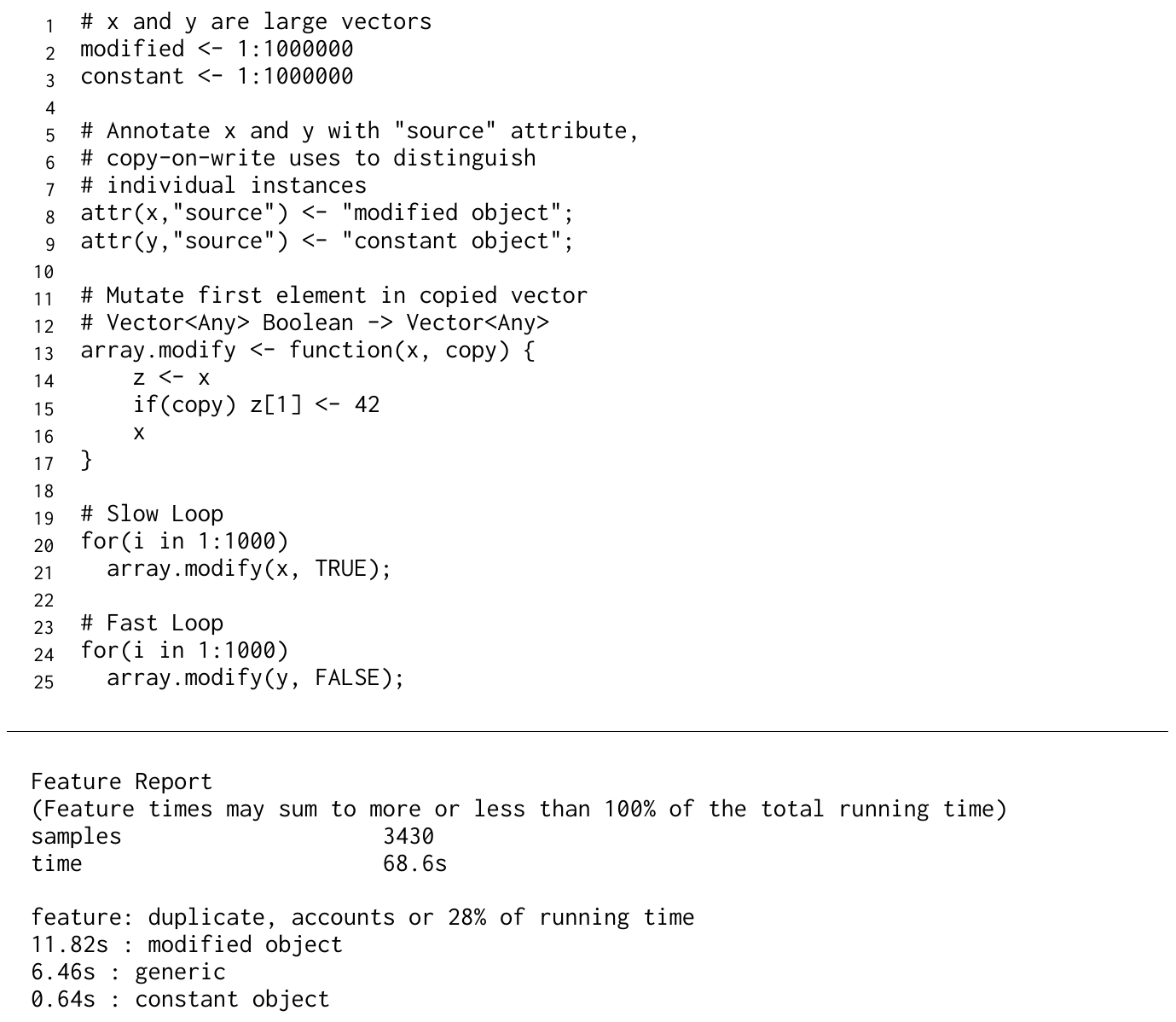}}}\end{FigureInside}\end{Centerfigure}

\Centertext{\Legend{\FigureTarget{\label{t:x28counter_x28x22figurex22_x22copyx2donx2dwritex2dsamplex22x29x29}\textsf{Fig.}~\textsf{24}. }{t:x28counter_x28x22figurex22_x22copyx2donx2dwritex2dsamplex22x29x29}\textsf{Copy{-}on{-}Write (top) and profile output (bottom)}}}\end{Figure}

The \RktMeta{array}\RktPn{{\hbox{\texttt{.}}}}\RktMeta{duplicate} function in figure~\hyperref[t:x28counter_x28x22figurex22_x22copyx2donx2dwritex2dsamplex22x29x29]{\FigureRef{24}{t:x28counter_x28x22figurex22_x22copyx2donx2dwritex2dsamplex22x29x29}}
illustrates the surprising impact of copy{-}on{-}write.  It duplicates the
vector only if the second parameter is true. The program has two loops: a
slow loop that causes the duplication of the array and a fast loop that does
not duplicate the array. Traditional profilers correctly identify
\RktMeta{array}\RktPn{{\hbox{\texttt{.}}}}\RktMeta{duplicate} as a bottleneck. Our profiler identifies array
duplication as the problem and furthermore identifies the duplication of a
specific vector.

\Ssubsubsectionstarx{Vector Subset}{Vector Subset}\label{t:x28part_x22Vectorx5fSubsetx22x29}

Vectors are the basic data structures in R. Even a number such as 42 is a
vector, which allows functions to operate over both vectors and other
objects seamlessly. The vector{-}subset feature retrieves elements from a
vector based on a vector of indices.  Subset occurs frequently and some of
their uses are more expensive than others.  The syntax for subset uses
square brackets, similar to array indexing. Traditional indexing is a
special case of subsetting where the argument is a singleton vector. For
example, the expression \RktMeta{c}\RktPn{(}\RktVal{2}\RktPn{,}\RktVal{4}\RktPn{,}\RktVal{6}\RktPn{)}\RktPn{[}\RktVal{2}\RktPn{]}\RktMeta{}, which uses the function \RktMeta{c} to
create a vector, evaluates to \RktMeta{}\RktVal{4}\RktMeta{}.

Figure~\hyperref[t:x28counter_x28x22figurex22_x22vectorx2dsubsetx2dsamplex22x29x29]{\FigureRef{25}{t:x28counter_x28x22figurex22_x22vectorx2dsubsetx2dsamplex22x29x29}} shows a code snippet with two subset
operations. The first retrieves every second element from the given
vector. The other retrieves every third element; it occurs roughly one
fourth as often as the first. Traditional profilers identify vector
subsetting as the primary bottleneck in the program. Unfortunately, these
profilers point to the implementation of subset, which is not enough
information to identify which subset operation is costly.  Our profiler
instead indicates that the first subset operation is the primary cost center.

\Ssubsection{Profiling Overhead}{Profiling Overhead}\label{t:x28part_x22rx2devalx22x29}

Figure~\hyperref[t:x28counter_x28x22figurex22_x22rx2dperformancex22x29x29]{\FigureRef{26}{t:x28counter_x28x22figurex22_x22rx2dperformancex22x29x29}} reports the overhead  our prototype imposes
on several benchmarks. These results are the mean of 30 executions on a
machine running OS X Yosemite with a 4 core Intel Core i7 clocked at 2.5 GHz
and 16 GB of 1600 MHz DDR3 ram. The error bars show the 95\% confidence
interval. The samples are collected with R build r69166,\NoteBox{\NoteContent{\href{https://github.com/LeifAndersen/R}{\Snolinkurl{https://github.com/LeifAndersen/R}}}} and the sampling interval is 20ms.

The benchmark programs are described in figure~\hyperref[t:x28counter_x28x22figurex22_x22rx2dperformancex22x29x29]{\FigureRef{26}{t:x28counter_x28x22figurex22_x22rx2dperformancex22x29x29}}.
They include two benchmarks from the
\href{http://benchmarksgame.alioth.debian.org/}{Computer Language
Benchmark Game} that use features our prototypes supports, the
five feature samples used earlier in the paper, and Oliver Keyes{'}s
{``}\href{https://github.com/oliver-papers/GoingPostel}{GoingPostel}{''}, a
program that aggregates information about IETF RFCs.

\begin{Figure}\begin{Centerfigure}\begin{FigureInside}\raisebox{-0.7427083333333369bp}{\makebox[392.00000000000006bp][l]{\includegraphics[trim=2.4000000000000004 2.4000000000000004 2.4000000000000004 2.4000000000000004]{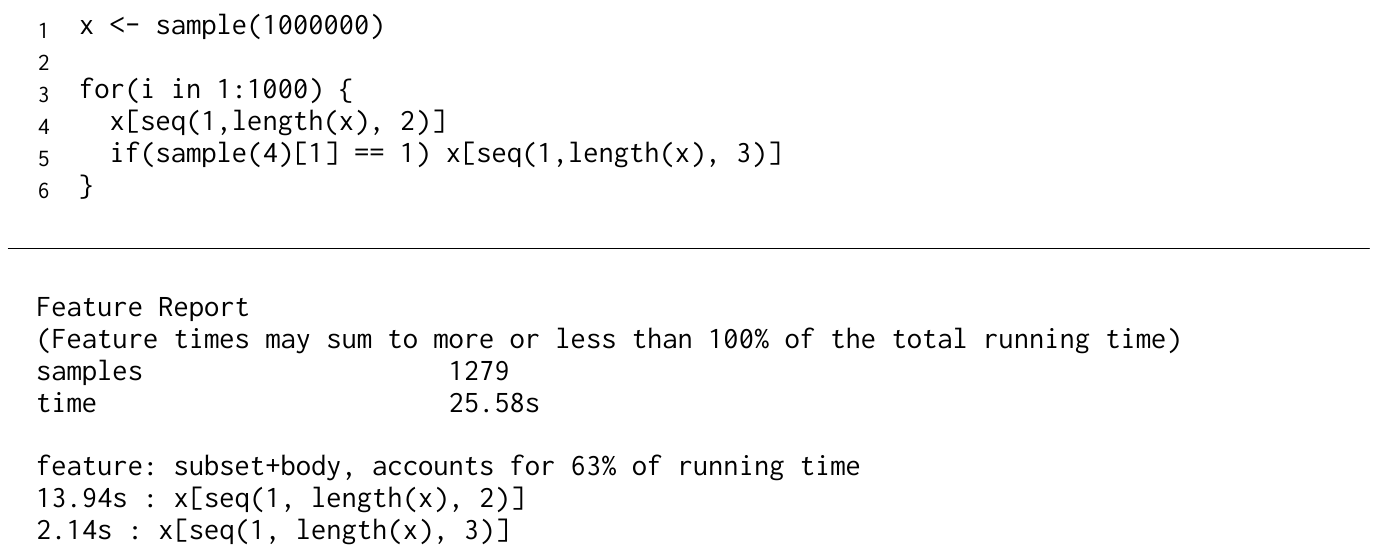}}}\end{FigureInside}\end{Centerfigure}

\Centertext{\Legend{\FigureTarget{\label{t:x28counter_x28x22figurex22_x22vectorx2dsubsetx2dsamplex22x29x29}\textsf{Fig.}~\textsf{25}. }{t:x28counter_x28x22figurex22_x22vectorx2dsubsetx2dsamplex22x29x29}\textsf{Vector Subset (top) and profile output (bottom)}}}\end{Figure}

We report runs of each program in three configurations:

\noindent \begin{itemize}\atItemizeStart

\item The first configuration corresponds to the program executing without
continuation marks or  profiler
in a build of R with all required packages installed.

\item The second configuration corresponds to the program executing in a
build of R with  continuation marks.
All of features that our profiler supports annotate
the stack with continuation marks, but the sampling is turned off.

\item The third configuration is like the second, but with profiling turned on.\end{itemize}

\begin{Figure}\begin{Centerfigure}\begin{FigureInside}\raisebox{-0.9156249999999773bp}{\makebox[392.00000000000006bp][l]{\includegraphics[trim=2.4000000000000004 2.4000000000000004 2.4000000000000004 2.4000000000000004]{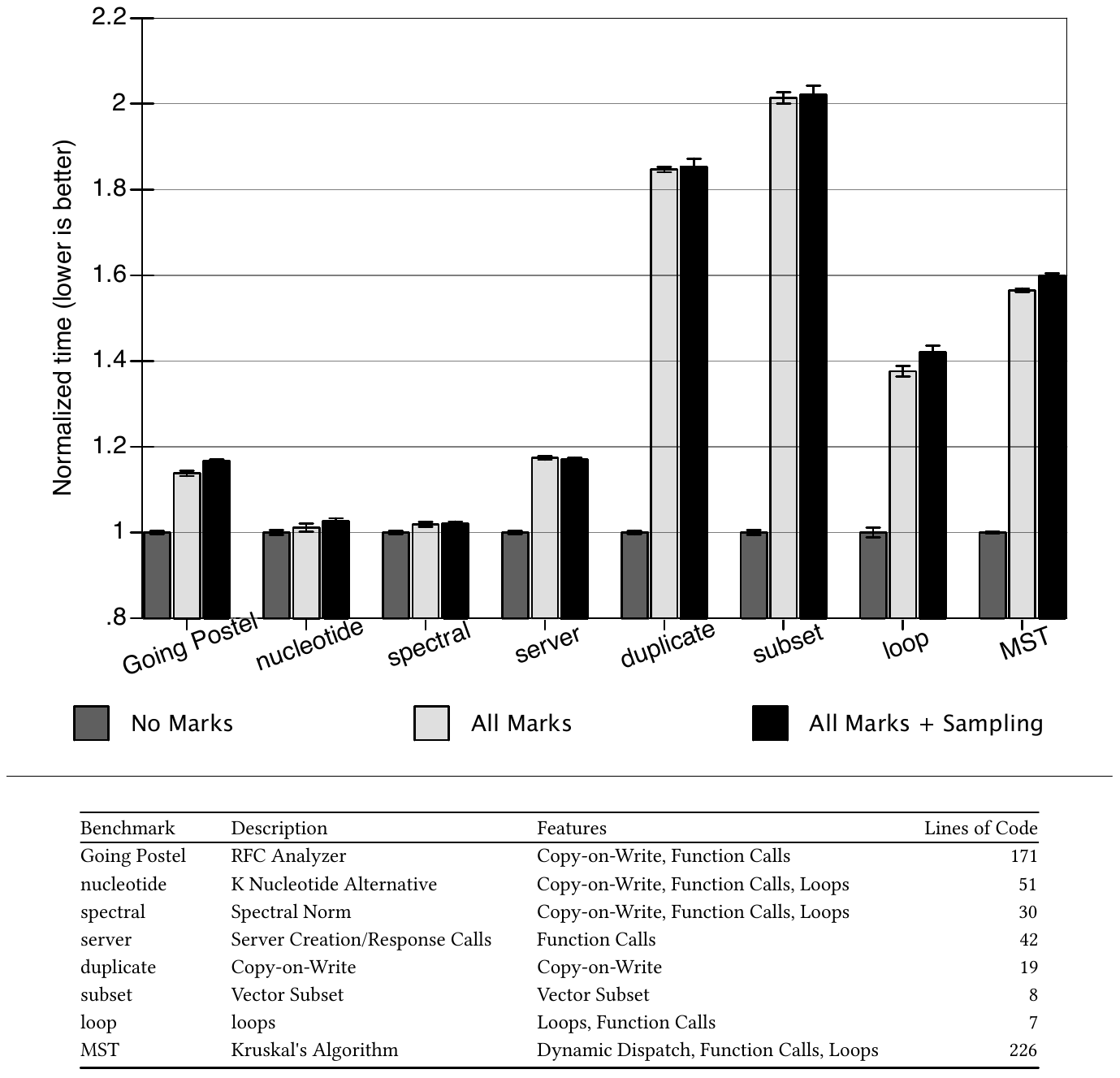}}}\end{FigureInside}\end{Centerfigure}

\Centertext{\Legend{\FigureTarget{\label{t:x28counter_x28x22figurex22_x22rx2dperformancex22x29x29}\textsf{Fig.}~\textsf{26}. }{t:x28counter_x28x22figurex22_x22rx2dperformancex22x29x29}\textsf{Instrumentation and Sampling Performance of the Going Postel (Left),Computer Language Benchmark Game Benchmarks (Center),and Feature Samples (Right)}}}\end{Figure}

With continuation marks and profiling, the overhead is lower than 20\% for
half of the programs and larger for the other half (85\%, 100\%, 42\%, and
59\%).  The latter four programs, however, are feature samples, which
essentially perform no work except exercise the relevant feature, and
therefore represent pathological worst cases.  In all cases the cost of
sampling is less than 2\%.  The primary cause of overhead comes from
continuation marks rather than the modified sampling profiler.  A threat to
validity comes from the fact that continuation mark overhead is concentrated
at feature annotations, which causes features to appear slower than they
are, thus skewing results. Nevertheless, we consider this experiment to
validate the viability of feature{-}specific profiling. While the overheads
are greater than in Racket, performance of the R profiler remains
acceptable. We conjecture that this prototype could be improved to match the
performance of the Racket implementation with careful tuning of the
implementation.

\sectionNewpage

\Ssection{Limitations}{Limitations}\label{t:x28part_x22limitationsx22x29}

Our approach to feature{-}specific profiling applies to some linguistic
features.  This section discusses limitations. We believe they are not
fundamental to the idea of feature{-}specific profiling and that they could be
addressed by different approaches to data gathering.

Because our instrumentation strategy relies on continuation marks, it does
not support features that interfere with marks. This rules out non{-}local
control features that unroll the stack, e.g. exception raising. This also
prevents us from profiling continuation marks themselves.

\identity{~\\}

The sampler must be able to observe a feature in order to profile it.  This
rules out uninterruptible features, e.g., allocation or FFI calls, which do
not allow the sampling thread to be scheduled during their execution.  Other
obstacles to observability include sampling
bias\Autobibref{~(\hyperref[t:x28autobib_x22Todd_Mytkowiczx2c_Amer_Diwanx2c_Matthias_Hauswirthx2c_and_Peter_Fx2e_SweeneyEvaluating_the_accuracy_of_Java_profilersIn_Procx2e_Programming_Langauges_Design_and_Implementationx2c_ppx2e_187x2dx2d1972010x22x29]{\AutobibLink{Mytkowicz et al\Sendabbrev{.}}} \hyperref[t:x28autobib_x22Todd_Mytkowiczx2c_Amer_Diwanx2c_Matthias_Hauswirthx2c_and_Peter_Fx2e_SweeneyEvaluating_the_accuracy_of_Java_profilersIn_Procx2e_Programming_Langauges_Design_and_Implementationx2c_ppx2e_187x2dx2d1972010x22x29]{\AutobibLink{2010}})} and instances that execute too quickly to
be sampled reliably.

Some non{-}syntactic language features, such as garbage collection, have costs
that cannot be attributed to a single source location in the program.
Frequently, these features have costs that are small and spread out, and are
thus difficult to capture with a sampling profiler.  An event{-}based
approach, such as \hyperref[t:x28autobib_x22Florxe9al_Morandatx2c_Brandon_Hillx2c_Leo_Osvaldx2c_and_Jan_VitekEvaluating_the_Design_of_the_R_LanguageIn_Procx2e_European_Conference_on_Objectx2dOriented_Programming2012httpsx3ax2fx2fdoix2eorgx2f10x2e1007x2f978x2d3x2d642x2d31057x2d7x5f6x22x29]{\AutobibLink{Morandat et al\Sendabbrev{.}}}{'}s (\hyperref[t:x28autobib_x22Florxe9al_Morandatx2c_Brandon_Hillx2c_Leo_Osvaldx2c_and_Jan_VitekEvaluating_the_Design_of_the_R_LanguageIn_Procx2e_European_Conference_on_Objectx2dOriented_Programming2012httpsx3ax2fx2fdoix2eorgx2f10x2e1007x2f978x2d3x2d642x2d31057x2d7x5f6x22x29]{\AutobibLink{2012}}), would fare better.

While our prototype profiles concurrent programs such as the Marketplace
described in \ChapRef{\SectionNumberLink{t:x28part_x22richx22x29}{5}}{Profiling Complex Features}, it cannot handle parallel programs.  We
conjecture that our approach could be extended to handle multi{-}threaded
programs but we have not tried.

Features have both direct costs and indirect costs. Direct costs come from
using a feature, while indirect costs are not imposed by the feature itself
but by lost opportunities due to a feature{'}s use. Profiliers only track
direct costs.

Finally, it is up to the feature authors to work out the
correctness of their annotations. While feature authors can
clearly make mistakes when annotating their libraries, in
our experience and that of our users, we have not found this
to be an issue at all. Because authors are familiar with
their libraries, they also tend to have a reasonable idea of
where adding annotations will be \textit{useful}.

\sectionNewpage

\Ssection{Related Work}{Related Work}\label{t:x28part_x22relatedx22x29}

Programmers already have access to a wide variety of
complementary performance tools. This section compares
feature{-}specific profiling to those approaches that are
closely related.

Profilers have been successfully used to diagnose performance issues for
decades.  They most commonly report on the consumption of time, space and
I/O resources.  Traditional profilers group costs according to program
organization, be it static{---}e.g., per function definition{---}or
dynamic{---}e.g., per HTTP request.  Each of these views is useful in
different contexts. For example, a feature{-}specific profiler{'}s view is most useful when
non{-}local feature costs make up a significant portion of a program{'}s running
time.  In contrast, traditional profilers may detect a broader range of
issues than feature{-}specific profilers, such as inefficient algorithms, which are
invisible to feature{-}specific profilers.

A vertical profiler\Autobibref{~(\hyperref[t:x28autobib_x22Matthias_Hauswirthx2c_Peter_Fx2e_Sweeneyx2c_Amer_Diwanx2c_and_Michael_HindVertical_profilingIn_Procx2e_Objectx2doriented_Programmingx2c_Systemsx2c_Languagesx2c_and_Applicationsx2c_ppx2e_251x2dx2d2692004x22x29]{\AutobibLink{Hauswirth et al\Sendabbrev{.}}} \hyperref[t:x28autobib_x22Matthias_Hauswirthx2c_Peter_Fx2e_Sweeneyx2c_Amer_Diwanx2c_and_Michael_HindVertical_profilingIn_Procx2e_Objectx2doriented_Programmingx2c_Systemsx2c_Languagesx2c_and_Applicationsx2c_ppx2e_251x2dx2d2692004x22x29]{\AutobibLink{2004}})} attempts to see through the use of
high{-}level language features. It therefore gathers information from
multiple layers{---}hardware performance counters, operating system, virtual
machine, libraries{---}and correlates them into a gestalt of performance.
Vertical profiling focuses on helping programmers understand how the
interaction between different layers of abstraction affects their program{'}s performance.
By comparison, feature{-}specific profiling focuses on helping them understand the cost of features
per se.
Feature{-}specific profiling also presents information in terms of features and feature instances,
which is accessible to non{-}expert programmers, whereas vertical profilers
report low{-}level information, which requires some understanding of the
compiler and run{-}time system.
Hauswirth et al.{'}s work introduces the notion of \textit{software performance
monitors}, which are analogous to hardware performance monitors but record
software{-}related performance events.
These monitors could possibly be used to implement feature{-}specific profiling by tracking the
execution of feature code.

\goAway{\Autobibref{~(\hyperref[t:x28autobib_x22Jeremy_Singer_and_Chris_KirkhamDynamic_analysis_of_Java_program_concepts_for_visualization_and_profilingScience_of_Computer_Programming_70x282x2d3x29x2c_ppx2e_111x2dx2d1262008x22x29]{\AutobibLink{Singer and Kirkham}} \hyperref[t:x28autobib_x22Jeremy_Singer_and_Chris_KirkhamDynamic_analysis_of_Java_program_concepts_for_visualization_and_profilingScience_of_Computer_Programming_70x282x2d3x29x2c_ppx2e_111x2dx2d1262008x22x29]{\AutobibLink{2008}})}}
\goAway{\Autobibref{~(\hyperref[t:x28autobib_x22Juan_Mx2e_Tamayox2c_Alex_Aikenx2c_Nathan_Bronsonx2c_and_Mooly_SagivUnderstanding_the_behavior_of_database_operations_under_program_controlIn_Procx2e_Objectx2doriented_Programmingx2c_Systemsx2c_Languagesx2c_and_Applicationsx2c_ppx2e_983x2dx2d9962012x22x29]{\AutobibLink{Tamayo et al\Sendabbrev{.}}} \hyperref[t:x28autobib_x22Juan_Mx2e_Tamayox2c_Alex_Aikenx2c_Nathan_Bronsonx2c_and_Mooly_SagivUnderstanding_the_behavior_of_database_operations_under_program_controlIn_Procx2e_Objectx2doriented_Programmingx2c_Systemsx2c_Languagesx2c_and_Applicationsx2c_ppx2e_983x2dx2d9962012x22x29]{\AutobibLink{2012}})}}
A number of profilers offer alternative views to the traditional attribution of
time costs to program locations.
Most of these views focus on particular aspects of program performance and are
complementary to the view offered by a feature{-}specific profiler.
Some recent examples include
Singer and Kirkham{'}s (2008) profiler, which assigns
costs to programmer{-}annotated code regions, listener latency
profiling\Autobibref{~(\hyperref[t:x28autobib_x22Milan_Jovic_and_Matthias_HauswirthListener_latency_profilingScience_of_Computer_Programming_19x284x29x2c_ppx2e_1054x2dx2d10722011x22x29]{\AutobibLink{Jovic and Hauswirth}} \hyperref[t:x28autobib_x22Milan_Jovic_and_Matthias_HauswirthListener_latency_profilingScience_of_Computer_Programming_19x284x29x2c_ppx2e_1054x2dx2d10722011x22x29]{\AutobibLink{2011}})}, which reports high{-}latency
operations, and Tamayo et al.{'}s (2012) tool, which provides
information about the cost of database operations.
One notable example, MAJOR\Autobibref{~(\hyperref[t:x28autobib_x22Walter_Binderx2c_Danilo_Ansalonix2c_Alex_Villazxf3nx2c_and_Philippe_MoretFlexible_and_efficient_profiling_with_aspectx2doriented_programmingIn_Procx2e_Concurrency_and_Computationx3a_Practice_and_Experiencex2c_ppx2e_1749x2dx2d17732011httpsx3ax2fx2fdoix2eorgx2f10x2e1002x2fcpex2e1760x22x29]{\AutobibLink{Binder et al\Sendabbrev{.}}} \hyperref[t:x28autobib_x22Walter_Binderx2c_Danilo_Ansalonix2c_Alex_Villazxf3nx2c_and_Philippe_MoretFlexible_and_efficient_profiling_with_aspectx2doriented_programmingIn_Procx2e_Concurrency_and_Computationx3a_Practice_and_Experiencex2c_ppx2e_1749x2dx2d17732011httpsx3ax2fx2fdoix2eorgx2f10x2e1002x2fcpex2e1760x22x29]{\AutobibLink{2011}})}, uses Aspect
Oriented Programming with inter{-}advice communication to
create these complementary views.

Dynamic instrumentation frameworks such as
Valgrind\Autobibref{~(\hyperref[t:x28autobib_x22Nicholas_Nethercote_and_Julian_SewardValgrindx3a_A_framework_for_heavyweight_dynamic_binaryx5cninstrumentationIn_Procx2e_Programming_Langauges_Design_and_Implementation2007httpsx3ax2fx2fdoix2eorgx2f10x2e1145x2f1273442x2e1250746x22x29]{\AutobibLink{Nethercote and Seward}} \hyperref[t:x28autobib_x22Nicholas_Nethercote_and_Julian_SewardValgrindx3a_A_framework_for_heavyweight_dynamic_binaryx5cninstrumentationIn_Procx2e_Programming_Langauges_Design_and_Implementation2007httpsx3ax2fx2fdoix2eorgx2f10x2e1145x2f1273442x2e1250746x22x29]{\AutobibLink{2007}})} or
Javana\Autobibref{~(\hyperref[t:x28autobib_x22Jonas_Maebex2c_Dries_Buytaertx2c_Lieven_Eeckhoutx2c_and_Koen_De_BosschereJavanax3a_A_System_for_Building_Customized_Java_Program_Analysis_ToolsIn_Procx2e_Objectx2doriented_Programmingx2c_Systemsx2c_Languagesx2c_and_Applications2006httpsx3ax2fx2fdoix2eorgx2f10x2e1145x2f1167515x2e1167487x22x29]{\AutobibLink{Maebe et al\Sendabbrev{.}}} \hyperref[t:x28autobib_x22Jonas_Maebex2c_Dries_Buytaertx2c_Lieven_Eeckhoutx2c_and_Koen_De_BosschereJavanax3a_A_System_for_Building_Customized_Java_Program_Analysis_ToolsIn_Procx2e_Objectx2doriented_Programmingx2c_Systemsx2c_Languagesx2c_and_Applications2006httpsx3ax2fx2fdoix2eorgx2f10x2e1145x2f1167515x2e1167487x22x29]{\AutobibLink{2006}})}
serve as the basis for profilers and other kinds of performance tools.
These frameworks resemble the use of continuation marks in our framework and
could potentially be used to build feature{-}specific profilers.
These frameworks are much more heavy{-}weight than continuation marks and, in turn,
allow more thorough instrumentation, e.g., of the memory
hierarchy, of hardware performance counters, etc. They have not been
used to measure the cost of individual linguistic features.

Like a feature{-}specific profiler, an optimization coach\Autobibref{~(\hyperref[t:x28autobib_x22Vincent_Stx2dAmourx2c_Sam_Tobinx2dHochstadtx2c_and_Matthias_FelleisenOptimization_coachingx3a_optimizers_learn_to_communicate_with_programmersIn_Procx2e_Objectx2doriented_Programmingx2c_Systemsx2c_Languagesx2c_and_Applicationsx2c_ppx2e_163x2dx2d1782012x22x29]{\AutobibLink{St{-}Amour et al\Sendabbrev{.}}} \hyperref[t:x28autobib_x22Vincent_Stx2dAmourx2c_Sam_Tobinx2dHochstadtx2c_and_Matthias_FelleisenOptimization_coachingx3a_optimizers_learn_to_communicate_with_programmersIn_Procx2e_Objectx2doriented_Programmingx2c_Systemsx2c_Languagesx2c_and_Applicationsx2c_ppx2e_163x2dx2d1782012x22x29]{\AutobibLink{2012}})} focuses on
enabling compiler optimizations through a feedback loop that involves the
developer. The two are complementary.  Optimization coaches operate at
compile time whereas feature{-}specific profilers, like other profilers, operate at run
time.  Because of this, feature{-}specific profilers require representative program input
to operate, whereas coaches do not.  Then again, by having access to run
time data, feature{-}specific profilers can target actual program hot spots, while existing
optimization coaches must rely on static heuristics to prioritize reports.

An important tool for measuring R programs is \Scribtexttt{tracemem}.  It is included
with the R tool suite, but requires programmers to rebuild R.  This tool
serves to track uses of copy{-}on{-}write during the execution of R programs.
It tracks the memory that is being copied, and the source location that is
responsible for causing the copy.  Also, it allows programmers to tag
individual objects they care about tracking, while ignoring everything else.

\sectionNewpage

\Ssection{Conclusion}{Conclusion}\label{t:x28part_x22conclusionx22x29}

\notitlesection\Sincsubsection\label{t:x28part_x22x22x29}Feature{-}specific profiling is a novel profiling technique
that supplements traditional cost{-}centers with
language{-}specific ones. These cost centers give a new
perspective on program performance, enabling developers to
tune their programs. Feature{-}specific profiling is
especially useful when programs use language features with
dispersed or non{-}local costs. Additionally, feature{-}specific
profiling is useful with languages that allow for the
programmatic creation of new features such as Racket, R, or
even C++. The implementation of a feature{-}specific profiler
is straightforward. If the host language supports stack
annotations and inspection, such as Racket, then
implementing is as simple as that of a sampling profiler.
Languages without this support, such as R, must be extended
by adding stack annotations. This paper shows that
modifications required are practical.

While using a feature{-}specific profiler requires little effort, it does
require more setup than traditional profilers. Either library authors must
add support for their code, or developers must modify the library{'}s source.
Fortunately, adding support is simple and generally requires only a few
lines of code.  The information provided by the profiler has the same
limitations as that of stack{-}based sampling profilers. This means that
language features that do not show up on the call stack cannot be measured.
The sampling nature of our profiler also means that it can only profile
interruptible features. Other profile designs, such as an event based
profiler, trade these limitations for a different set. The idea of
feature{-}specific profiling itself is not limited to the architecture
designed in this paper.  We conjecture that other architectures can also
support feature{-}specific profiling.

\Ssubsubsectionstarx{Acknowledgements}{Acknowledgements}\label{t:x28part_x22Acknowledgementsx22x29}

 Tony
Garnock{-}Jones implemented the Marketplace plug{-}in and helped
with the SSH case study. Stephen Chang assisted with the
Parsack plug{-}in and the Markdown case study. Christos
Dimoulas and Scott Moore collaborated on the Shill plug{-}in
and the grading script experiment. Robby Findler provided
assistance with the contract system. Oliver Keyes
implemented Going Postel. We thank Eli Barzilay, Matthew
Flatt, Asumu Takikawa, Sam Tobin{-}Hochstadt, Benjamin Chung,
Helena Kotthaus, Tomas Kalibera, Oli Fl\"{u}ckiger, Kyle Bemis,
Olga Vitek, and Luke Tierney for helpful discussions. This
work was partially supported by the National Science
Foundation (NSF) under Grants SHF 1544542 and 1518{-}844, as
well as the European Research Council (ERC) under the
European Union{'}s Horizon 2020 research and innovation
program (grant agreement 695412), and finally the Office of
Navel Research (ONR) award 503353. Any opinions, findings,
and conclusions expressed in this material may be those of
the authors and likely do not reflect the views of our
funding agencies.

\sectionNewpage

\Ssectionstarx{Bibliography}{Bibliography}\label{t:x28part_x22docx2dbibliographyx22x29}

\begin{AutoBibliography}\begin{SingleColumn}\label{t:x28autobib_x22Gene_Mx2e_AmdahlValidity_of_the_Single_Processor_Approach_to_Achieving_Large_Scale_Computing_CapabilitiesIn_Procx2e_Spring_Joint_Computer_Conference1967x22x29}\Autobibentry{Gene M. Amdahl. Validity of the Single Processor Approach to Achieving Large Scale Computing Capabilities. In \textit{Proc. Spring Joint Computer Conference}, 1967.}

\label{t:x28autobib_x22Walter_Binderx2c_Danilo_Ansalonix2c_Alex_Villazxf3nx2c_and_Philippe_MoretFlexible_and_efficient_profiling_with_aspectx2doriented_programmingIn_Procx2e_Concurrency_and_Computationx3a_Practice_and_Experiencex2c_ppx2e_1749x2dx2d17732011httpsx3ax2fx2fdoix2eorgx2f10x2e1002x2fcpex2e1760x22x29}\Autobibentry{Walter Binder, Danilo Ansaloni, Alex Villaz\'{o}n, and Philippe Moret. Flexible and efficient profiling with aspect{-}oriented programming. In \textit{Proc. Concurrency and Computation: Practice and Experience}, pp. 1749{--}1773, 2011. \href{https://doi.org/10.1002/cpe.1760}{\Snolinkurl{https://doi.org/10.1002/cpe.1760}}}

\label{t:x28autobib_x22John_Clementsx2c_Matthew_Flattx2c_and_Matthias_FelleisenModeling_an_algebraic_stepperIn_Procx2e_European_Symposium_on_Programmingx2c_ppx2e_320x2dx2d3342001x22x29}\Autobibentry{John Clements, Matthew Flatt, and Matthias Felleisen. Modeling an algebraic stepper. In \textit{Proc. European Symposium on Programming}, pp. 320{--}334, 2001.}

\label{t:x28autobib_x22John_Clementsx2c_Ayswarya_Sundaramx2c_and_David_HermanImplementing_continuation_marks_in_JavaScriptIn_Procx2e_Scheme_and_Functional_Programming_Workshopx2c_ppx2e_1x2dx2d102008x22x29}\Autobibentry{John Clements, Ayswarya Sundaram, and David Herman. Implementing continuation marks in JavaScript. In \textit{Proc. Scheme and Functional Programming Workshop}, pp. 1{--}10, 2008.}

\label{t:x28autobib_x22Rx2e_Kent_DybvigChez_Scheme_Version_8_Userx27s_GuideCadence_Research_Systems2009x22x29}\Autobibentry{R. Kent Dybvig. \textit{Chez Scheme Version 8 User{'}s Guide}. Cadence Research Systems, 2009.}

\label{t:x28autobib_x22Rx2e_Kent_Dybvigx2c_Robert_Hiebx2c_and_Carl_BruggemanSyntax_Abstracton_in_SchemeIn_Procx2e_Lisp_and_Symbolic_Computation1993x22x29}\Autobibentry{R. Kent Dybvig, Robert Hieb, and Carl Bruggeman. Syntax Abstracton in Scheme. In \textit{Proc. Lisp and Symbolic Computation}, 1993.}

\label{t:x28autobib_x22Robert_Bruce_Findlerx2c_John_Clementsx2c_Cormac_Flanaganx2c_Matthew_Flattx2c_Shriram_Krishnamurthix2c_Paul_Stecklerx2c_and_Matthias_FelleisenDrSchemex3a_a_programming_environment_for_SchemeJornal_of_Functional_Programming_12x282x29x2c_ppx2e_159x2dx2d1822002x22x29}\Autobibentry{Robert Bruce Findler, John Clements, Cormac Flanagan, Matthew Flatt, Shriram Krishnamurthi, Paul Steckler, and Matthias Felleisen. DrScheme: a programming environment for Scheme. \textit{Jornal of Functional Programming} 12(2), pp. 159{--}182, 2002.}

\label{t:x28autobib_x22Robert_Bruce_Findler_and_Matthias_FelleisenContracts_for_Higherx2dorder_FunctionsIn_Procx2e_International_Conference_on_Functional_Programming2002httpsx3ax2fx2fdoix2eorgx2f10x2e1145x2f581478x2e581484x22x29}\Autobibentry{Robert Bruce Findler and Matthias Felleisen. Contracts for Higher{-}order Functions. In \textit{Proc. International Conference on Functional Programming}, 2002. \href{https://doi.org/10.1145/581478.581484}{\Snolinkurl{https://doi.org/10.1145/581478.581484}}}

\label{t:x28autobib_x22Matthew_Flatt_and_Eli_BarzilayKeyword_and_Optional_Arguments_in_PLT_SchemeIn_Procx2e_Workshop_on_Scheme_and_Functional_Programming2009x22x29}\Autobibentry{Matthew Flatt and Eli Barzilay. Keyword and Optional Arguments in PLT Scheme. In \textit{Proc. Workshop on Scheme and Functional Programming}, 2009.}

\label{t:x28autobib_x22Matthew_Flatt_and_PLTReferencex3a_RacketPLT_Incx2ex2c_PLTx2dTRx2d2010x2d12010httpx3ax2fx2fracketx2dlangx2eorgx2ftr1x2fx22x29}\Autobibentry{Matthew Flatt and PLT. Reference: Racket. PLT Inc., PLT{-}TR{-}2010{-}1, 2010. \href{http://racket-lang.org/tr1/}{\Snolinkurl{http://racket-lang.org/tr1/}}}

\label{t:x28autobib_x22Tony_Garnockx2dJonesx2c_Sam_Tobinx2dHochstadtx2c_and_Matthias_FelleisenThe_network_as_a_language_constructIn_Procx2e_European_Symposium_on_Programming_Languagesx2c_ppx2e_473x2dx2d4922014x22x29}\Autobibentry{Tony Garnock{-}Jones, Sam Tobin{-}Hochstadt, and Matthias Felleisen. The network as a language construct. In \textit{Proc. European Symposium on Programming Languages}, pp. 473{--}492, 2014.}

\label{t:x28autobib_x22Matthias_Hauswirthx2c_Peter_Fx2e_Sweeneyx2c_Amer_Diwanx2c_and_Michael_HindVertical_profilingIn_Procx2e_Objectx2doriented_Programmingx2c_Systemsx2c_Languagesx2c_and_Applicationsx2c_ppx2e_251x2dx2d2692004x22x29}\Autobibentry{Matthias Hauswirth, Peter F. Sweeney, Amer Diwan, and Michael Hind. Vertical profiling. In \textit{Proc. Object{-}oriented Programming, Systems, Languages, and Applications}, pp. 251{--}269, 2004.}

\label{t:x28autobib_x22Carl_Hewittx2c_Peter_Bishopx2c_and_Richard_SteigerA_Universal_Modular_ACTOR_Formalism_for_Artificial_IntelligenceIn_Procx2e_International_Joint_Conference_on_Artificial_Intelligence1973x22x29}\Autobibentry{Carl Hewitt, Peter Bishop, and Richard Steiger. A Universal Modular ACTOR Formalism for Artificial Intelligence. In \textit{Proc. International Joint Conference on Artificial Intelligence}, 1973.}

\label{t:x28autobib_x22Milan_Jovic_and_Matthias_HauswirthListener_latency_profilingScience_of_Computer_Programming_19x284x29x2c_ppx2e_1054x2dx2d10722011x22x29}\Autobibentry{Milan Jovic and Matthias Hauswirth. Listener latency profiling. \textit{Science of Computer Programming} 19(4), pp. 1054{--}1072, 2011.}

\label{t:x28autobib_x22Jonas_Maebex2c_Dries_Buytaertx2c_Lieven_Eeckhoutx2c_and_Koen_De_BosschereJavanax3a_A_System_for_Building_Customized_Java_Program_Analysis_ToolsIn_Procx2e_Objectx2doriented_Programmingx2c_Systemsx2c_Languagesx2c_and_Applications2006httpsx3ax2fx2fdoix2eorgx2f10x2e1145x2f1167515x2e1167487x22x29}\Autobibentry{Jonas Maebe, Dries Buytaert, Lieven Eeckhout, and Koen De Bosschere. Javana: A System for Building Customized Java Program Analysis Tools. In \textit{Proc. Object{-}oriented Programming, Systems, Languages, and Applications}, 2006. \href{https://doi.org/10.1145/1167515.1167487}{\Snolinkurl{https://doi.org/10.1145/1167515.1167487}}}

\label{t:x28autobib_x22Simon_Marlowx2c_Josxe9_Iborrax2c_Bernard_Popex2c_and_Andy_GillA_lightweight_interactive_debugger_for_HaskellIn_Procx2e_Haskell_Workshopx2c_ppx2e_13x2dx2d242007x22x29}\Autobibentry{Simon Marlow, Jos\'{e} Iborra, Bernard Pope, and Andy Gill. A lightweight interactive debugger for Haskell. In \textit{Proc. Haskell Workshop}, pp. 13{--}24, 2007.}

\label{t:x28autobib_x22Jay_McCarthyThe_twox2dstate_solutionx3a_native_and_serializable_continuations_accordIn_Procx2e_Objectx2doriented_Programmingx2c_Systemsx2c_Languagesx2c_and_Applicationsx2c_ppx2e_567x2dx2d5822010x22x29}\Autobibentry{Jay McCarthy. The two{-}state solution: native and serializable continuations accord. In \textit{Proc. Object{-}oriented Programming, Systems, Languages, and Applications}, pp. 567{--}582, 2010.}

\label{t:x28autobib_x22Scott_Moorex2c_Christos_Dimoulasx2c_Dan_Kingx2c_and_Stephen_ChongSHILLx3a_a_secure_shell_scripting_languageIn_Procx2e_USENIX_Symposium_on_Operating_Systems_Design_and_Implementation2014httpsx3ax2fx2fwwwx2eusenixx2eorgx2fconferencex2fosdi14x2ftechnicalx2dsessionsx2fpresentationx2fmoorex22x29}\Autobibentry{Scott Moore, Christos Dimoulas, Dan King, and Stephen Chong. SHILL: a secure shell scripting language. In \textit{Proc. USENIX Symposium on Operating Systems Design and Implementation}, 2014. \href{https://www.usenix.org/conference/osdi14/technical-sessions/presentation/moore}{\Snolinkurl{https://www.usenix.org/conference/osdi14/technical-sessions/presentation/moore}}}

\label{t:x28autobib_x22Florxe9al_Morandatx2c_Brandon_Hillx2c_Leo_Osvaldx2c_and_Jan_VitekEvaluating_the_Design_of_the_R_LanguageIn_Procx2e_European_Conference_on_Objectx2dOriented_Programming2012httpsx3ax2fx2fdoix2eorgx2f10x2e1007x2f978x2d3x2d642x2d31057x2d7x5f6x22x29}\Autobibentry{Flor\'{e}al Morandat, Brandon Hill, Leo Osvald, and Jan Vitek. Evaluating the Design of the R Language. In \textit{Proc. European Conference on Object{-}Oriented Programming}, 2012. \href{https://doi.org/10.1007/978-3-642-31057-7_6}{\Snolinkurl{https://doi.org/10.1007/978-3-642-31057-7_6}}}

\label{t:x28autobib_x22Todd_Mytkowiczx2c_Amer_Diwanx2c_Matthias_Hauswirthx2c_and_Peter_Fx2e_SweeneyEvaluating_the_accuracy_of_Java_profilersIn_Procx2e_Programming_Langauges_Design_and_Implementationx2c_ppx2e_187x2dx2d1972010x22x29}\Autobibentry{Todd Mytkowicz, Amer Diwan, Matthias Hauswirth, and Peter F. Sweeney. Evaluating the accuracy of Java profilers. In \textit{Proc. Programming Langauges Design and Implementation}, pp. 187{--}197, 2010.}

\label{t:x28autobib_x22Nicholas_Nethercote_and_Julian_SewardValgrindx3a_A_framework_for_heavyweight_dynamic_binaryx5cninstrumentationIn_Procx2e_Programming_Langauges_Design_and_Implementation2007httpsx3ax2fx2fdoix2eorgx2f10x2e1145x2f1273442x2e1250746x22x29}\Autobibentry{Nicholas Nethercote and Julian Seward. Valgrind: A framework for heavyweight dynamic binary
instrumentation. In \textit{Proc. Programming Langauges Design and Implementation}, 2007. \href{https://doi.org/10.1145/1273442.1250746}{\Snolinkurl{https://doi.org/10.1145/1273442.1250746}}}

\label{t:x28autobib_x22Greg_Pettyjohnx2c_John_Clementsx2c_Joe_Marshallx2c_Shriram_Krishnamurthix2c_and_Matthias_FelleisenContinuations_from_generalized_stack_inspectionIn_Procx2e_International_Conference_on_Functional_Programmingx2c_ppx2e_216x2dx2d2272005x22x29}\Autobibentry{Greg Pettyjohn, John Clements, Joe Marshall, Shriram Krishnamurthi, and Matthias Felleisen. Continuations from generalized stack inspection. In \textit{Proc. International Conference on Functional Programming}, pp. 216{--}227, 2005.}

\label{t:x28autobib_x22R_Development_Core_TeamR_Language_DefinitionR_Development_Core_Teamx2c_3x2e3x2e12016httpx3ax2fx2fwebx2emitx2eedux2fx7erx2fcurrentx2farchx2famd64x5flinux26x2flibx2fRx2fdocx2fmanualx2fRx2dlangx2epdfx22x29}\Autobibentry{R Development Core Team. R Language Definition. R Development Core Team, 3.3.1, 2016. \href{http://web.mit.edu/~r/current/arch/amd64_linux26/lib/R/doc/manual/R-lang.pdf}{\Snolinkurl{http://web.mit.edu/~r/current/arch/amd64_linux26/lib/R/doc/manual/R-lang.pdf}}}

\label{t:x28autobib_x22Jeremy_Singer_and_Chris_KirkhamDynamic_analysis_of_Java_program_concepts_for_visualization_and_profilingScience_of_Computer_Programming_70x282x2d3x29x2c_ppx2e_111x2dx2d1262008x22x29}\Autobibentry{Jeremy Singer and Chris Kirkham. Dynamic analysis of Java program concepts for visualization and profiling. \textit{Science of Computer Programming} 70(2{-}3), pp. 111{--}126, 2008.}

\label{t:x28autobib_x22Vincent_Stx2dAmourx2c_Leif_Andersenx2c_and_Matthias_FelleisenFeaturex2dspecific_ProfilingIn_Procx2e_International_Conference_on_Compiler_Construction2015httpsx3ax2fx2fdoix2eorgx2f10x2e1007x2f978x2d3x2d662x2d46663x2d6x5f3x22x29}\Autobibentry{Vincent St{-}Amour, Leif Andersen, and Matthias Felleisen. Feature{-}specific Profiling. In \textit{Proc. International Conference on Compiler Construction}, 2015. \href{https://doi.org/10.1007/978-3-662-46663-6_3}{\Snolinkurl{https://doi.org/10.1007/978-3-662-46663-6_3}}}

\label{t:x28autobib_x22Vincent_Stx2dAmourx2c_Sam_Tobinx2dHochstadtx2c_and_Matthias_FelleisenOptimization_coachingx3a_optimizers_learn_to_communicate_with_programmersIn_Procx2e_Objectx2doriented_Programmingx2c_Systemsx2c_Languagesx2c_and_Applicationsx2c_ppx2e_163x2dx2d1782012x22x29}\Autobibentry{Vincent St{-}Amour, Sam Tobin{-}Hochstadt, and Matthias Felleisen. Optimization coaching: optimizers learn to communicate with programmers. In \textit{Proc. Object{-}oriented Programming, Systems, Languages, and Applications}, pp. 163{--}178, 2012.}

\label{t:x28autobib_x22Juan_Mx2e_Tamayox2c_Alex_Aikenx2c_Nathan_Bronsonx2c_and_Mooly_SagivUnderstanding_the_behavior_of_database_operations_under_program_controlIn_Procx2e_Objectx2doriented_Programmingx2c_Systemsx2c_Languagesx2c_and_Applicationsx2c_ppx2e_983x2dx2d9962012x22x29}\Autobibentry{Juan M. Tamayo, Alex Aiken, Nathan Bronson, and Mooly Sagiv. Understanding the behavior of database operations under program control. In \textit{Proc. Object{-}oriented Programming, Systems, Languages, and Applications}, pp. 983{--}996, 2012.}

\label{t:x28autobib_x22Sam_Tobinx2dHochstadt_and_Matthias_FelleisenThe_design_and_implementation_of_Typed_SchemeIn_Procx2e_Principles_of_Programming_Languagesx2c_ppx2e_395x2dx2d4062008x22x29}\Autobibentry{Sam Tobin{-}Hochstadt and Matthias Felleisen. The design and implementation of Typed Scheme. In \textit{Proc. Principles of Programming Languages}, pp. 395{--}406, 2008.}

\label{t:x28autobib_x22Hadley_WickhamAdvanced_RFirst_editionx2e_Chapman_and_Hallx2fCRC2014httpx3ax2fx2fadvx2drx2ehadx2ecox2enzx2fx22x29}\Autobibentry{Hadley Wickham. Advanced R. First edition. Chapman and Hall/CRC, 2014. \href{http://adv-r.had.co.nz/}{\Snolinkurl{http://adv-r.had.co.nz/}}}\end{SingleColumn}\end{AutoBibliography}

\postDoc
\end{document}